\documentclass{aa}

\usepackage{graphics}
\usepackage{natbib,xtab}

\begin{document}

\title{On the sizes of stellar X-ray coronae}

 \author{J.-U.~Ness\inst{1}
	\and
        M.~G\"udel\inst{2}
	\and
        J.H.M.M.~Schmitt\inst{1}
	\and
        M.~Audard\inst{3}
	\and
	A.~Telleschi\inst{2}
}

\institute{
 Universit\"at Hamburg, Gojenbergsweg 112, D-21029 Hamburg, Germany
 \and
Paul Scherrer Institut, W\"urenlingen \& Villingen, 5232 Villingen PSI, Switzerland
 \and
Columbia Astrophysics Laboratory, 550 West 120th Street, New York, NY 10027, USA
}

\authorrunning{Ness, G\"udel, et al.}
\titlerunning{Sizes of X-ray coronae}

\offprints{J.-U.\ Ness}
\mail{jness@hs.uni-hamburg.de}
\date{Received 25 March 2004 / Accepted 5 July 2004}

\abstract{
Spatial information from stellar X-ray coronae cannot be assessed directly,
but scaling laws from the solar corona make it possible to estimate sizes of
stellar coronae from the physical parameters temperature and density.
While coronal plasma temperatures have long been available, we concentrate
on the newly available density measurements from line fluxes of X-ray lines
measured for a large sample of stellar coronae with the Chandra and XMM-Newton
gratings.
We compiled a set of 64 grating spectra of 42 stellar coronae.
Line counts of strong H-like and He-like ions and Fe\,{\sc xxi} lines were
measured with the CORA single-purpose line fitting tool by \cite{newi02}. 
Densities are estimated from He-like f/i flux ratios of O\,{\sc vii} and
Ne\,{\sc ix} representing the cooler (1-6\,MK) plasma
components. The densities scatter between $\log n_e \approx 9.5-11$ from the
O\,{\sc vii} triplet and between  $\log n_e \approx 10.5-12$ from the Ne\,{\sc
ix} triplet, but we caution that the latter triplet may be biased by
contamination from Fe\,{\sc xix} and Fe\,{\sc xxi} lines. We find
that low-activity stars (as parameterized by the characteristic temperature
derived from H- and He-like line flux ratios) tend to show densities
derived from O\,{\sc vii} of no more than a few times $10^{10}$\,cm$^{-3}$,
whereas no definitive trend is found for the more active stars. Investigating
the densities of the hotter plasma with various Fe\,{\sc xxi} line ratios, we
found that none of the spectra consistently indicates the presence of very high
densities. We argue that our measurements are compatible with the low-density
limit for the respective ratios ($\approx 5\times 10^{12}$\,cm$^{-3}$). These
upper limits are in line with constant pressure in the emitting active regions.

We focus on the commonly used \cite{rtv} scaling law to derive loop lengths
from temperatures and densities assuming loop-like structures as identical
building blocks.
We derive the emitting volumes from direct measurements of ion-specific emission
measures and densities. Available volumes are calculated from the
loop-lengths and stellar radii, and are compared with the emitting volumes to
infer filling factors. For all stages of activity we find similar filling
factors up to 0.1.
}

\maketitle

\section{Introduction}
\label{intro}

 The magnetic outer atmosphere of the Sun, the corona, was recognized in radio
and X-ray emission. While the radio emission is associated with bremsstrahlung
and cyclotron emission from free electrons in the hot plasma, the X-ray
emission is produced by bremsstrahlung and line emission. Stellar coronal
activity is therefore investigated primarily in these two bands of the
electromagnetic spectrum. We focus on the potentials offered by X-ray
spectroscopy of stellar coronae. Systematic measurements with Einstein, ROSAT,
ASCA, and other satellite-based X-ray missions revealed that all late-type
main sequence stars from type M to F have coronal X-ray emission. \cite{schm97}
found X-ray surface fluxes covering four orders of magnitude. Coronal activity
appears thus to be a universal process for cool stars of spectral types F-M,
but no correlation between spectral type and degree of activity could be
established. The only fundamental stellar parameter found to be correlated
with the X-ray luminosity is the rotational velocity $v\sin i$ \citep[and thus
age, e.g.,][]{pal81}, suggesting that a magnetic dynamo process is involved
in producing the X-ray coronae. Some information on the spatial
distribution of coronal plasma was inferred by indirect means such as modeling
of eclipses and rotational modulation
\citep[e.g.,][]{white90,schmkue,gued95,gued03,siark96}, but such analyses can
only be carried out for very special systems with advantageous geometries. X-ray
spectra allow us to gather a more general insight into the physical properties
of a large variety of stellar coronae. The analysis of X-ray coronae has in the
past been possible only with very limited spectral resolution or with low
sensitivity. Plasma densities (as the subject of this work) could not be
measured from these spectra, because information from spectral lines was not
available. Nevertheless, temperature distributions (or emission measure
distributions EMD) and coronal abundances could be estimated from low-resolution
spectra by the application of global fit approaches. A model spectrum
composed of a continuum and all known emission line fluxes formed under assumed 
temperature conditions and with assumed elemental abundances (mostly only scaled
to solar abundances) is iterated using one or two temperature components left
free to vary. The information on bremsstrahlung continuum and emission lines is
extracted from atomic databases containing line emissivities as a function of
plasma temperature under assumptions of solar elemental composition and
collisionally ionized plasma. A spectral model is thus composed as the sum of
bremsstrahlung and all lines formed under the assumption of thermal equilibrium,
and model parameters are the equilibrium temperature and elemental abundances.
The first detailed survey of low resolution X-ray spectra for a large sample of
130 late-type stars ($B-V$ colors redder than 0.0) was presented by
\cite{schm90} using Einstein data. For each spectrum the temperature structure
was obtained with global spectral models. Different approaches were tested
ranging from isothermal plasma with one or two (absorbed) temperature
components to continuous emission measure distributions.
 A much smaller sample concentrating on a sample covering the Sun in time was
analyzed by \cite{guedel97} using ROSAT and ASCA data. Their sample
consisted of nine G stars in different stages of evolution.
From MEKAL and Raymond-Smith models they found that the older stars (with slow
rotation) to contained only a single cool temperature component in the
emission measure distribution while the younger, more active stars had a
bimodal emission measure distribution with a similar cool component
and an additional hot component apparently independent of the cooler component
suggesting an additional heating mechanism for the more active stars. The
hotter temperature component $T_{\rm hot}$ was found to scale with the X-ray
luminosity $L_X$:
\begin{equation}
\label{thot}
L_X\approx 55\,T_{\rm hot}^4 \ \ {\rm (in\ cgs\ units).}
\end{equation}
The other crucial parameter that determines heating, cooling, and geometric
properties of a stellar corona is the electron density. It is responsible
for the time scale of physical changes (e.g., via the sound and Alfv\'en
velocity), and, together with the temperature, controls the emissivity of a
plasma. Since the densities $n_e$ are the missing parameter linking emitting
volumes $V$ and emission measures EM$=n_e^2V$, these measurements are
very important for calculating the sizes of X-ray emitting regions. Scaling laws
derived from the solar corona relate spatial structure to temperatures and
densities. Independently determined densities and temperatures thus make it
possible to access semi-loop lengths. The two structural parameters loop length
and emitting volume can only be obtained with measurements of emission measures,
temperatures, and densities. In contrast to rare and often extreme stellar
systems like eclipsing binaries, density measurements open up a new avenue
to coronal structure
recognition since we can apply them to all stars of sufficient brightness.\\

Despite relatively good spatial resolution with previous X-ray satellites,
the technology of their photon counting detectors reached only moderate
spectral resolution in the X-ray regime. Very few attempts were made to use
dispersive gratings converting spatial resolution into spectral resolution
with the potential to resolve individual emission lines. For grating
spectroscopy sufficient light is needed and it is therefore only feasible for
the brightest sources with long exposure times. In the extreme ultraviolet
range this technique has been successfully applied, e.g., with the EUVE mission.
Many low-temperature Fe lines from stages Fe\,{\sc x} to Fe\,{\sc xvi} can be
measured with EUVE and are sensitive to densities \citep{schmitt94}. Also, at
higher temperatures density-sensitive Fe lines can be measured with EUVE
\citep{dupr93}, but they are only
sensitive for relatively high ranges of density $n_e>10^{11}$\,cm$^{-3}$. A
summary of results from EUVE measurements is presented by \cite{bowyer}.
Since the density information is inferred from mostly weak lines the results
tend to be ambiguous. The hotter plasma regions of Capella and AB\,Dor were
investigated by \cite{dupr93,jsf03} using Fe\,{\sc xix-xxii} line ratios.
Densities as high as $10^{13}$\,cm$^{-3}$ were reported, suggesting very
compact emission regions. However, later Chandra LETGS measurements of Capella
contradict these results \citep{mewe01}. Because of low sensitivity, EUVE
data could only be obtained for some bright sources such as Capella or AB\,Dor.
The apertures of the new missions Chandra and
XMM-Newton are large enough to allow grating spectroscopy for many stellar
coronae, and X-ray spectra of unprecedented spectral resolution are available.
We are able now to measure densities and temperatures from line flux ratios.
In this paper we describe the formalism for calculating densities from selected
emission line fluxes in Sect.~\ref{analysis}. We then present the results from
density-sensitive ratios representative for O\,{\sc vii} and Ne\,{\sc ix}
plasma as well as carbon-like Fe\,{\sc xxi} lines in Sect.~\ref{results}. In
Sect.~\ref{discussion} we discuss our results.

\section{Analysis}
\label{analysis}

\subsection{Density measurement}

Spectroscopic information on coronal plasma densities for
stars other than the Sun first became possible with the advent of high
resolution EUVE spectra \citep[$\lambda/\Delta\lambda \sim\,200$;][]{euve_cal}
that allowed the separation of individual spectral lines.
Even with this resolution the available diagnostics often tended to be
ambiguous because of the poor signal-to-noise ratio of observed spectra or due
to blended lines. Studies of density-sensitive lines of Fe\,{\sc xix}
to Fe\,{\sc xxii} in EUVE spectra of the RS~CVn system Capella revealed some
evidence for high densities of $n_e\sim\,10^{12}$ to $10^{13}$\,cm$^{-3}$
at coronal temperatures near $10^7$\,K reported by \cite{dupr93}, but often
the densities derived from different lines or ions varied greatly
despite the similar formation temperatures. More recent analyses of EUVE
Fe\,{\sc xix-xxii} lines \citep{jsf03} and Chandra HETGS Fe\,{\sc xxi} lines
for AB\,Dor \citep{jsf03AB} also returned high densities $\log n_e>12.3$, but
again, not consistently for all considered lines. Since many of the
lines used in the analyses are intrinsically faint, even the HETGS data suffer
from rather low signal-to-noise. Also the consequences of unidentified blends
can be more severe for faint lines, which could be the reason for the
discrepant densities derived from Fe lines in the same ionization stage.

The spectrometers on board the X-ray telescopes Chandra and XMM-Newton make it
possible
to measure emission lines at wavelengths shorter than 95\,\AA\ with far better
signal to noise and spectral resolution. In particular, the Fe L-shell and
M-shell lines and lines of the He-like ions from carbon to silicon are available
for density measurements. A few of the density-sensitive Fe lines measurable
with EUVE in the 120\,\AA\ range can also be measured with the Low Energy
Transmission Grating Spectrometer (LETGS) on board {\it Chandra} but only upper
limits were found by \cite{mewe01} ($n_e < 2-5 \times 10^{12}$\,cm$^{-3}$).
They point out that the Fe\,{\sc xix} to Fe\,{\sc xxii} line ratios are only
sensitive above $\sim 10^{11}$\,cm$^{-3}$, so that no tracer for low
densities for the hotter plasma component is available, neither with EUVE nor
with {\it Chandra} or XMM-Newton.\\

\subsection{Theoretical background}

\begin{figure*}[!ht]
 \resizebox{\hsize}{!}{\includegraphics{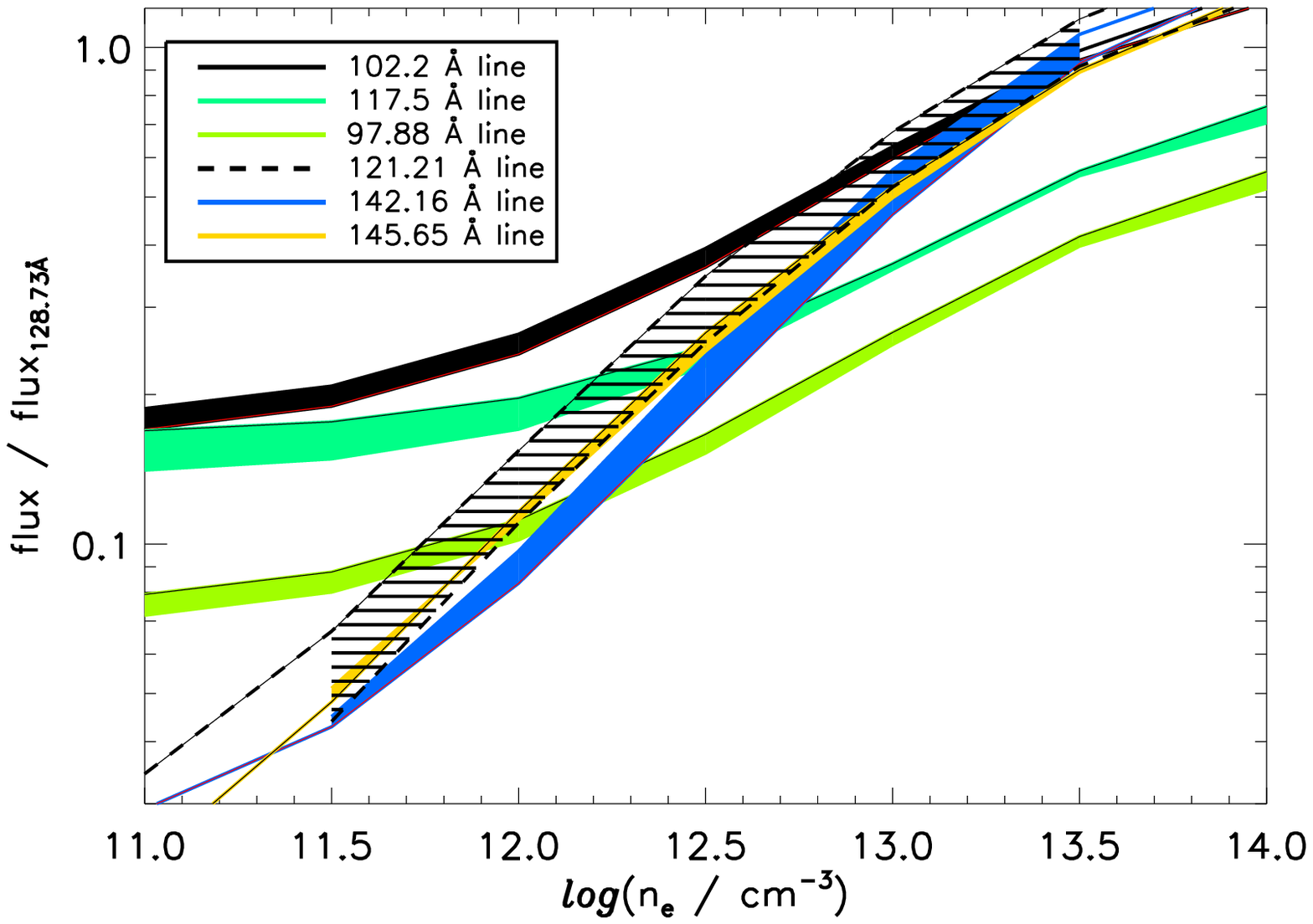}\includegraphics{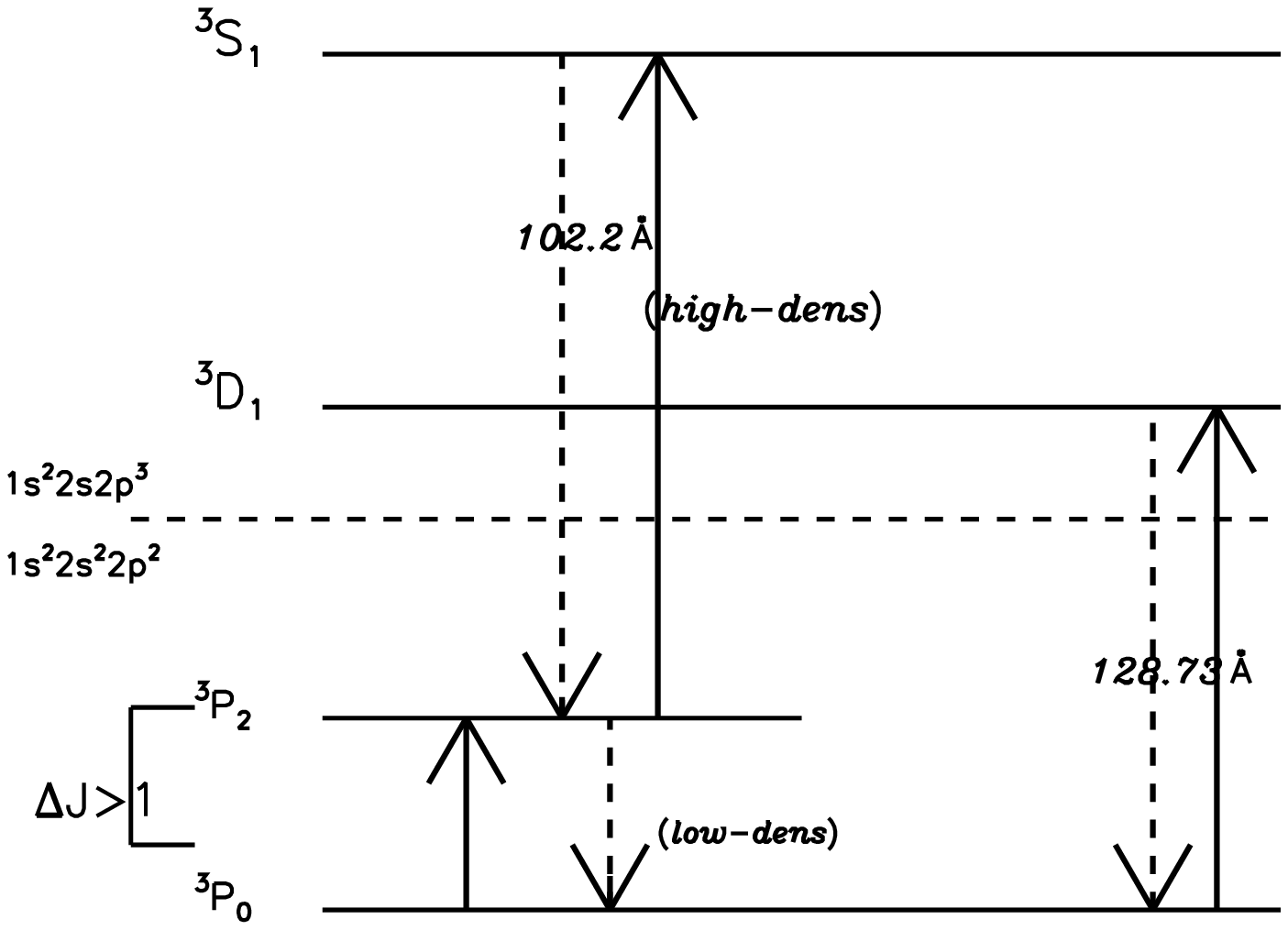}}
\caption[]{\label{fe21mod}Density-sensitive Fe\,{\sc xxi} line flux ratios as
predicted by the APEC database for the temperature range between log$(T)=6.6$
(lower borders) and 7.6 (upper borders). In the right panel we show the term
diagram explaining the formation of the 102.22\,\AA\ line as an example.}
\end{figure*}

The XMM-Newton and {\it Chandra} grating spectra allow high precision
measurements of individual line fluxes and line flux ratios.
Line fluxes are used to compute emission measures in specific lines,
which can be combined to differential emission measure distributions
\citep[e.g.,][]{nebr,abun}. Ratios of certain line fluxes are sensitive to
temperatures or densities and can be used to describe local conditions in
stellar coronae describing the physical conditions in the line-forming regions.
In this paper we analyse the He-like triplets of oxygen and neon that probe the
low-temperature component of the plasma, and four density-sensitive
Fe\,{\sc xxi} lines that probe the hotter component of the plasma in a larger
sample of stars.

\subsubsection{Density diagnostics with carbon-like ions}
\label{fe21_theo}

With six electrons (1s$^2$2s$^2$2p$^2$) the Fe\,{\sc xxi} ion is carbon-like.
Its ground configuration is split into the states $^3$P$_0$ (ground state),
$^3$P$_1$, $^3$P$_2$, $^1$D$_2$, and $^1$S$_0$. Transitions within the
ground configuration, which are naturally forbidden by definition, do occur;
for example, the $^3$P$_1$-$^3$P$_0$ transition is located at 1354\,\AA\
observable with HST \citep[e.g.,][]{hst1354,linsky1354}, the highest energy
transition is $^1$S$_0$-$^3$P$_1$ at
335\,\AA. Fe\,{\sc xxi} at UV and X-ray wavelengths involve transitions
from the $^3$S$_1$ and $^3$D$_1$ states of the excited configuration
1s$^2$2s2p$^3$ (see term diagram in right panel of Fig.~\ref{fe21mod}).

Collisional excitations from the ground state $^3$P$_0$ predominantly occur into
the excited $^3$D$_1$ state, producing the strong extreme ultraviolet (XUV)
Fe\,{\sc xxi} at 128.73\,\AA, which is essentially independent of
density and can be used as a reference line reflecting the Fe abundance and the
ionization balance for the stage Fe\,{\sc xxi}. Collisional excitation from the
ground state into the $^3$S$_1$ state of the excited configuration is much less
likely than collisional excitation from the ``excited'' ground state
$^3$P$_2$-$^3$S$_1$. The population of
the state $^3$S$_1$ of the excited configuration
therefore depends critically on the population of
the state $^3$P$_2$. In a low-density plasma the population of the latter
will be small and eventually decay radiatively into the $^3$P$_0$ level; in
a high density plasma the $^3$P$_2$ state will be collisionally depopulated
into the $^3$S$_1$ level of the excited configuration, which in turn decays
radiatively into $^3$S$_1$--$^3$P$_2$ (102.22\,\AA) or
$^3$S$_1$--$^3$P$_1$ (97.88\,\AA). The latter two transitions are therefore
density-sensitive, because they depend on the density-sensitive population of
the $^3$P$_2$ state. Similar considerations apply to the ground- and excited
levels $^3$P$_1$ and $^3$P$_2$ \citep{mason} leading to transitions at
$^3$P$_1$--$^3$P$_1$ (117.5\,\AA), and $^3$P$_2$--$^3$P$_2$ (121.21\,\AA).
All these lines are in the band pass of the {\it Chandra} LETGS and can be
used to estimate densities of the Fe\,{\sc xxi} emitting plasma at
$T\sim 10$\,MK. In Fig.~\ref{fe21mod} we show predicted line flux ratios
as a function of density \citep[theoretical emissivities were taken from the
APEC database, e.g.,][]{smith01}\footnote{Version 1.2; available at
http://cxc.harvard.edu/atomdb}. Although these lines are all in the same
ionization stage, the line flux ratios depend slightly on the plasma
temperature. This is illustrated by the associated curves for a low temperature
log$(T)=6.6$ and a high temperature log$(T)=7.6$ for each ratio in
Fig.~\ref{fe21mod}. Clearly, temperature primarily affects the low-density
limit. To be conservative we will use the high-temperature theoretical ratios
for comparison with our measured ratios, yielding higher theoretical line flux
ratios.\\

\subsubsection{Density diagnostics with Helium-like ions}

\begin{figure}[!ht]
 \resizebox{\hsize}{!}{\includegraphics{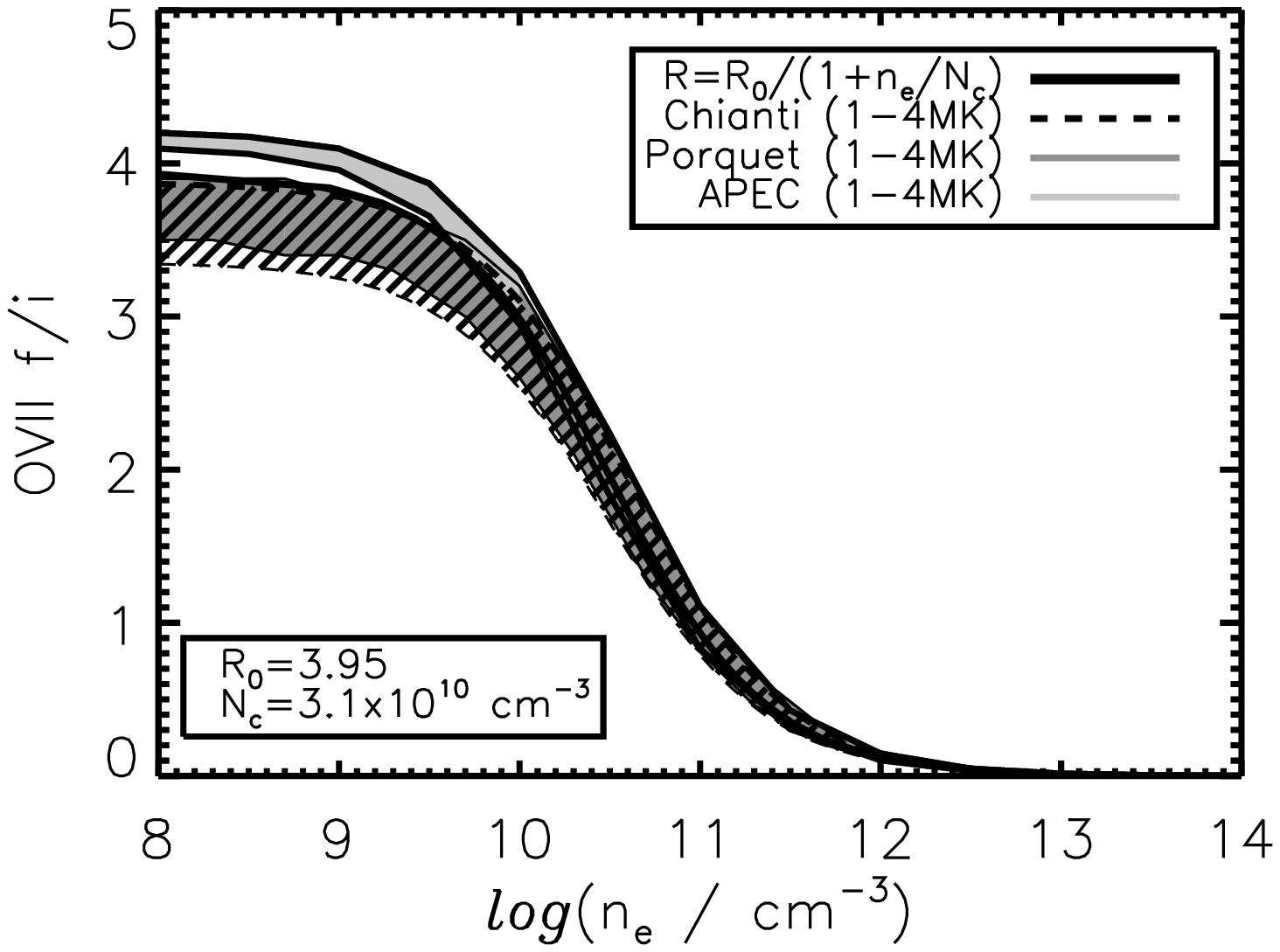}}
 \resizebox{\hsize}{!}{\includegraphics{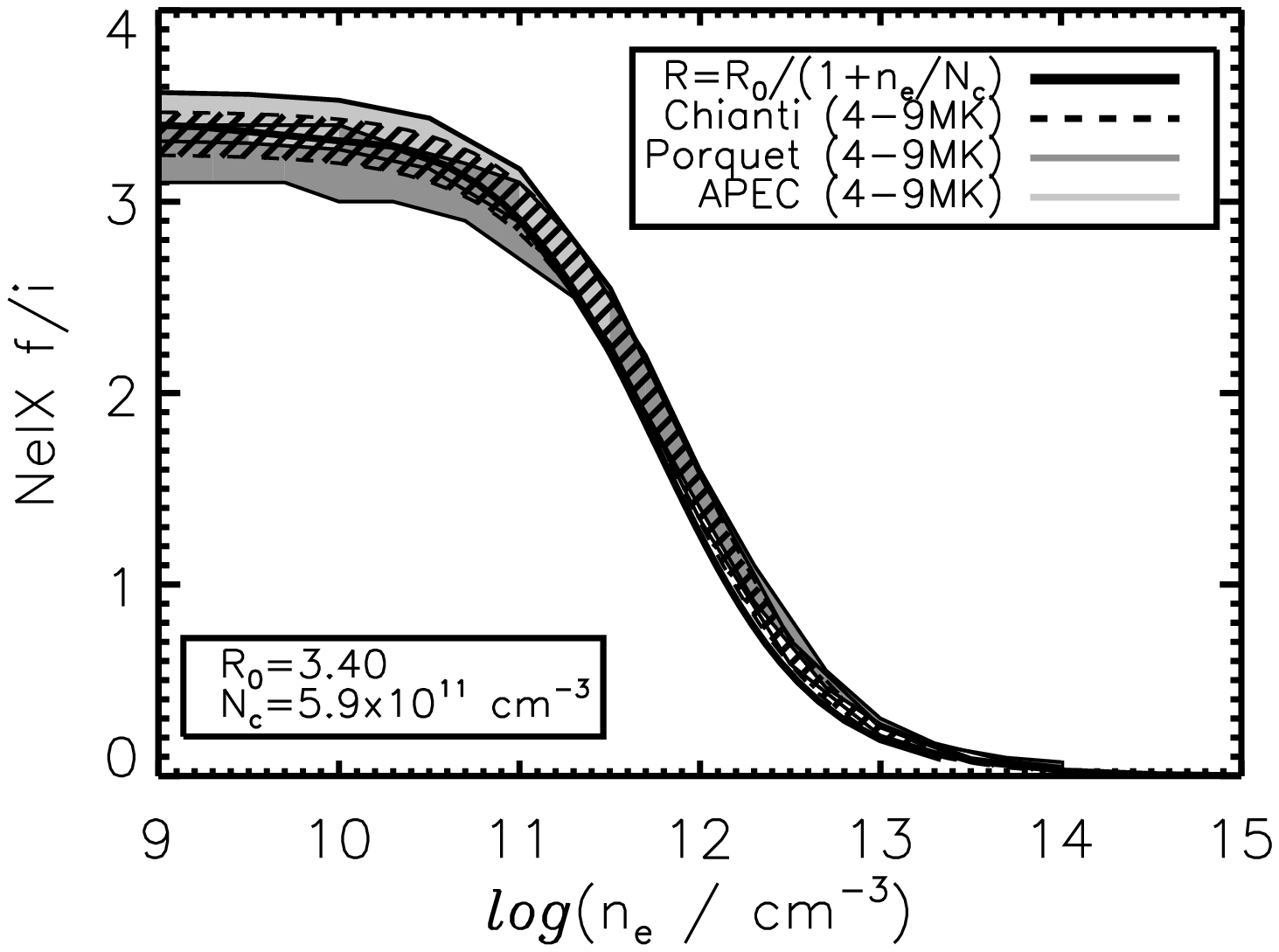}}
\caption[]{\label{fi_cmp}Comparison of theoretical predictions for density
sensitive f/i ratios for O\,{\sc vii} (top panel) and Ne\,{\sc ix} (bottom
panel) from Eq.~(\ref{fidens}), the CHIANTI database, \cite{porq}, and the APEC
database. Filled areas represent the varying electron temperatures as given
in the legends (hatched: CHIANTI, dark-grey shaded: \cite{porq},
light-shaded: APEC). Lower temperatures yield lower f/i ratios.}
\end{figure}

The derivation of densities with He-like triplets originated
in solar observations \citep{gj69}. The excited state transitions
$^1$P$_1$, $^3$P$_1$, and $^3$S$_1$ to the ground state
$^1$S$_0$ are by convention named resonance line (r), intercombination line
(i), and forbidden line (f), respectively. The ratio f/i is density-sensitive
due to collisional excitations $^3$S$_1\rightarrow ^3$P$_1$ in high-density
plasmas. These transitions compete with radiative transitions induced by
possible external radiation sources, namely the stellar surface. An analytical
description was given by, e.g., \cite{gj69,blum72}:
\begin{equation}
\label{fidens}
\frac{\rm f}{\rm i}=\frac{R_0}{1+n_e/N_c+\phi/\phi_c}
\end{equation}
with $R_0$ being the low density limit (an atomic parameter derived from a
weighted ratio of Einstein A-coefficients), $N_c$ the critical density where
f/i drops to half the low-density limit $R_0$, and $\phi/\phi_c$ describing the
influence from external radiation fields. The values for these coefficients
depend on the atomic number $Z$ and slightly on plasma temperature. For our
analysis we use the ions from $Z=8$ (i.e., O\,{\sc vii}) and $Z=10$
(Ne\,{\sc ix}). The parameters in Eq.~(\ref{fidens}) are taken from
\cite{ps} with $R_0=3.95$ and $N_c=3.1\times 10^{10}$\,cm$^{-3}$
for O\,{\sc vii} and $R_0=3.4$ and $N_c=5.9\times 10^{11}$\,cm$^{-3}$ for
Ne\,{\sc ix}. The parameter $\phi/\phi_c$ is commonly neglected, since most
active stars have no critical levels of photospheric UV emission; nevertheless
we will derive this parameter for the hottest photospheres in our sample
according to \cite{ness_StC_p}. The
functional dependence of the f/i ratio according to Eq.~(\ref{fidens}) with
varying electron density is illustrated in Fig.~\ref{fi_cmp} in comparison
with predictions from the CHIANTI database version 4.0 with ionization
balances from \cite{ar} \citep{dere01,young02}, the APEC database,
and \cite{porq} for the temperature range 1--4\,MK for O\,{\sc vii} and
4--9\,MK for Ne\,{\sc ix}. It can be seen that within the given temperature
range the analytical determination of densities according to
Eq.~(\ref{fidens}) is consistent with the other predictions.\\

\subsubsection {Previous work}

Since the gratings were to a large extent designed to measure the He-like
triplets of N\,{\sc vii} up to Si\,{\sc xiii} it is not surprising that such
analyses have been carried out for quite a few individual sources
\citep[e.g.,][]{aud01,gued01,ness_cap,ness_alg,nebr,stelz04}. Especially the
O\,{\sc vii} triplet has been analyzed, because the lines are strongest and
least blended. From the measured f/i-ratios densities were calculated. For
Capella and Procyon the densities were found to be at the lower end of the
sensitivity range \citep{ness_cap}. The first study of f/i-ratios in a
sample of stellar coronae was carried out by \cite{ness_10}, who measured
f/i-ratios for all He-like ions measurable with the LETGS for a sample of
ten stellar coronae. For inactive stars (with low $L_X\sim 10^{27}$\,erg/s)
only low density limits were found, while for the active stars in their
sample higher densities were encountered, although a little surprisingly
only low-densitly limits were measured  for some RS CVn stars.
\cite{ness_10} concluded that for the
high-temperature plasma LETGS data offer no conclusive tracer for densities
because of blending problems with the Ne\,{\sc ix} triplet, which
is better measured with the HETGS \citep[see also][]{nebr}.
Especially for the more active stars only a very small fraction of the X-ray emitting
plasma is produced in the temperature range where the O\,{\sc vii} triplet
is produced.
\\

The purpose of this paper is the analysis of He-like f/i ratios
for a large sample of X-ray spectra obtained with the Reflection Grating
Spectrometer (RGS) on board XMM-Newton and the Low Energy Transmission Grating
(LETGS) and the High Energy
Transmission Grating Spectrometers (HETGS; consisting of the Medium Energy
Grating MEG and the High Energy Grating HEG) on board {\it Chandra}. The LETGS
spectra are also used to measure density-sensitive Fe\,{\sc xxi} lines.
The aspect of resonant line scattering in stellar coronae has been addressed
with a large sample of grating spectra \citep{ness_opt} and strong evidence is
found that opacity effects can generally be neglected in coronal plasmas. We
apply the same procedures for data reduction as in \cite{ness_opt}. For
internal consistency, the LETGS spectra presented by \cite{ness_10}
are re-analyzed and included in our sample. We focus on density measurements
with the O\,{\sc vii} and Ne\,{\sc ix} triplets and Fe\,{\sc xxi} ratios,
and estimate systematic emitting volumes and filling factors for the different coronae.\\

\subsection{Measurement of line fluxes}
\label{cora}

We prefer to measure line counts with a program developed specifically for this
task named named CORA \citep{newi02}. The {\it lheasoft} package XSPEC can
also do the job, but for a large number of different spectra the CORA program
is more efficient.

\subsubsection{Systematic errors from line profile modeling}

The CORA program measures line
counts from the raw spectrum, i.e., the instrumental background is not
subtracted and the background spectrum is instead added to the iterated model
spectrum. While the measurement of line counts for the Chandra gratings is
straightforward \citep{ness_opt}, more difficulties arise for RGS spectra.
For Capella and AB\,Dor we measured RGS line counts with both CORA and XSPEC
(convolving a $\delta$-function profile with the line spread function) and we
found consistent results within the $1\sigma$ errors. However, line fluxes
obtained with the CORA program were found to be systematically lower than the
results
obtained with XSPEC. A possible reason for these systematic discrepancies could
be that XSPEC uses a wavelength-dependent response matrix, which provides a more
accurate instrument description than CORA (which uses only approximate
analytical line profiles). We tested the effects arising from different
treatments of the instrumental line profiles by exporting the line profiles
used for the XSPEC fits into CORA and found similar results compared to what
we obtained using the analytical Lorentzian profile function, without any
systematic trend. Obviously the Lorentzian used by CORA is an adequate
representation for the RGS line profiles for our purposes.\\

\subsubsection{Systematic errors from background modeling}

Another source of systematic errors is the estimation of source background.
As source background we consider the combination of continuum emission and
the sum of unresolved weak lines above the measured instrumental background.
This problem has been extensively described by \cite{ness_opt} who developed
a modified median routine providing a parameterized method to determine a source
continuum value. We present an alternative approach, viz. a $\chi^2$
optimization of the source continuum.\\
In the CORA program the continuum is assumed to be constant in sufficiently
small wavelength regions, which is justified for individual line fitting
with nearby lines. However, use of the median value as a source background
value as applied in the CORA program is
only valid as long as more than 50\% of the bins belong to the desired
background. A significant difference between the LETGS, for which the CORA
program was originally developed, and the RGS spectra in our sample is the
profile function. The wings of the RGS line profile are wider and contribute
to the continuum value obtained with the median function and CORA will therefore
return an overestimated continuum and thus underestimated line counts. Our
modified method uses the median value as a start value and we minimize the
$\chi^2$ value iterating only the source background value. The other
parameters of the model spectrum (wavelengths and line widths) are kept fixed
to the given initial values but the line counts are optimized with the
implemented likelihood method in each iteration step.\\
We tested the new procedure and found it stable and particularly useful
for wavelength regions with few emission lines. For RGS spectra this
method returned systematically lower source background values, while for LETGS
spectra these values were consistent with the median values. The discrepancies
with the XSPEC results are significantly reduced. We point
out that even without the new procedure the discrepancies are not significant
within the errors. We also tested the $\chi^2$ fit procedure for the
Fe\,{\sc xvii} lines measured by \cite{ness_opt}, but found that it did not
work well.
We attribute these difficulties to the large number of lines in the 15\,\AA\
region. The disadvantage of our $\chi^2$ approach is that all line features
not selected to be measured as emission lines increase the source background
value in the attempt to minimize the $\chi^2$ value in those wavelength
regions. We thus conclude that the parameterized median value must be used
for line-crowded regions, while the $\chi^2$ approach represents a
non-parameterized procedure for measuring in wavelength regions where all line
features are selected to be fitted.\\

\section{Observations}

For our study of stellar coronae we selected a sample of coronal sources
as large as possible. We gathered 64 grating spectra of 42 stellar coronae;
we specifically discuss 22 RGS spectra, 16 LETGS spectra, and 26 HETGS spectra.
The reduction of the spectra has been carried out with standard SAS and CIAO
routines and is described by \cite{ness_opt}. The details of the observations
details with exposure times and derived luminosities are summarized in Table~1
of \cite{ness_opt}. X-ray luminosities, averaged over the complete observations
(i.e., including flares), are obtained by summing up all
dispersed photons in the wavelength range 5.15--38.19\,\AA\ after
consideration of $A_{\rm eff}(\lambda)$. In order to
have the largest possible data sample for our systematic analysis of
densities we extracted all additional stellar spectra that were publicly
available by 31 January 2004 from the Chandra archive. These additional
observations are listed in Table~\ref{newstars} using the same format as in
\cite{ness_opt}. For internal consistency the LETGS observations of Algol and
Capella were reanalyzed for this work and the results are also listed in
Table~\ref{newstars} for comparison. We extracted effective areas for flux
conversion from the Capella observations
and used these areas for all observations. Since this paper deals only with
line ratios (and therefore use only ratios of effective areas), this procedure
is sufficiently accurate. Nevertheless we compared these effective
areas with individually extracted areas and found sufficient agreement for all
instruments. The complete stellar sample used for this work is described in
Sect.~\ref{sample} and all stellar properties relevant for this paper are
listed in Table~\ref{sprop}.

\begin{table*}
\begin{flushleft}
\renewcommand{\arraystretch}{1.2}
\caption{\label{newstars}Properties of observations not already described in \cite{ness_opt}}
{\scriptsize
\begin{tabular}{lccc|ccccccc}
 star& \multicolumn{3}{c|}{exposure time [ks]} & \multicolumn{7}{c}{L$_X$(10$^{28}$\,erg/s)}\\
 & RGS1,2 & LETG & HEG/& \multicolumn{2}{c}{RGS1$^a$} & \multicolumn{2}{c}{RGS2$^a$} & LETG$^a$ & MEG$^a$ & HEG$^b$\\
 &&&\ MEG&1$^{\rm st}$order&2$^{\rm nd}$order&1$^{\rm st}$order&2$^{\rm nd}$order&&\\
           AD\,Leo&   36.25&   48.50&   45.16&    3.50&    3.01&    3.07&    2.03&    3.93&    3.19&    1.88\\
    $\lambda$\,And&   31.83&   98.59&   81.91&  249.61&  238.61&  295.03&  188.31&  935.90&  198.33&  121.17\\
   $\sigma^2$\,CrB&   19.31&    --&   83.72&  322.59&  338.72&  307.83&  312.74&    --&  305.60&  154.81\\
           Capella&   52.92&  226.42&  154.68&  145.38&  157.22&  168.06&  132.10&  186.56&  153.08&  127.43\\
             Algol&   52.66&   79.53&   51.73&  283.73&    --&    --&    --&  944.99&  673.63&  557.72\\
           II\,Peg&    --&    --&   42.74&    --&    --&    --&    --&    --& 1318.60& 1045.50\\
           44\,Boo&    --&    --&   59.14&    --&    --&    --&    --&    --&   45.35&   41.36\\
         Prox\,Cen&    --&    --&   42.39&    --&    --&    --&    --&    --&    0.03&    0.03\\
           ER\,Vul&    --&    --&  112.01&    --&    --&    --&    --&    --&  240.10&  258.20\\
           TY\,Pyx&    --&    --&   49.13&    --&    --&    --&    --&    --&  532.60&  550.70\\
           24\,UMa&    --&    --&   88.90&    --&    --&    --&    --&    --&  125.04&   86.16\\
        $\xi$\,UMa&    --&    --&   70.93&    --&    --&    --&    --&    --&   14.87&   16.45\\
         V824\,Ara&    --&    --&   94.23&    --&    --&    --&    --&    --&  254.66&  212.00\\
           31\,Com&    --&    --&  130.19&    --&    --&    --&    --&    --& 6461.70&12301.80\\
        HD\,223460&    --&    --&   95.66&    --&    --&    --&    --&    --& 3289.00& 3551.00\\
           Canopus&    --&    --&   94.55&    --&    --&    --&    --&    --&  322.00&  316.10\\
        $\mu$\,Vel&    --&    --&  134.06&    --&    --&    --&    --&    --&  123.40&  146.45\\
           IM\,Peg&    --&    --&   95.03&    --&    --&    --&    --&    --& 2520.70& 2316.90\\
       Speedy\,Mic&    --&    --&   69.00&    --&    --&    --&    --&    --&  223.00&  100.40\\
         V471\,Tau&    --&   87.49&    --&    --&    --&    --&    --&  112.34&    --&    --\\
           VW\,Cep&    --&  101.17&    --&    --&    --&    --&    --&   70.42&    --&    --\\
\hline
\end{tabular}
\\
$^a$5.15--38.19\,\AA\ \hspace{2cm}$^b$5.15--21.5\,\AA\
}
\renewcommand{\arraystretch}{1}
\end{flushleft}
\end{table*}

\subsection{Measured line counts}

For this paper we measure the O\,{\sc vii} triplet (21.6/21.8/22.1\,\AA,
log$(T_m/{\rm K})=6.3$) and the Ly$_\alpha$ line of O\,{\sc viii} (18.97\,\AA,
log$(T_m/{\rm K})=6.5$) with the RGS1, the LETGS, and the MEG. The RGS2
cannot measure O\,{\sc vii} because of chip failure, and the HEG does not cover
the O\,{\sc vii} triplet. The Ne\,{\sc ix} triplet (13.44/13.55/13.7\,\AA,
log$(T_m/{\rm K})=6.6$) is severely blended with highly ionized Fe lines
\citep{nebr} and the blends can only be resolved with the HEG and the MEG.
Although the RGS and LETGS cover
the 13.5\,\AA\ region we do not analyze those data here, because de-blending
is too complicated and must be done in a future paper. The measured counts
and derived f/i ratios \citep[using effective areas as described in][]{ness_opt}
are listed in Tables~\ref{fi}/\ref{fi_rs}. The densities derived from
Eq.~(\ref{fidens}) are also listed with additional description in
Sect.~\ref{hedens}. For the hot sources for which LETGS spectra are available,
we measured the lines at 128.73\,\AA, 117.5\,\AA,
102.22\,\AA, and 97.87\,\AA. The measured line counts are listed in
Table~\ref{fe21} and discussed in Sect.~\ref{fe21dens}.

\subsection{Description of the stellar sample}
\label{sample}

\begin{table*}[!ht]
\begin{flushleft}
\renewcommand{\arraystretch}{1.2}
\caption{\label{sprop}Description of the stellar sample}
{\scriptsize
\begin{tabular}{lrrcccccccc}
 star  & HD/Gl & Spectr. Type$^a$ & distance$^a$ & V$^a$ & B.C.$^b$ & $T_{\rm eff}^b$ & log($L_{\rm bol}$) & R$_\star^c$&R$_\star^d$ &$L_X^e$\\
&&&pc&mag&mag&K&erg/s&[$R_\odot$]&[$R_\odot$]&10$^{28}$\,erg/s\\
           24\,UMa&    82210&   G4.0III-IV&    32.37&     4.57&    -0.07& 5666&   34.72&     3.2&     3.82&   218.00\\
           31\,Com&   111812&         K2.0&   335.60&     9.26&    -2.17& 4780&   34.94&    6.38&     6.89&     0.00\\
           44\,Boo&   133640&  G2.0V/G2.0V&    12.76&     4.76&    -2.46& 5780&   33.83&0.87/0.6&     1.31&    49.60\\
           47\,Cas&    12230& G2.0V/F0.0Vn&    33.56&     5.27&    -0.19& 5780&   34.46&     1.0&     2.73&   236.07\\
           AB\,Dor&    36705&     K1.0IIIp&    14.94&     6.93&    -0.06& 5010&   33.15&     1.0&     0.81&   150.00\\
  $\alpha$\,Cen\,A&   128620&        G2.0V&     1.34&     0.01&    -0.06& 5780&   33.77&    1.23&     1.23&     0.13\\
  $\alpha$\,Cen\,B&   128621&        K0.0V&     1.34&     1.34&    -0.10& 5240&   33.28&     0.8&     0.85&     0.20\\
           AD\,Leo&    Gl388&        M3.5V&     4.70&     9.43&    -0.15& 3295&   31.93&     0.5&     0.45&     7.22\\
             Algol&    19356&B8.0V/K2.0III&    28.00&     2.12&    -0.04&13400&   35.88&     3.5&     2.59&   661.00\\
           AR\,Lac&   210334&G2.0IV/K0.0IV&    42.03&     6.13&    -0.19& 5780&   34.31&1.54/2.8&     2.30&  1050.00\\
           AT\,Mic&   196982&   dM4.5/M4.5&    10.22&    10.25&    -0.22& 3175&   32.39&      --&     0.84&    29.40\\
           AU\,Mic&   197481&         M0.0&     9.94&     8.61&    -0.07& 3920&   32.52&    0.67&     0.63&    55.49\\
      $\beta$\,Cet&     4128&      K0.0III&    29.38&     2.04&    -1.20& 5240&   35.69&    11.6&    13.58&   268.52\\
           Canopus&    45348&       F0.0II&    95.88&    -0.72&    -0.06& 7240&   37.77&      --&    78.66&     0.00\\
           Capella&    34029&G1.0III/K0.0I&    12.94&     0.08&    -2.17& 5850&   35.71&  9.2/13&    11.17&   419.16\\
     $\chi^1$\,Ori&    39587&        G0.0V&     8.66&     4.41&    -2.17& 5920&   33.63&     1.1&     0.99&    12.00\\
           EK\,Dra&   129333&       dG0.0e&    33.94&     7.61&    -0.10& 5920&   33.53&      --&     0.89&   102.07\\
   $\epsilon$\,Eri&    22049&        K2.0V&     3.22&     3.73&    -0.07& 4780&   33.11&    0.81&     0.84&     2.10\\
           EQ\,Peg&    Gl896&    M3.5/M4.5&     6.25&    10.32&    -0.10& 3295&   31.82&0.4/0.26&     0.40&     3.97\\
           ER\,Vul&   200391&    G0.0V/G5V&    49.85&     7.36&    -2.46& 5920&   33.97&    1.07&     1.47&   376.00\\
           EV\,Lac&    Gl873&         M3.5&     5.05&    10.09&    -0.85& 3295&   31.73&    0.34&     0.36&    12.25\\
        HD\,223460&   223460&      G1.0III&   134.95&     5.90&    -0.04& 5850&   35.42&      --&     7.99&  1691.10\\
          HR\,1099&    22468&G5.0IV/K1.0IV&    28.97&     5.91&    -0.25& 5610&   34.09& 3.9/1.3&     1.89&  1512.10\\
           II\,Peg&   224085&        K0.0V&    42.34&     7.37&    -1.34& 5240&   33.87&     4.5&     1.68&   650.00\\
           IM\,Peg&   216489&  K1.5II-IIIe&    96.80&     5.90&    -0.07& 4895&   35.20&    26.2&     8.85&  2756.10\\
     $\kappa$\,Cet&    20630&     G5.0Vvar&     9.16&     4.83&    -0.19& 5610&   33.52&    0.94&     0.98&     7.72\\
    $\lambda$\,And&   222107&      G8.0III&    25.81&     3.82&    -0.19& 5490&   34.85&     7.5&     4.70&   335.00\\
        $\mu$\,Vel&    93497&     G5.0III/&    35.50&     2.72&    -0.07& 5610&   35.54&     10.&    10.04&     0.00\\
      $\pi^1$\,UMa&    72905&       G1.5Vb&    14.27&     5.64&    -2.88& 5815&   33.57&     1.0&     0.96&    12.83\\
           Procyon&    61421&     F5.0IV-V&     3.50&     0.34&    -0.06& 6540&   34.46&    2.06&     2.12&     1.90\\
         Prox\,Cen&   Gl551C&       M5.5Ve&     1.29&    11.05&    -0.10& 3043&   30.44&    0.15&     0.10&     0.17\\
   $\sigma^2$\,CrB&  146361J&  F6.0V/G0.0V&    21.70&     5.64&    -0.09& 6450&   33.92&     1.1&     1.18&   460.61\\
       Speedy\,Mic&   197890&        K0.0V&    44.40&     9.44&    -0.06& 5240&   33.08&    0.73&     0.68&     0.00\\
           TY\,Pyx&    77137&G5.0IV/G5.0IV&    55.83&     6.90&    -0.22& 5610&   34.27&1.59/1.6&     2.31&   463.00\\
           UX\,Ari&    21242& G5.0V/K0.0IV&    50.23&     6.47&    -0.25& 5610&   34.34&4.7/0.93&     2.53&  1205.00\\
         V471\,Tau&    17962&         K0.0&    46.79&     9.48&    -0.06& 5240&   33.11&    0.85&     0.70&   185.10\\
         V824\,Ara&   155555&       K1.0Vp&    31.42&     6.88&    -0.08& 5010&   33.82&      --&     1.74&   449.30\\
           VW\,Cep&   197433&     K0.0Vvar&    27.65&     7.38&    -0.10& 5240&   33.50&    0.88&     1.09&   105.00\\
           VY\,Ari&    17433&         K0.0&    43.99&     6.76&    -0.23& 5240&   34.15&     1.9&     2.31&  1243.63\\
        $\xi$\,UMa&    98239&        G0.0V&     8.80&     3.78&    -0.19& 5920&   33.89&    0.94&     1.35&    30.50\\
           YY\,Gem&   60179C&  M0.5V/M0.5V&    15.80&     9.07&    -0.19& 3800&   32.80&    0.66&     0.93&    82.37\\
           YZ\,CMi&    Gl285&      M4.5V:e&     5.93&    11.12&    -0.19& 3175&   31.57&    0.36&     0.33&     4.44\\
\hline
\end{tabular}
\\
$^a${From Simbad}\hspace{1cm}$^b${From \cite{kaler}}\hspace{1cm}
$^c${Literature}\\
$^d${From $L_{\rm bol}=4\pi R_\star^2\sigma T_{\rm eff}^4$} (used for analysis)\hspace{.5cm}
$^e${measured with ROSAT (5.2--124\,\AA)}
}
\renewcommand{\arraystretch}{1}
\end{flushleft}
\end{table*}

\begin{figure}[!ht]
 \resizebox{\hsize}{!}{\includegraphics{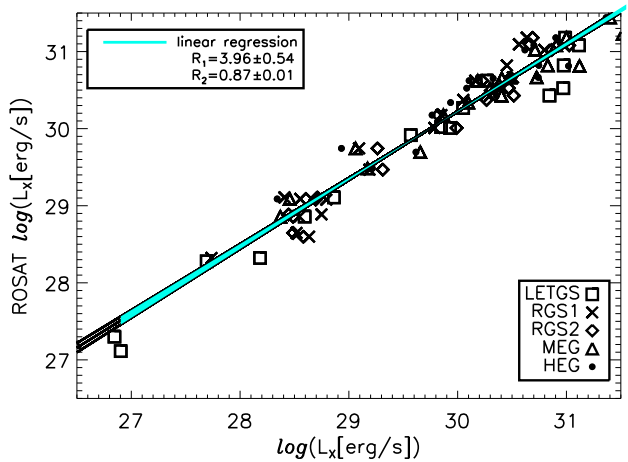}}
\caption[]{\label{lx}\bf Comparison of ROSAT luminosities ($5.2-124$\,\AA)
listed in Table~\ref{sprop} with $L_X$ measurements for RGS1, RGS2, LETGS, MEG
($5.15-38.2$\,\AA), and HEG ($5.15-21.5$\,\AA) from \cite{ness_opt} and
Table~\ref{newstars}. The regression curve $\log L_X$(ROSAT)$=$R$_1+$R$_2
\log L_X$(Chandra/XMM) with parameters R$_1$ and R$_2$ as
given in the legend is overplotted. The slope R$_2$ is near unity, indicating
good cross-calibration between instruments.}
\end{figure}

In Table~\ref{sprop} we list all relevant stellar parameters. The spectral
type information has been taken from the Simbad
database\footnote{http://simbad.u-strasbg.fr/}. It can be seen that a broad
range of coronae is included in the sample. The sample covers stars with
extremely high flare activity, RS~CVn systems, and other double systems. The
distances are also from Simbad and are based on Hipparcos parallaxes. An
important parameter for our analysis is the stellar radius which is needed for
scaling the derived
coronal loop sizes to typical geometries. For some stars in the sample radii are not available in the literature, and we adopted a procedure
to calculate stellar radii from the apparent visual magnitudes V and distances
(and thus absolute magnitudes) and spectral types, listed in
Table~\ref{sprop}. We estimate bolometric corrections and effective
temperatures $T_{\rm eff}$ from the spectral type interpolating tables from
\cite{kaler}. We use the bolometric luminosity $L_{\rm bol}$, calculated from
the absolute luminosity and the bolometric correction to calculate the
stellar radius for each star from $L_{\rm bol}=4\pi R_\star^2\sigma
T_{\rm eff}^4$ with $\sigma$ the Stefan-Boltzmann constant.
In Table~\ref{sprop} we list the effective temperatures thus derived,
bolometric luminosities, and stellar radii in comparison to what we
found in the literature. Good agreement with most values from the literature
is found and we use our derived radii for further analysis. In the last column
we list the X-ray luminosity from ROSAT \citep[e.g.,][ wavelength range
$5.2-124$\,\AA]{huensch99}. In Fig.~\ref{lx} we compare these values with
X-ray luminosities obtained from the new spectra (wavelength range
$5.15-38.2$\,\AA) listed in Tables~\ref{lxem}/\ref{lxem_rs}.
The regression fit has a slope slightly lower than one, indicating
that the ROSAT fluxes for the more active stars have been underestimated
With the best-fit regression
parameters the discrepancies are no higher than a factor of two.

\section{Results}
\label{results}

\begin{table*}[!ht]
\begin{flushleft}
\renewcommand{\arraystretch}{1.1}
\caption{\label{lxem}X-ray luminosities, H-like and He-like line fluxes, and ion-specific luminosities and emission measures.}
\vspace{-.3cm}
{\scriptsize
\begin{tabular}{lccccccccc}
&&$L_X$&\multicolumn{2}{c}{line flux ($10^{-13}$\,erg\,cm$^{-2}$\,s$^{-1}$)}&$L_{X, O\,\mathsc{vii}}$&$L_{X, Ne\,\mathsc{ix}}$&\multicolumn{2}{c}{log(EM/cm$^{-3}$)}\\
 star & Instr.& [10$^{28}$erg/s]& O\,{\sc viii} & O\,{\sc vii}(r)&[10$^{28}$erg/s] & [10$^{28}$erg/s] & O\,{\sc vii}$^a$&Ne\,{\sc ix}$^b$\\
         24\,UMa&   MEG& 125.04&   2.44\,$\pm$\,0.16&   0.39\,$\pm$\,0.09&   0.94\,$\pm$\,0.63&   1.43\,$\pm$\,0.40&  51.18\,$\pm$\,0.22&  52.11\,$\pm$\,0.10\\
 &   HEG&  86.16&   --&   --&   --&   1.41\,$\pm$\,0.83&   --&  52.12\,$\pm$\,0.19\\
         44\,Boo&   MEG&  45.35&  12.61\,$\pm$\,0.46&   2.59\,$\pm$\,0.27&   1.00\,$\pm$\,0.30&   1.53\,$\pm$\,0.24&  51.20\,$\pm$\,0.10&  52.11\,$\pm$\,0.06\\
 &   HEG&  41.36&   --&   --&   --&   1.50\,$\pm$\,0.41&   --&  52.07\,$\pm$\,0.10\\
         47\,Cas&  RGS1& 112.88&   4.66\,$\pm$\,0.15&   0.69\,$\pm$\,0.07&   1.77\,$\pm$\,0.50&   --&  51.46\,$\pm$\,0.09&   --\\
         AB\,Dor&  RGS1&  70.87&  12.58\,$\pm$\,0.23&   2.68\,$\pm$\,0.12&   1.31\,$\pm$\,0.17&   --&  51.35\,$\pm$\,0.04&   --\\
 &   MEG&  73.87&  14.89\,$\pm$\,0.53&   1.33\,$\pm$\,0.22&   0.85\,$\pm$\,0.37&   2.18\,$\pm$\,0.28&  51.04\,$\pm$\,0.16&  52.29\,$\pm$\,0.05\\
 &   HEG&  58.21&   --&   --&   --&   2.16\,$\pm$\,0.55&   --&  52.25\,$\pm$\,0.09\\
$\alpha$\,Cen\,A&  LETG&   0.08&   0.32\,$\pm$\,0.05&   1.00\,$\pm$\,0.09&   --&   --&  48.82\,$\pm$\,0.08&   --\\
$\alpha$\,Cen\,B&  RGS1&   0.00&   5.27\,$\pm$\,0.41&   4.49\,$\pm$\,0.40&   --&   --&  49.48\,$\pm$\,0.08&   --\\
 &  LETG&   0.07&   0.69\,$\pm$\,0.07&   1.16\,$\pm$\,0.10&   0.01\,$\pm$\,0.00&   --&  48.89\,$\pm$\,0.08&   --\\
         AD\,Leo&  RGS1&   3.50&  11.59\,$\pm$\,0.27&   3.89\,$\pm$\,0.17&   0.18\,$\pm$\,0.02&   --&  50.50\,$\pm$\,0.04&   --\\
 &  LETG&   3.93&  10.94\,$\pm$\,0.32&   3.20\,$\pm$\,0.21&   0.16\,$\pm$\,0.03&   --&  50.42\,$\pm$\,0.06&   --\\
 &   MEG&   3.19&   8.70\,$\pm$\,0.43&   2.27\,$\pm$\,0.28&   0.11\,$\pm$\,0.04&   0.13\,$\pm$\,0.02&  50.27\,$\pm$\,0.12&  51.07\,$\pm$\,0.06\\
 &   HEG&   1.88&   --&   --&   --&   0.12\,$\pm$\,0.04&   --&  51.08\,$\pm$\,0.12\\
           Algol&  RGS1& 283.73&   8.34\,$\pm$\,0.21&   1.15\,$\pm$\,0.11&   1.79\,$\pm$\,0.63&   --&  51.52\,$\pm$\,0.09&   --\\
 &  LETG& 944.99&  17.06\,$\pm$\,0.33&   2.53\,$\pm$\,0.19&   4.95\,$\pm$\,1.03&   --&  51.87\,$\pm$\,0.07&   --\\
 &   MEG& 673.63&  11.60\,$\pm$\,0.48&   2.15\,$\pm$\,0.30&   4.69\,$\pm$\,1.75&   5.89\,$\pm$\,0.97&  51.80\,$\pm$\,0.13&  52.64\,$\pm$\,0.06\\
 &   HEG& 557.72&   --&   --&   --&   5.49\,$\pm$\,1.85&   --&  52.60\,$\pm$\,0.12\\
         AT\,Mic&  RGS1&  15.47&  11.20\,$\pm$\,0.30&   3.56\,$\pm$\,0.19&   0.79\,$\pm$\,0.12&   --&  51.14\,$\pm$\,0.05&   --\\
         AU\,Mic&  RGS1&  12.43&  10.70\,$\pm$\,0.20&   3.19\,$\pm$\,0.12&   0.75\,$\pm$\,0.07&   --&  51.07\,$\pm$\,0.03&   --\\
 &   MEG&  11.52&   7.39\,$\pm$\,0.35&   1.43\,$\pm$\,0.20&   0.38\,$\pm$\,0.14&   0.56\,$\pm$\,0.09&  50.72\,$\pm$\,0.29&  51.67\,$\pm$\,0.06\\
 &   HEG&   8.55&   --&   --&   --&   0.63\,$\pm$\,0.18&   --&  51.74\,$\pm$\,0.10\\
    $\beta$\,Cet&  RGS1& 198.67&   8.49\,$\pm$\,0.40&   1.97\,$\pm$\,0.22&   3.67\,$\pm$\,1.22&   --&  51.80\,$\pm$\,0.10&   --\\
 &  LETG& 699.24&   8.63\,$\pm$\,0.20&   1.31\,$\pm$\,0.12&   2.69\,$\pm$\,0.70&   --&  51.63\,$\pm$\,0.08&   --\\
 &   MEG& 253.19&   6.10\,$\pm$\,0.26&   0.85\,$\pm$\,0.13&   1.53\,$\pm$\,0.71&   3.79\,$\pm$\,0.62&  51.44\,$\pm$\,0.13&  52.50\,$\pm$\,0.06\\
 &   HEG& 227.48&   --&   --&   --&   4.11\,$\pm$\,1.17&   --&  52.49\,$\pm$\,0.11\\
         Canopus&   MEG& 322.00&   1.16\,$\pm$\,0.11&   0.30\,$\pm$\,0.07&   6.46\,$\pm$\,4.70&   8.27\,$\pm$\,2.72&   --&  52.90\,$\pm$\,0.11\\
 &   HEG& 316.10&   --&   --&   --&   8.96\,$\pm$\,5.72&   --&  52.92\,$\pm$\,0.21\\
   $\chi^1$\,Ori&  RGS1&   6.21&   3.23\,$\pm$\,0.16&   1.23\,$\pm$\,0.11&   0.20\,$\pm$\,0.05&   --&  50.54\,$\pm$\,0.02&   --\\
         EK\,Dra&  RGS1&  61.25&   2.35\,$\pm$\,0.13&   0.50\,$\pm$\,0.07&   1.44\,$\pm$\,0.53&   --&  51.33\,$\pm$\,0.08&   --\\
 &  LETG&  85.01&   2.07\,$\pm$\,0.12&   0.47\,$\pm$\,0.08&   1.31\,$\pm$\,0.67&   --&  51.30\,$\pm$\,0.17&   --\\
 $\epsilon$\,Eri&  RGS1&   0.55&   7.13\,$\pm$\,0.35&   3.27\,$\pm$\,0.25&   0.08\,$\pm$\,0.02&   --&  50.10\,$\pm$\,0.12&   --\\
 &  LETG&   1.53&   8.04\,$\pm$\,0.18&   3.80\,$\pm$\,0.15&   0.09\,$\pm$\,0.01&   --&  50.17\,$\pm$\,0.04&   --\\
         EQ\,Peg&  RGS1&   4.31&   9.10\,$\pm$\,0.37&   3.53\,$\pm$\,0.25&   0.28\,$\pm$\,0.06&   --&  50.71\,$\pm$\,0.07&   --\\
         ER\,Vul&   MEG& 240.10&   2.57\,$\pm$\,0.15&   0.43\,$\pm$\,0.08&   2.38\,$\pm$\,1.52&   5.13\,$\pm$\,1.06&  51.60\,$\pm$\,0.25&  52.60\,$\pm$\,0.08\\
 &   HEG& 258.20&   --&   --&   --&   4.99\,$\pm$\,2.06&   --&  52.61\,$\pm$\,0.15\\
         EV\,Lac&  RGS1&   3.58&   9.19\,$\pm$\,0.25&   2.89\,$\pm$\,0.16&   0.17\,$\pm$\,0.02&   --&  50.44\,$\pm$\,0.06&   --\\
 &   MEG&   2.86&   6.64\,$\pm$\,0.25&   1.86\,$\pm$\,0.17&   0.11\,$\pm$\,0.03&   0.10\,$\pm$\,0.01&  50.25\,$\pm$\,0.03&  51.00\,$\pm$\,0.05\\
 &   HEG&   2.19&   --&   --&   --&   0.12\,$\pm$\,0.03&   --&  51.04\,$\pm$\,0.09\\
      HD\,223460&   MEG&3289.00&   1.82\,$\pm$\,0.14&   0.32\,$\pm$\,0.08&  14.38\,$\pm$\,11.75&   --&  52.34\,$\pm$\,0.19&   --\\
   $\kappa$\,Cet&  RGS1&   5.61&   2.93\,$\pm$\,0.13&   1.12\,$\pm$\,0.09&   0.22\,$\pm$\,0.05&   --&  50.54\,$\pm$\,0.07&   --\\
      $\mu$\,Vel&   MEG& 123.40&   2.29\,$\pm$\,0.13&   0.45\,$\pm$\,0.08&   1.35\,$\pm$\,0.67&   1.23\,$\pm$\,0.35&  51.33\,$\pm$\,0.31&  52.04\,$\pm$\,0.10\\
 &   HEG& 146.45&   --&   --&   --&   2.36\,$\pm$\,0.91&   --&  52.28\,$\pm$\,0.13\\
    $\pi^1$\,UMa&  RGS1&   2.56&   1.43\,$\pm$\,0.08&   0.51\,$\pm$\,0.05&   0.24\,$\pm$\,0.07&   --&  50.59\,$\pm$\,0.08&   --\\
         Procyon&  LETG&   0.49&   2.05\,$\pm$\,0.08&   3.06\,$\pm$\,0.12&   0.10\,$\pm$\,0.01&   --&  50.14\,$\pm$\,0.03&   --\\
       Prox\,Cen&   MEG&   0.03&   2.19\,$\pm$\,0.22&   0.76\,$\pm$\,0.17&   --&   --&  48.67\,$\pm$\,0.12&  49.25\,$\pm$\,0.14\\
 &   HEG&   0.03&   --&   --&   --&   --&   --&  49.24\,$\pm$\,0.35\\
     Speedy\,Mic&   MEG& 223.00&   1.26\,$\pm$\,0.13&   0.20\,$\pm$\,0.07&   1.25\,$\pm$\,1.36&   1.66\,$\pm$\,0.65&  51.17\,$\pm$\,0.22&  52.16\,$\pm$\,0.14\\
 &   HEG& --40&   --&   --&   --&   1.36\,$\pm$\,1.40&   --&  51.89\,$\pm$\,0.47\\
       V471\,Tau&  LETG& 112.34&   1.46\,$\pm$\,0.09&   0.30\,$\pm$\,0.06&   1.85\,$\pm$\,1.06&   --&  51.39\,$\pm$\,0.20&   --\\
         VW\,Cep&  LETG&  70.42&   3.74\,$\pm$\,0.13&   0.83\,$\pm$\,0.09&   1.62\,$\pm$\,0.46&   --&  51.37\,$\pm$\,0.10&   --\\
      $\xi$\,UMa&   MEG&  14.87&  11.45\,$\pm$\,0.40&   3.43\,$\pm$\,0.28&   0.59\,$\pm$\,0.13&   0.64\,$\pm$\,0.07&  50.99\,$\pm$\,0.14&  51.75\,$\pm$\,0.04\\
 &   HEG&  16.45&   --&   --&   --&   0.60\,$\pm$\,0.14&   --&  51.72\,$\pm$\,0.08\\
         YY\,Gem&  LETG&  37.12&   7.06\,$\pm$\,0.23&   1.93\,$\pm$\,0.15&   1.07\,$\pm$\,0.24&   --&  51.25\,$\pm$\,0.07&   --\\
         YZ\,CMi&  RGS1&   3.30&   5.48\,$\pm$\,0.21&   1.99\,$\pm$\,0.13&   0.16\,$\pm$\,0.03&   --&  50.41\,$\pm$\,0.09&   --\\
\hline
\end{tabular}
\\
$^a$at T=2\,MK\ \ \ \
$^b$at T=4\,MK\\
}
\renewcommand{\arraystretch}{1}
\end{flushleft}
\end{table*}

\begin{table*}[!ht]
\begin{flushleft}
\renewcommand{\arraystretch}{1.1}
\caption{\label{fi}Measured line counts for O\,{\sc vii} and Ne\,{\sc ix} intercombination (i) and forbidden (f) lines, corresponding f/i ratios (corrected for A$_{\rm eff}$), and plasma densities $n_e$ derived from Eq.~\ref{fidens}. All errors are 1\,$\sigma$ errors.}
\vspace{-.4cm}
{\scriptsize
\begin{tabular}{p{1.cm}p{.5cm}cccccccc}
&&\multicolumn{4}{c}{O\,{\sc vii}}&\multicolumn{4}{c}{Ne\,{\sc ix}}\\
 star & Instr.& i [cts] & f [cts] & f/i & log($n_e$) & i [cts] & f [cts] & f/i& log($n_e$)\\
         24\,UMa&   MEG&   2.55\,$\pm$\,1.97&  14.52\,$\pm$\,4.00&   6.37\,$\pm$\,5.22&          $<\,11.9$&  18.84\,$\pm$\,5.70&  80.49\,$\pm$\,10.10&   4.69\,$\pm$\,1.53&          $<\,11.8$\\
 &   HEG&   --&   --&   --&                 --&   8.83\,$\pm$\,3.32&  13.64\,$\pm$\,3.99&   1.61\,$\pm$\,0.77&  $11.8\,\pm\,0.42$\\
         44\,Boo&   MEG&  32.92\,$\pm$\,6.46&  49.79\,$\pm$\,7.70&   1.69\,$\pm$\,0.42&  $10.6\,\pm\,0.19$& 132.90\,$\pm$\,15.01& 352.56\,$\pm$\,25.21&   2.91\,$\pm$\,0.39&  $11.0\,\pm\,0.52$\\
 &   HEG&   --&   --&   --&                 --&  40.76\,$\pm$\,7.17&  84.48\,$\pm$\,10.46&   2.17\,$\pm$\,0.47&  $11.5\,\pm\,0.27$\\
         47\,Cas&  RGS1&  55.19\,$\pm$\,12.64& 105.41\,$\pm$\,14.67&   2.00\,$\pm$\,0.54&  $10.5\,\pm\,0.24$&   --&   --&   --&                 --\\
         AB\,Dor&  RGS1& 230.94\,$\pm$\,24.70& 407.50\,$\pm$\,27.21&   1.83\,$\pm$\,0.23&  $10.5\,\pm\,0.10$&   --&   --&   --&                 --\\
 &   MEG&  15.60\,$\pm$\,5.02&  36.98\,$\pm$\,6.75&   2.65\,$\pm$\,0.98&  $10.2\,\pm\,0.74$& 137.61\,$\pm$\,13.53& 290.84\,$\pm$\,18.73&   2.32\,$\pm$\,0.27&  $11.4\,\pm\,0.16$\\
 &   HEG&   --&   --&   --&                 --&  33.50\,$\pm$\,6.11&  77.68\,$\pm$\,9.19&   2.42\,$\pm$\,0.53&  $11.4\,\pm\,0.37$\\
$\alpha$\,Cen\,A&  LETG&  38.41\,$\pm$\,6.99& 109.60\,$\pm$\,11.06&   2.82\,$\pm$\,0.59&  $10.1\,\pm\,0.39$&   --&   --&   --&                 --\\
$\alpha$\,Cen\,B&  RGS1&  20.34\,$\pm$\,6.49& 111.23\,$\pm$\,11.60&   5.68\,$\pm$\,1.91&          $<\,10.2$&   --&   --&   --&                 --\\
 &  LETG&  36.09\,$\pm$\,7.09& 154.54\,$\pm$\,13.06&   4.23\,$\pm$\,0.90&          $<\,10.8$&   --&   --&   --&                 --\\
         AD\,Leo&  RGS1& 166.09\,$\pm$\,17.75& 356.13\,$\pm$\,22.02&   2.23\,$\pm$\,0.27&  $10.4\,\pm\,0.12$&   --&   --&   --&                 --\\
 &  LETG&  53.01\,$\pm$\,9.21& 170.31\,$\pm$\,14.52&   3.17\,$\pm$\,0.61&  $9.92\,\pm\,0.74$&   --&   --&   --&                 --\\
 &   MEG&  18.56\,$\pm$\,4.45&  30.68\,$\pm$\,5.65&   1.85\,$\pm$\,0.56&  $10.5\,\pm\,0.25$&  53.71\,$\pm$\,7.85& 171.40\,$\pm$\,14.05&   3.50\,$\pm$\,0.59&          $<\,12.0$\\
 &   HEG&   --&   --&   --&                 --&  12.77\,$\pm$\,3.61&  26.91\,$\pm$\,5.26&   2.20\,$\pm$\,0.76&  $11.5\,\pm\,0.50$\\
           Algol&  RGS1& 104.99\,$\pm$\,25.25&  90.89\,$\pm$\,24.58&   0.90\,$\pm$\,0.33&  $10.6\,\pm\,1.41$&   --&   --&   --&                 --\\
 &  LETG& 186.85\,$\pm$\,22.27& 184.01\,$\pm$\,22.30&   0.97\,$\pm$\,0.17&  $10.4\,\pm\,0.47$&   --&   --&   --&                 --\\
 &   MEG&  51.46\,$\pm$\,8.54&  31.58\,$\pm$\,7.29&   0.69\,$\pm$\,0.19&  $10.9\,\pm\,0.29$& 150.82\,$\pm$\,15.43& 236.96\,$\pm$\,18.28&   1.72\,$\pm$\,0.22&  $11.7\,\pm\,0.11$\\
 &   HEG&   --&   --&   --&                 --&  35.73\,$\pm$\,6.99&  53.05\,$\pm$\,8.12&   1.55\,$\pm$\,0.39&  $11.8\,\pm\,0.20$\\
         AT\,Mic&  RGS1& 134.91\,$\pm$\,17.58& 238.57\,$\pm$\,18.83&   1.84\,$\pm$\,0.28&  $10.5\,\pm\,0.12$&   --&   --&   --&                 --\\
         AU\,Mic&  RGS1& 209.47\,$\pm$\,21.95& 688.37\,$\pm$\,31.28&   3.44\,$\pm$\,0.39&  $9.70\,\pm\,0.67$&   --&   --&   --&                 --\\
 &   MEG&  10.05\,$\pm$\,3.52&  45.99\,$\pm$\,6.97&   5.11\,$\pm$\,1.95&          $<\,10.9$&  78.25\,$\pm$\,9.95& 214.19\,$\pm$\,15.80&   3.00\,$\pm$\,0.44&  $10.9\,\pm\,1.00$\\
 &   HEG&   --&   --&   --&                 --&  24.99\,$\pm$\,5.08&  52.33\,$\pm$\,7.29&   2.19\,$\pm$\,0.54&  $11.5\,\pm\,0.32$\\
    $\beta$\,Cet&  RGS1&  19.48\,$\pm$\,8.40&  79.44\,$\pm$\,11.74&   4.23\,$\pm$\,1.93&          $<\,11.3$&   --&   --&   --&                 --\\
 &  LETG&  58.12\,$\pm$\,16.27& 178.69\,$\pm$\,20.17&   3.03\,$\pm$\,0.92&          $<\,11.0$&   --&   --&   --&                 --\\
 &   MEG&   5.49\,$\pm$\,3.00&  23.61\,$\pm$\,5.25&   4.80\,$\pm$\,2.83&          $<\,11.5$& 127.79\,$\pm$\,14.00& 211.92\,$\pm$\,17.14&   1.82\,$\pm$\,0.25&  $11.7\,\pm\,0.12$\\
 &   HEG&   --&   --&   --&                 --&  32.41\,$\pm$\,6.07&  64.72\,$\pm$\,8.30&   2.09\,$\pm$\,0.47&  $11.6\,\pm\,0.26$\\
         Canopus&   MEG&   1.20\,$\pm$\,1.41&  13.42\,$\pm$\,3.87&  12.47\,$\pm$\,15.09&                 --&  23.29\,$\pm$\,5.74&  42.20\,$\pm$\,7.13&   1.99\,$\pm$\,0.59&  $11.6\,\pm\,0.33$\\
 &   HEG&   --&   --&   --&                 --&   4.68\,$\pm$\,2.28&  13.03\,$\pm$\,3.66&   2.91\,$\pm$\,1.64&          $<\,12.0$\\
   $\chi^1$\,Ori&  RGS1&  22.27\,$\pm$\,8.18& 123.42\,$\pm$\,13.40&   5.75\,$\pm$\,2.20&          $<\,10.5$&   --&   --&   --&                 --\\
         EK\,Dra&  RGS1&  30.25\,$\pm$\,8.06&  59.68\,$\pm$\,9.82&   2.05\,$\pm$\,0.64&  $10.4\,\pm\,0.29$&   --&   --&   --&                 --\\
 &  LETG&  23.82\,$\pm$\,7.19&  31.03\,$\pm$\,7.53&   1.29\,$\pm$\,0.50&  $10.8\,\pm\,0.28$&   --&   --&   --&                 --\\
 $\epsilon$\,Eri&  RGS1&  41.69\,$\pm$\,8.90& 160.21\,$\pm$\,14.34&   3.99\,$\pm$\,0.92&          $<\,10.9$&   --&   --&   --&                 --\\
 &  LETG& 153.61\,$\pm$\,14.85& 453.99\,$\pm$\,22.78&   2.92\,$\pm$\,0.32&  $10.0\,\pm\,0.20$&   --&   --&   --&                 --\\
         EQ\,Peg&  RGS1&  46.25\,$\pm$\,9.95& 142.72\,$\pm$\,13.99&   3.20\,$\pm$\,0.76&          $<\,10.8$&   --&   --&   --&                 --\\
         ER\,Vul&   MEG&   4.14\,$\pm$\,3.03&  18.06\,$\pm$\,5.05&   4.88\,$\pm$\,3.83&          $<\,11.9$&  51.41\,$\pm$\,9.56& 168.72\,$\pm$\,14.75&   3.60\,$\pm$\,0.74&          $<\,12.1$\\
 &   HEG&   --&   --&   --&                 --&  16.96\,$\pm$\,4.53&  32.71\,$\pm$\,6.17&   2.02\,$\pm$\,0.66&  $11.6\,\pm\,0.37$\\
         EV\,Lac&  RGS1& 128.92\,$\pm$\,16.34& 291.00\,$\pm$\,20.31&   2.34\,$\pm$\,0.34&  $10.3\,\pm\,0.16$&   --&   --&   --&                 --\\
 &   MEG&  40.31\,$\pm$\,6.59&  54.92\,$\pm$\,7.58&   1.52\,$\pm$\,0.33&  $10.7\,\pm\,0.15$&  85.95\,$\pm$\,10.78& 222.32\,$\pm$\,16.34&   2.84\,$\pm$\,0.41&  $11.1\,\pm\,0.48$\\
 &   HEG&   --&   --&   --&                 --&  24.32\,$\pm$\,5.20&  68.82\,$\pm$\,8.55&   2.96\,$\pm$\,0.73&          $<\,12.0$\\
      HD\,223460&   MEG&  10.64\,$\pm$\,3.89&   7.37\,$\pm$\,3.36&   0.77\,$\pm$\,0.45&  $11.1\,\pm\,0.43$&   --&   --&   --&                 --\\
   $\kappa$\,Cet&  RGS1&  63.15\,$\pm$\,11.60& 153.99\,$\pm$\,15.00&   2.55\,$\pm$\,0.53&  $10.2\,\pm\,0.28$&   --&   --&   --&                 --\\
      $\mu$\,Vel&   MEG&   4.26\,$\pm$\,2.96&  27.60\,$\pm$\,5.73&   7.24\,$\pm$\,5.24&          $<\,11.5$&  50.20\,$\pm$\,8.94&  59.62\,$\pm$\,9.23&   1.30\,$\pm$\,0.31&  $11.9\,\pm\,0.17$\\
 &   HEG&   --&   --&   --&                 --&  24.12\,$\pm$\,5.49&  31.65\,$\pm$\,5.93&   1.37\,$\pm$\,0.40&  $11.9\,\pm\,0.22$\\
    $\pi^1$\,UMa&  RGS1&  30.39\,$\pm$\,8.22&  94.81\,$\pm$\,11.76&   3.24\,$\pm$\,0.96&          $<\,10.8$&   --&   --&   --&                 --\\
         Procyon&  LETG& 203.00\,$\pm$\,16.80& 652.40\,$\pm$\,27.40&   3.17\,$\pm$\,0.29&  $9.90\,\pm\,0.25$&   --&   --&   --&                 --\\
       Prox\,Cen&   MEG&   5.73\,$\pm$\,2.45&  18.19\,$\pm$\,4.33&   3.55\,$\pm$\,1.74&          $<\,10.5$&   5.54\,$\pm$\,2.89&  18.02\,$\pm$\,5.03&   3.57\,$\pm$\,2.11&          $<\,12.9$\\
 &   HEG&   --&   --&   --&                 --&   2.93\,$\pm$\,1.73&  12.81\,$\pm$\,4.92&   4.57\,$\pm$\,3.22&          $<\,12.9$\\
     Speedy\,Mic&   MEG&   3.75\,$\pm$\,2.22&   8.56\,$\pm$\,3.14&   2.55\,$\pm$\,1.77&          $<\,11.2$&  11.57\,$\pm$\,3.68&  37.35\,$\pm$\,6.82&   3.54\,$\pm$\,1.30&          $<\,12.5$\\
 &   HEG&   --&   --&   --&                 --&   2.77\,$\pm$\,1.74&  10.59\,$\pm$\,3.32&   3.99\,$\pm$\,2.80&          $<\,13.0$\\
       V471\,Tau&  LETG&  16.11\,$\pm$\,7.37&  43.43\,$\pm$\,9.19&   2.66\,$\pm$\,1.34&          $<\,11.2$&   --&   --&   --&                 --\\
         VW\,Cep&  LETG&  72.06\,$\pm$\,12.18&  89.71\,$\pm$\,12.90&   1.23\,$\pm$\,0.27&  $10.8\,\pm\,0.15$&   --&   --&   --&                 --\\
      $\xi$\,UMa&   MEG&  39.15\,$\pm$\,6.40&  72.85\,$\pm$\,8.65&   2.08\,$\pm$\,0.42&  $10.4\,\pm\,0.19$& 138.57\,$\pm$\,13.41& 354.90\,$\pm$\,20.13&   2.81\,$\pm$\,0.32&  $11.1\,\pm\,0.31$\\
 &   HEG&   --&   --&   --&                 --&  29.59\,$\pm$\,5.73&  81.17\,$\pm$\,9.20&   2.87\,$\pm$\,0.64&          $<\,12.1$\\
         YY\,Gem&  LETG&  50.64\,$\pm$\,9.63& 115.75\,$\pm$\,12.44&   2.26\,$\pm$\,0.49&  $10.4\,\pm\,0.23$&   --&   --&   --&                 --\\
         YZ\,CMi&  RGS1&  88.70\,$\pm$\,13.07& 169.54\,$\pm$\,15.80&   2.00\,$\pm$\,0.35&  $10.5\,\pm\,0.15$&   --&   --&   --&                 --\\
\hline
\end{tabular}
}
\renewcommand{\arraystretch}{1}
\end{flushleft}
\end{table*}

\begin{table*}[!ht]
\begin{flushleft}
\renewcommand{\arraystretch}{1.1}
\caption{\label{lxem_rs}Same as Table~\ref{lxem} for RS\,CVn systems in our sample.}
\vspace{-.3cm}
{\scriptsize
\begin{tabular}{lccccccccc}
&&$L_X$&\multicolumn{2}{c}{line flux ($10^{-13}$\,erg\,cm$^{-2}$\,s$^{-1}$)}&$L_{X, O\,\mathsc{vii}}$&$L_{X, Ne\,\mathsc{ix}}$&\multicolumn{2}{c}{log(EM/cm$^{-3}$)}\\
 star & Instr.& [10$^{28}$erg/s]& O\,{\sc viii} & O\,{\sc vii}(r)&[10$^{28}$erg/s] & [10$^{28}$erg/s] & O\,{\sc vii}$^a$&Ne\,{\sc ix}$^b$\\
         AR\,Lac&  RGS1& 627.14&   9.35\,$\pm$\,0.28&   1.15\,$\pm$\,0.12&   4.59\,$\pm$\,1.50&   --&  51.88\,$\pm$\,0.10&   --\\
 &   MEG& 514.49&   7.39\,$\pm$\,0.48&   0.80\,$\pm$\,0.23&   2.34\,$\pm$\,3.31&  10.31\,$\pm$\,2.31&   --&  52.87\,$\pm$\,0.09\\
 &   HEG& 415.12&   --&   --&   --&   7.62\,$\pm$\,4.18&   --&  52.88\,$\pm$\,0.17\\
         Capella&  RGS1& 145.38&  28.64\,$\pm$\,0.36&   9.96\,$\pm$\,0.23&   3.54\,$\pm$\,0.22&   --&   --&   --\\
 &  LETG& 186.56&  30.56\,$\pm$\,0.25&   8.87\,$\pm$\,0.16&   3.33\,$\pm$\,0.16&   --&  51.74\,$\pm$\,0.02&   --\\
 &   MEG& 153.08&  30.89\,$\pm$\,0.44&   8.15\,$\pm$\,0.29&   3.24\,$\pm$\,0.30&   2.67\,$\pm$\,0.16&  51.71\,$\pm$\,0.14&  52.37\,$\pm$\,0.02\\
 &   HEG& 127.43&   --&   --&   --&   2.91\,$\pm$\,0.31&   --&  52.39\,$\pm$\,0.04\\
        HR\,1099&  RGS1& 432.49&  16.22\,$\pm$\,0.40&   2.72\,$\pm$\,0.20&   5.40\,$\pm$\,1.15&   --&  51.93\,$\pm$\,0.05&   --\\
 &  LETG& 973.74&  24.54\,$\pm$\,0.35&   2.84\,$\pm$\,0.17&   5.09\,$\pm$\,0.88&   --&  51.95\,$\pm$\,0.05&   --\\
 &   MEG&1000.85&  28.99\,$\pm$\,0.56&   3.42\,$\pm$\,0.27&   5.98\,$\pm$\,1.41&  13.44\,$\pm$\,1.07&  52.03\,$\pm$\,0.09&  53.11\,$\pm$\,0.03\\
 &   HEG& 798.38&   --&   --&   --&  16.25\,$\pm$\,2.12&   --&  53.20\,$\pm$\,0.04\\
         II\,Peg&   MEG&1318.60&  19.90\,$\pm$\,0.68&   1.54\,$\pm$\,0.28&  10.43\,$\pm$\,4.29&  19.72\,$\pm$\,2.71&  52.01\,$\pm$\,0.25&  53.26\,$\pm$\,0.05\\
 &   HEG&1045.50&   --&   --&   --&  12.80\,$\pm$\,4.19&   --&  53.02\,$\pm$\,0.12\\
         IM\,Peg&   MEG&2520.70&   3.37\,$\pm$\,0.19&   0.28\,$\pm$\,0.08&   6.38\,$\pm$\,6.84&  19.34\,$\pm$\,4.41&  51.98\,$\pm$\,0.07&  53.20\,$\pm$\,0.08\\
 &   HEG&2316.90&   --&   --&   --&  20.76\,$\pm$\,8.66&   --&  53.25\,$\pm$\,0.14\\
  $\lambda$\,And&  RGS1& 249.61&  12.12\,$\pm$\,0.31&   1.67\,$\pm$\,0.14&   2.21\,$\pm$\,0.61&   --&  51.62\,$\pm$\,0.07&   --\\
 &  LETG& 935.90&  23.22\,$\pm$\,0.35&   2.25\,$\pm$\,0.17&   3.38\,$\pm$\,0.77&   --&  51.75\,$\pm$\,0.07&   --\\
 &   MEG& 198.33&  10.74\,$\pm$\,0.36&   1.41\,$\pm$\,0.18&   1.93\,$\pm$\,0.75&   3.51\,$\pm$\,0.51&  51.54\,$\pm$\,0.17&  52.55\,$\pm$\,0.05\\
 &   HEG& 121.17&   --&   --&   --&   1.92\,$\pm$\,0.73&   --&  52.22\,$\pm$\,0.13\\
 $\sigma^2$\,CrB&  RGS1& 322.59&  20.56\,$\pm$\,0.52&   3.23\,$\pm$\,0.25&   3.24\,$\pm$\,0.76&   --&  51.75\,$\pm$\,0.09&   --\\
 &   MEG& 305.60&  19.17\,$\pm$\,0.47&   2.59\,$\pm$\,0.23&   3.25\,$\pm$\,0.75&   5.78\,$\pm$\,0.54&  51.66\,$\pm$\,0.16&  52.73\,$\pm$\,0.03\\
 &   HEG& 154.81&   --&   --&   --&   5.73\,$\pm$\,1.00&   --&  52.68\,$\pm$\,0.06\\
         TY\,Pyx&   MEG& 532.60&   3.64\,$\pm$\,0.27&   0.73\,$\pm$\,0.16&   4.64\,$\pm$\,3.35&   8.64\,$\pm$\,2.34&  51.93\,$\pm$\,0.08&  52.93\,$\pm$\,0.09\\
 &   HEG& 550.70&   --&   --&   --&   7.19\,$\pm$\,4.54&   --&  52.78\,$\pm$\,0.21\\
         UX\,Ari&  RGS1& 953.03&  13.85\,$\pm$\,0.34&   1.59\,$\pm$\,0.15&   7.80\,$\pm$\,2.51&   --&  52.17\,$\pm$\,0.07&   --\\
 &  LETG&1306.36&  12.62\,$\pm$\,0.23&   1.57\,$\pm$\,0.12&   9.15\,$\pm$\,1.91&   --&  52.17\,$\pm$\,0.07&   --\\
 &   MEG& 804.46&  10.36\,$\pm$\,0.46&   1.69\,$\pm$\,0.25&   8.31\,$\pm$\,4.07&  21.45\,$\pm$\,3.09&  52.20\,$\pm$\,0.39&  53.26\,$\pm$\,0.05\\
 &   HEG& 502.40&   --&   --&   --&   8.87\,$\pm$\,4.10&   --&  52.97\,$\pm$\,0.15\\
       V824\,Ara&   MEG& 254.66&   8.63\,$\pm$\,0.30&   1.02\,$\pm$\,0.14&   2.39\,$\pm$\,0.92&   5.44\,$\pm$\,0.72&  51.57\,$\pm$\,0.21&  52.67\,$\pm$\,0.05\\
 &   HEG& 212.00&   --&   --&   --&   5.49\,$\pm$\,1.34&   --&  52.72\,$\pm$\,0.08\\
         VY\,Ari&  RGS1& 366.79&   7.66\,$\pm$\,0.24&   1.22\,$\pm$\,0.12&   5.62\,$\pm$\,1.55&   --&  51.94\,$\pm$\,0.09&   --\\
\hline
\end{tabular}
\\
$^a$at T=2\,MK\ \ \ \
$^b$at T=4\,MK\\
}
\renewcommand{\arraystretch}{1}
\end{flushleft}
\end{table*}

\begin{table*}[!ht]
\begin{flushleft}
\renewcommand{\arraystretch}{1.1}
\caption{\label{fi_rs}Same as Table~\ref{fi} for RS\,CVn systems in our sample.}
\vspace{-.4cm}
{\scriptsize
\begin{tabular}{p{.8cm}p{.5cm}cccccccc}
&&\multicolumn{4}{c}{O\,{\sc vii}}&\multicolumn{4}{c}{Ne\,{\sc ix}}\\
 star & Instr.& i [cts] & f [cts] & f/i & log($n_e$) & i [cts] & f [cts] & f/i& log($n_e$)\\
         AR\,Lac&  RGS1&  48.70\,$\pm$\,14.75& 110.15\,$\pm$\,16.82&   2.35\,$\pm$\,0.80&  $10.3\,\pm\,0.42$&   --&   --&   --&                 --\\
 &   MEG&   1.60\,$\pm$\,2.58&   3.80\,$\pm$\,2.86&   2.65\,$\pm$\,4.71&                 --&  48.98\,$\pm$\,8.38& 140.80\,$\pm$\,13.32&   3.15\,$\pm$\,0.62&          $<\,11.8$\\
 &   HEG&   --&   --&   --&                 --&  10.09\,$\pm$\,3.67&  13.90\,$\pm$\,4.23&   1.44\,$\pm$\,0.68&  $11.9\,\pm\,0.40$\\
         Capella&  RGS1& 339.67\,$\pm$\,30.84&1618.64\,$\pm$\,48.15&   4.95\,$\pm$\,0.47&                 --&   --&   --&   --&                 --\\
 &  LETG& 645.44\,$\pm$\,33.14&2348.32\,$\pm$\,53.78&   3.59\,$\pm$\,0.20&  $9.51\,\pm\,0.40$&   --&   --&   --&                 --\\
 &   MEG& 167.92\,$\pm$\,13.94& 505.48\,$\pm$\,23.16&   3.36\,$\pm$\,0.32&  $9.76\,\pm\,0.38$& 709.92\,$\pm$\,32.20&1373.17\,$\pm$\,42.87&   2.12\,$\pm$\,0.12&  $11.5\,\pm\,0.06$\\
 &   HEG&   --&   --&   --&                 --& 190.47\,$\pm$\,14.76& 373.48\,$\pm$\,20.09&   2.05\,$\pm$\,0.19&  $11.6\,\pm\,0.10$\\
        HR\,1099&  RGS1&  82.61\,$\pm$\,19.30& 254.27\,$\pm$\,23.89&   3.20\,$\pm$\,0.80&          $<\,10.9$&   --&   --&   --&                 --\\
 &  LETG& 114.78\,$\pm$\,18.72& 254.79\,$\pm$\,22.49&   2.19\,$\pm$\,0.41&  $10.4\,\pm\,0.18$&   --&   --&   --&                 --\\
 &   MEG&  36.05\,$\pm$\,8.90&  93.87\,$\pm$\,11.80&   2.91\,$\pm$\,0.81&  $10.0\,\pm\,0.75$& 304.42\,$\pm$\,24.39& 834.17\,$\pm$\,34.81&   3.01\,$\pm$\,0.27&  $10.9\,\pm\,0.38$\\
 &   HEG&   --&   --&   --&                 --& 100.55\,$\pm$\,11.62& 228.75\,$\pm$\,16.25&   2.38\,$\pm$\,0.32&  $11.4\,\pm\,0.20$\\
         II\,Peg&   MEG&  33.61\,$\pm$\,6.95&  44.83\,$\pm$\,7.82&   1.49\,$\pm$\,0.40&  $10.7\,\pm\,0.20$& 103.95\,$\pm$\,13.27& 268.87\,$\pm$\,18.85&   2.84\,$\pm$\,0.41&  $11.1\,\pm\,0.48$\\
 &   HEG&   --&   --&   --&                 --&  19.99\,$\pm$\,4.55&  46.00\,$\pm$\,6.95&   2.41\,$\pm$\,0.66&  $11.4\,\pm\,0.50$\\
         IM\,Peg&   MEG&   4.46\,$\pm$\,3.49&  10.70\,$\pm$\,4.16&   2.68\,$\pm$\,2.34&          $<\,11.2$&  49.64\,$\pm$\,9.16& 127.99\,$\pm$\,13.33&   2.83\,$\pm$\,0.60&  $11.1\,\pm\,1.05$\\
 &   HEG&   --&   --&   --&                 --&  12.86\,$\pm$\,4.01&  30.64\,$\pm$\,5.83&   2.49\,$\pm$\,0.91&  $11.3\,\pm\,1.14$\\
  $\lambda$\,And&  RGS1&  39.10\,$\pm$\,15.08& 131.08\,$\pm$\,17.88&   3.48\,$\pm$\,1.42&          $<\,10.6$&   --&   --&   --&                 --\\
 &  LETG&  72.16\,$\pm$\,21.57& 259.98\,$\pm$\,26.03&   3.56\,$\pm$\,1.12&          $<\,10.5$&   --&   --&   --&                 --\\
 &   MEG&  11.47\,$\pm$\,4.64&  33.55\,$\pm$\,6.64&   3.27\,$\pm$\,1.47&          $<\,10.8$&  81.33\,$\pm$\,11.89& 219.34\,$\pm$\,17.80&   2.96\,$\pm$\,0.49&  $11.0\,\pm\,1.12$\\
 &   HEG&   --&   --&   --&                 --&  12.84\,$\pm$\,3.92&  35.64\,$\pm$\,6.22&   2.90\,$\pm$\,1.02&          $<\,12.0$\\
 $\sigma^2$\,CrB&  RGS1&  61.46\,$\pm$\,15.44& 173.44\,$\pm$\,20.53&   2.93\,$\pm$\,0.81&  $10.0\,\pm\,0.80$&   --&   --&   --&                 --\\
 &   MEG&  50.90\,$\pm$\,7.91&  94.28\,$\pm$\,10.29&   2.07\,$\pm$\,0.39&  $10.4\,\pm\,0.17$& 296.10\,$\pm$\,20.73& 521.80\,$\pm$\,26.92&   1.93\,$\pm$\,0.17&  $11.6\,\pm\,0.08$\\
 &   HEG&   --&   --&   --&                 --&  88.11\,$\pm$\,9.85& 130.60\,$\pm$\,11.74&   1.55\,$\pm$\,0.22&  $11.8\,\pm\,0.11$\\
         TY\,Pyx&   MEG&   6.52\,$\pm$\,2.82&   7.52\,$\pm$\,2.98&   1.29\,$\pm$\,0.76&  $10.8\,\pm\,0.49$&  26.51\,$\pm$\,7.03&  72.69\,$\pm$\,9.73&   3.01\,$\pm$\,0.89&          $<\,11.9$\\
 &   HEG&   --&   --&   --&                 --&   7.09\,$\pm$\,3.13&  16.96\,$\pm$\,4.54&   2.50\,$\pm$\,1.29&          $<\,12.3$\\
         UX\,Ari&  RGS1&  30.31\,$\pm$\,15.89& 117.54\,$\pm$\,18.79&   4.03\,$\pm$\,2.21&          $<\,11.5$&   --&   --&   --&                 --\\
 &  LETG&  53.55\,$\pm$\,14.68& 227.42\,$\pm$\,20.40&   4.19\,$\pm$\,1.21&          $<\,11.0$&   --&   --&   --&                 --\\
 &   MEG&   7.02\,$\pm$\,3.76&  20.91\,$\pm$\,5.42&   3.33\,$\pm$\,1.98&          $<\,10.8$& 120.16\,$\pm$\,12.36& 237.38\,$\pm$\,17.32&   2.17\,$\pm$\,0.27&  $11.5\,\pm\,0.15$\\
 &   HEG&   --&   --&   --&                 --&   7.99\,$\pm$\,3.23&  18.57\,$\pm$\,4.60&   2.43\,$\pm$\,1.15&          $<\,12.4$\\
       V824\,Ara&   MEG&  22.35\,$\pm$\,5.24&  30.23\,$\pm$\,6.02&   1.51\,$\pm$\,0.46&  $10.7\,\pm\,0.23$& 157.09\,$\pm$\,15.50& 298.00\,$\pm$\,19.26&   2.08\,$\pm$\,0.25&  $11.6\,\pm\,0.13$\\
 &   HEG&   --&   --&   --&                 --&  29.84\,$\pm$\,5.90&  66.01\,$\pm$\,8.44&   2.31\,$\pm$\,0.54&  $11.4\,\pm\,0.35$\\
         VY\,Ari&  RGS1&  52.58\,$\pm$\,14.52& 144.81\,$\pm$\,17.34&   2.86\,$\pm$\,0.86&  $10.1\,\pm\,0.79$&   --&   --&   --&                 --\\
\hline
\end{tabular}
}
\renewcommand{\arraystretch}{1}
\end{flushleft}
\end{table*}

\subsection{Temperatures from H-like and He-like line ratios}

\begin{figure*}[!ht]
 \resizebox{\hsize}{!}{\includegraphics{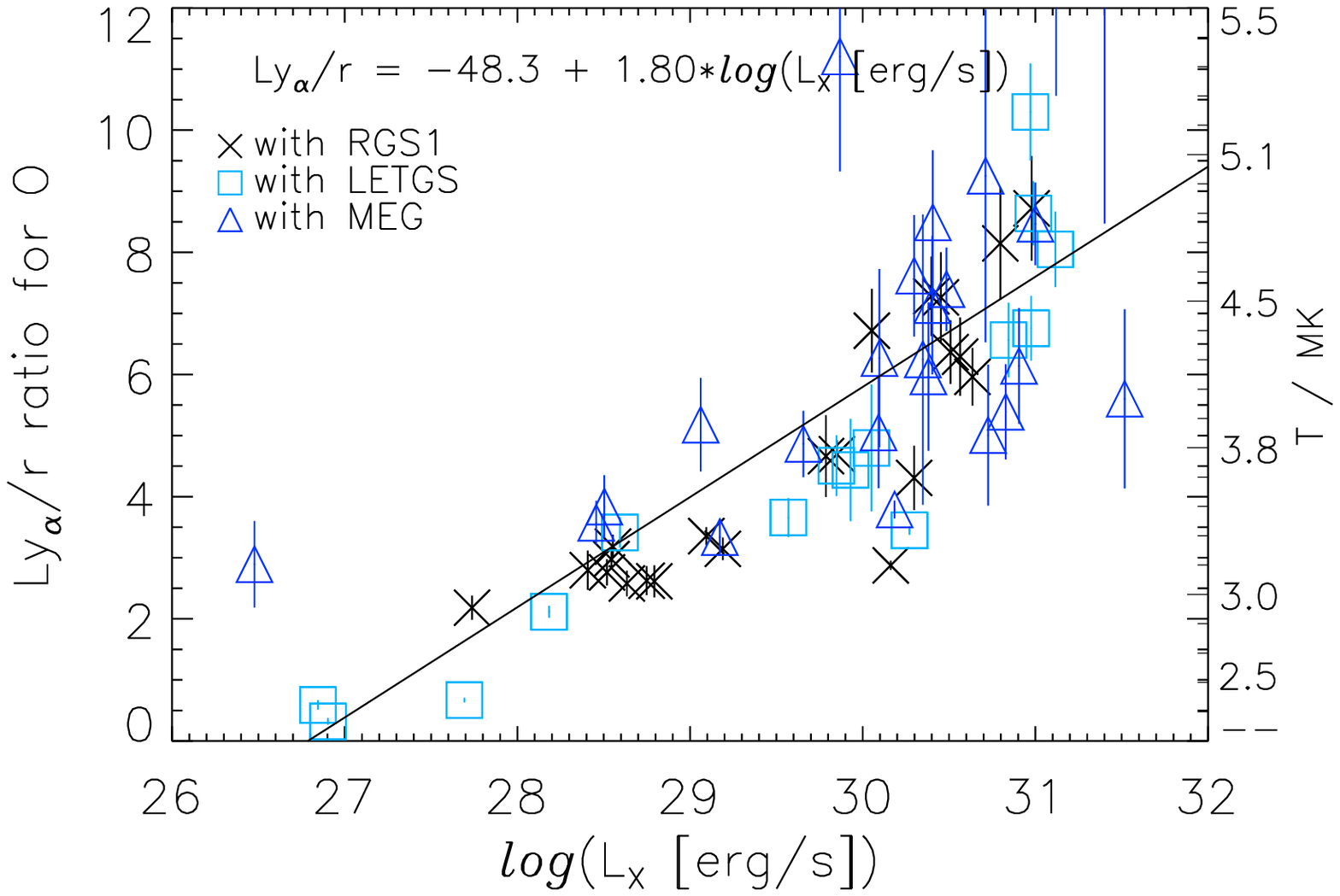}\includegraphics{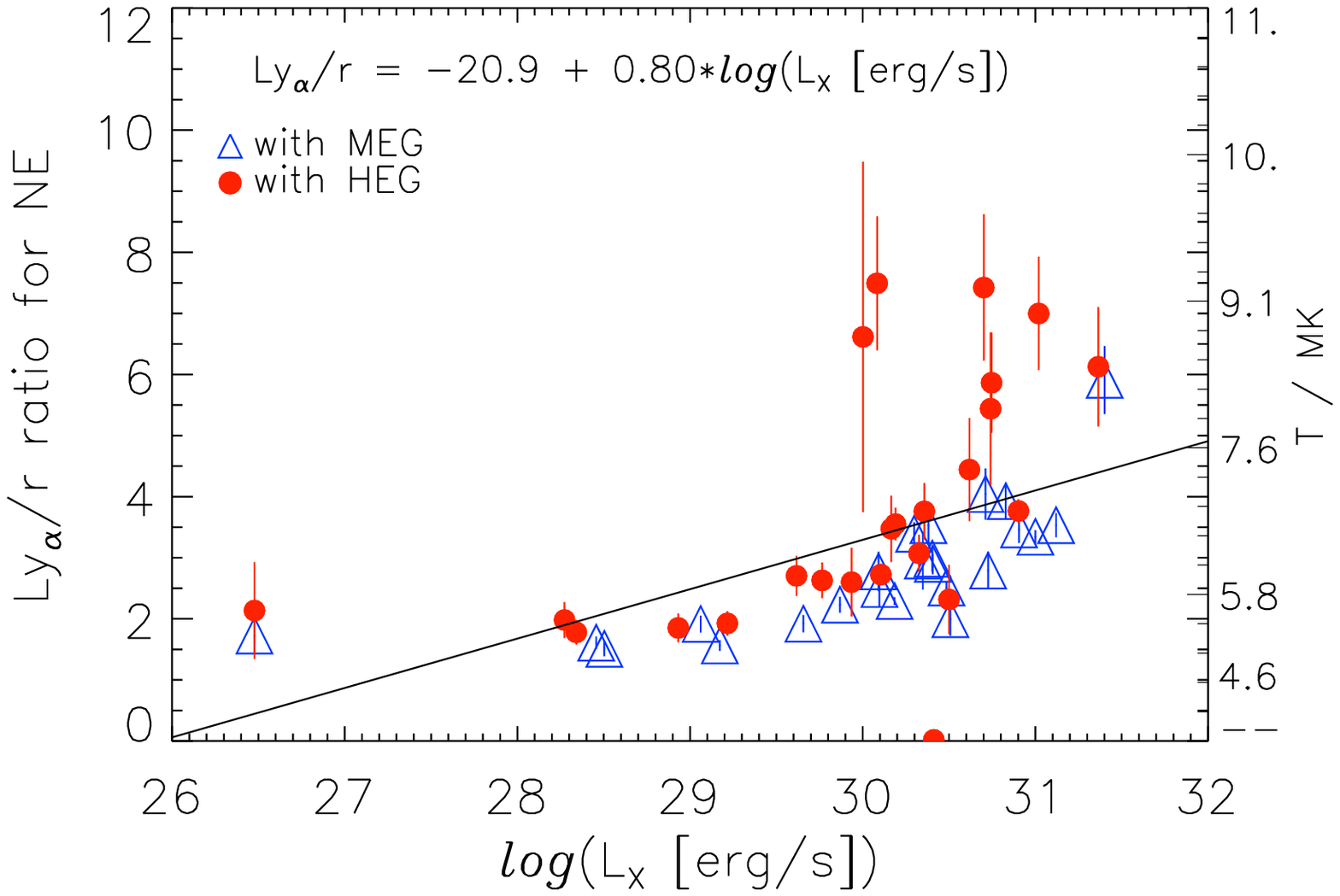}}
 \resizebox{\hsize}{!}{\includegraphics{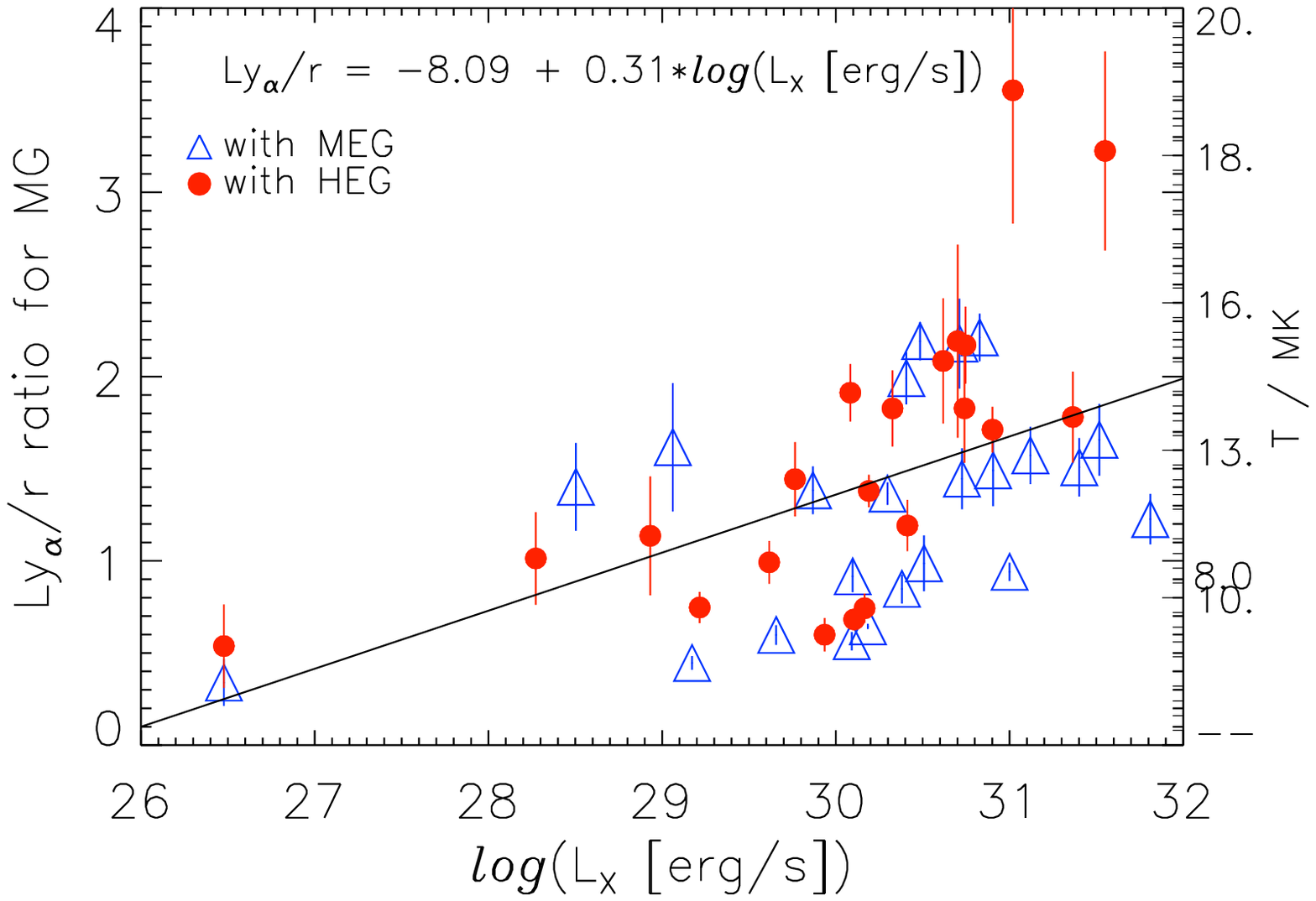}\includegraphics{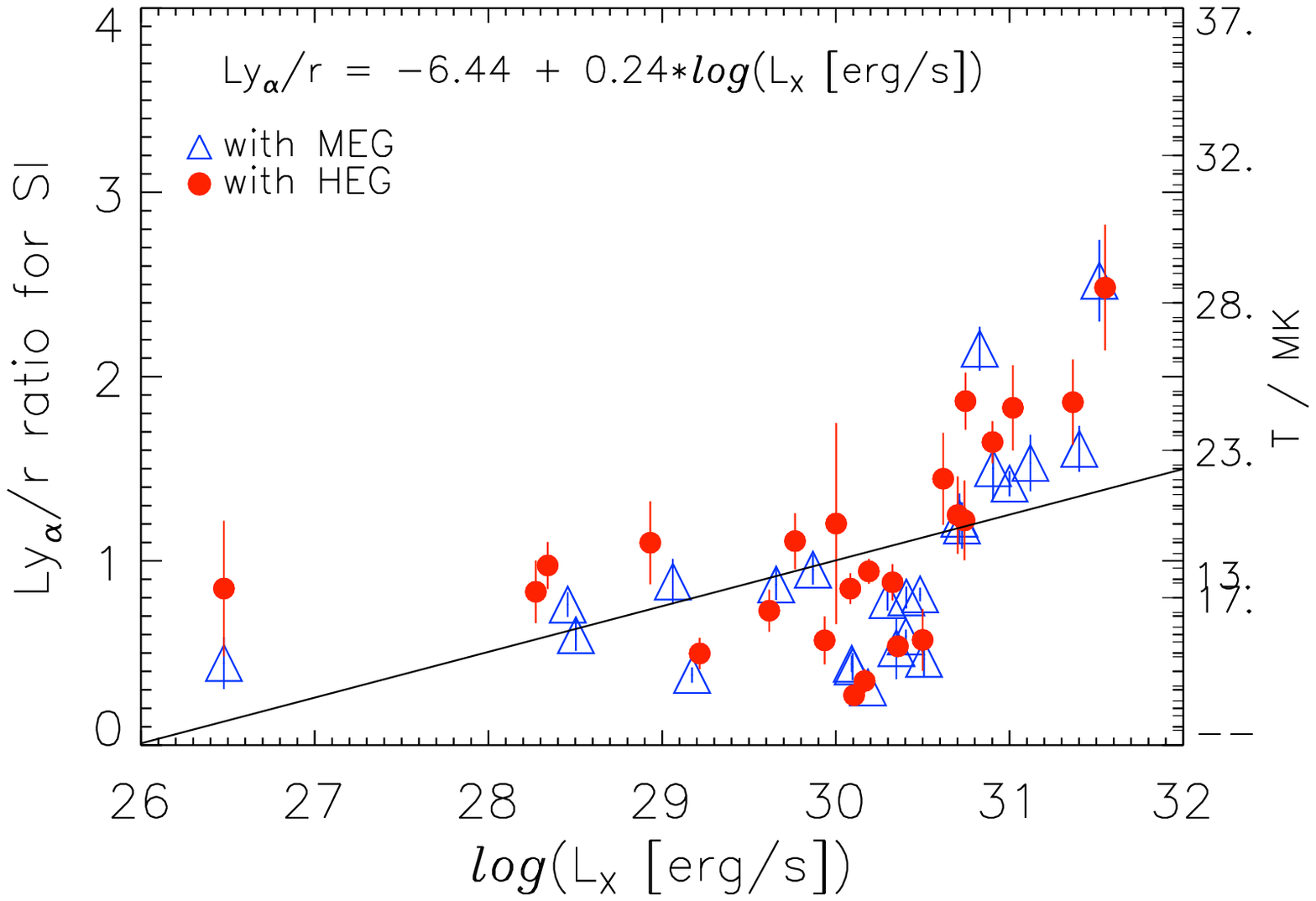}}
\caption[]{\label{lxly}Line flux ratios of H-like (Ly$_\alpha$) and
He-like (r) lines vs. activity indicator $L_X$ for the respective ions of
oxygen, neon, magnesium, and silicon. Linear fits yield good approximations
with decreasing slopes for higher-Z ions. The right axes give average
temperatures derived from comparison of the measured line ratios with
temperature-sensitive predictions from the APEC line database.}
\end{figure*}

In the past, coronal temperatures were estimated via global spectral fits from
low-resolution spectra. With high-resolution spectra now available we can
additionally determine average coronal temperatures by considering individual
line fluxes, that yield a temperature characterizing the formation of, e.g., the
O\,{\sc vii} triplet. The ratios of line fluxes originating from adjacent
ionization stages of the same chemical element allow the determination of
temperatures independent of the respective elemental abundances.
Since for the conversion of measured line
fluxes to plasma temperatures and for the construction of synthetic spectra from
global plasma emission models the same databases are used, the systematic
errors are basically the same. However, a careful choice of strong lines for the
calculation of line ratios can reduce systematic errors, since strong lines
are mostly not as much affected by uncertainties as weaker lines. Also, line
blends from unidentified weaker lines cannot harm strong lines as much as weaker
lines. A detailed discussion of the interpretation of line ratios and global
fit approaches is given in \cite{guedel04}.\\
The strongest lines in the grating spectra are the H-like and He-like lines of
carbon, nitrogen, oxygen, neon, magnesium, and silicon. We measured the line
fluxes of these lines in order to calculate temperature-sensitive ratios for
these ions for each star in our sample. In Fig.~\ref{lxly} we show our results
plotted vs. the total X-ray luminosities (cf. Tables~\ref{lxem}/\ref{lxem_rs}
for oxygen); a clear trend can be recognized for all ions. We overplotted
best-fit linear regressions and find decreasing slopes for increasing Z, thus
increasing formation temperatures of the H-like and He-like states. A
correlation of plasma temperature with the degree of activity has long been
known \citep[e.g.,][]{schrijver84,schm90}. In terms of a putative emission
measure distribution, this trend would indicate a predominance of higher
emission
measure at higher temperature in stars with a higher activity level. When
considering H- and He-like lines of individual elements, this trend should be
reflected in a larger ratio of H-like to He-like line fluxes in the more active
stars, and this is what we now recover from our analysis.\\
A continuous temperature
distribution suggests a mixture of temperatures making the adjacent lines
an ideal means for defining interpolation points fixing the shape of the
temperature distribution. Such an approach has been introduced by,
e.g., \cite{abun} for finding abundance-independent emission measure
distributions. For our purposes we can conclude that the temperatures derived
from the line ratios of H-like and He-like lines characterize the plasma
conditions around the plasma that produces the He-like lines used for the
density analysis.

\subsection{Densities from line flux ratios}

We measured line fluxes from both density-sensitive
He-like lines and carbon-like Fe\,{\sc xxi} lines. With the He-like lines we
probe only the 'cool' plasma component, while with the Fe\,{\sc xxi}
densities we probe the 'hotter' coronal components.

\subsubsection{Densities from Fe\,{\sc xxi} line flux ratios}
\label{fe21dens}

\begin{table}[!ht]
\begin{flushleft}
\renewcommand{\arraystretch}{1.1}
\caption{\label{fe21}Measured Fe\,{\sc xxi} counts obtained from the available LETGS spectra}
\vspace{-.6cm}
{\scriptsize
\begin{tabular}{p{.9cm}rrrr}
& 128.73\,\AA & 117.5\,\AA & 102.22\,\AA & 98.87\,\AA \\
 $transm.^a$ &      0.8050&      0.8491&      0.8993&      0.9115\\
 A$_{\rm eff}$/cm$^2$&  3.67&  6.17&  6.71&  7.24\\
    $\beta$\,Cet& 587.4\,$\pm$\,31.6& 178.8\,$\pm$\,23.1& 198.9\,$\pm$\,24.7& 98.20\,$\pm$\,17.6\\
  $\lambda$\,And& 471.4\,$\pm$\,29.4& 139.5\,$\pm$\,20.7& 144.3\,$\pm$\,24.9& 109.7\,$\pm$\,18.9\\
         AD\,Leo& 61.17\,$\pm$\,14.2& 15.48\,$\pm$\,11.7& 21.64\,$\pm$\,12.9& 14.89\,$\pm$\,9.14\\
           Algol& 354.0\,$\pm$\,25.4& 120.3\,$\pm$\,19.4& 145.3\,$\pm$\,19.9& 77.86\,$\pm$\,14.3\\
         Capella& 1086.\,$\pm$\,43.1& 501.7\,$\pm$\,33.9& 362.0\,$\pm$\,36.6& 69.97\,$\pm$\,21.1\\
         EK\,Dra& 52.02\,$\pm$\,10.4& 28.27\,$\pm$\,11.5& 9.360\,$\pm$\,12.1&                -- \\
          HR1099& 386.8\,$\pm$\,25.6& 156.4\,$\pm$\,19.6& 142.3\,$\pm$\,22.2& 63.43\,$\pm$\,15.4\\
         UX\,Ari& 167.6\,$\pm$\,19.3& 71.58\,$\pm$\,17.6& 35.31\,$\pm$\,17.0& 55.07\,$\pm$\,13.0\\
         YY\,Gem& 46.10\,$\pm$\,10.5& 29.74\,$\pm$\,10.7& 8.531\,$\pm$\,9.01& 17.86\,$\pm$\,7.87\\
\hline
\end{tabular}
\\
$^a$ISM transmission using N(H{\sc i}) $= 10^{18}$, N(He{\sc ii})/N(H{\sc i}) = 0.09, N(He{\sc ii})/N(H{\sc i}) = 0.01
}
\renewcommand{\arraystretch}{1}
\end{flushleft}
\end{table}

\begin{figure*}[!ht]
 \resizebox{\hsize}{!}{\includegraphics{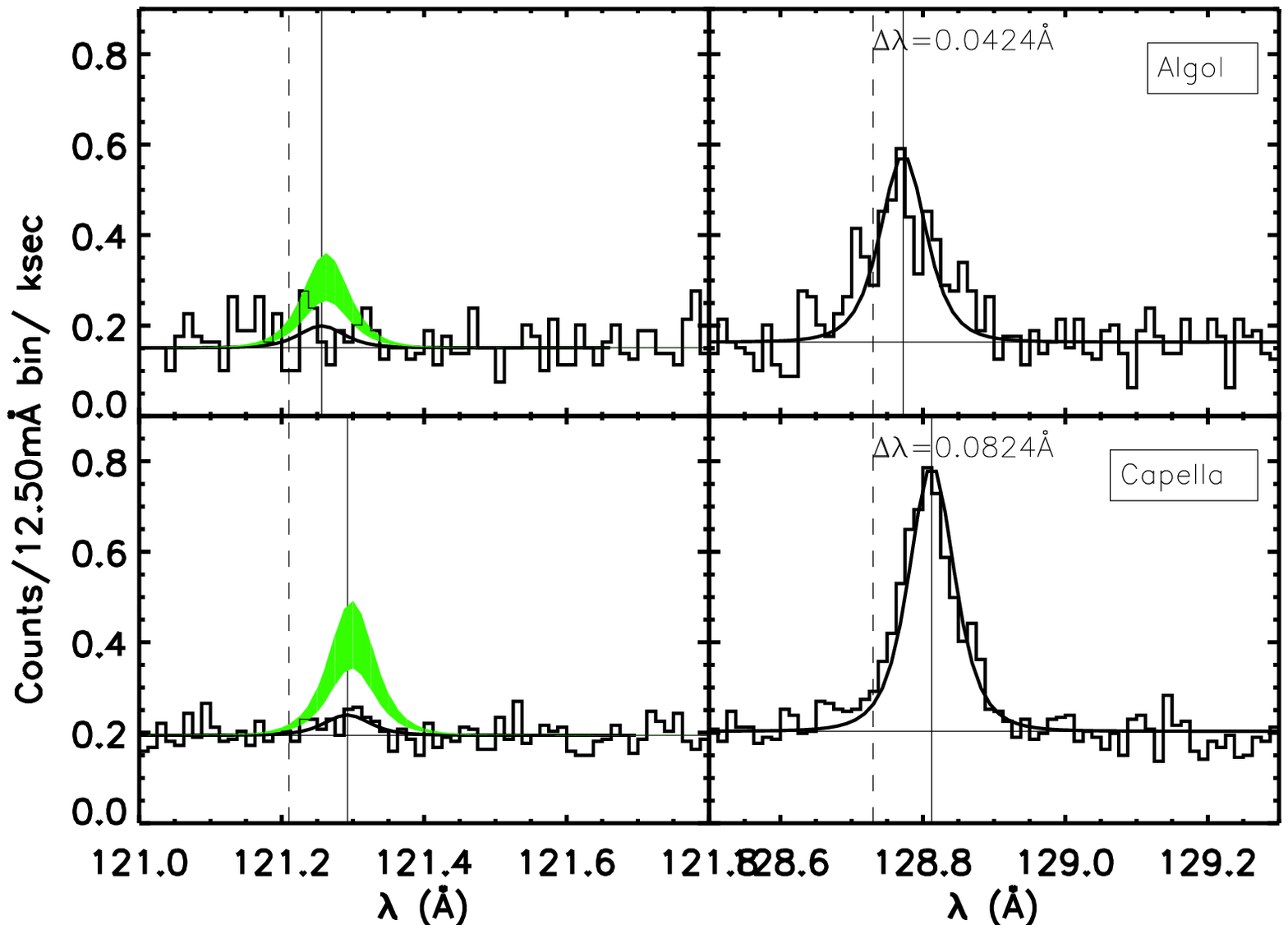}\includegraphics{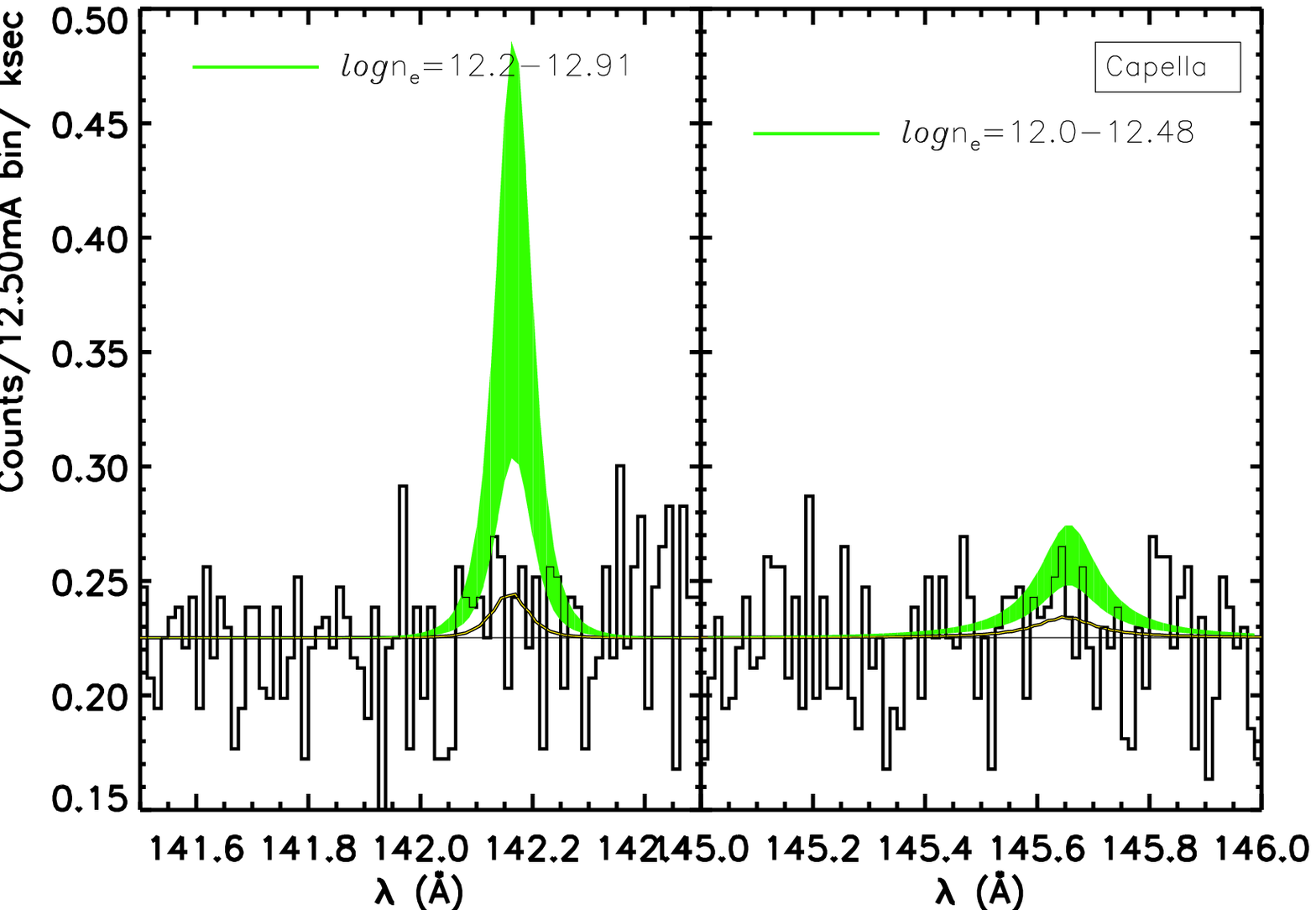}}
\caption[]{\label{fe21_ac}Measurement of Fe\,{\sc xxi} lines for Algol and
Capella.
At most marginal evidence for the presence of Fe\,{\sc xxi} lines at
121.21\,\AA\ ({\bf left panel}) and at 142.16\,\AA\ as well as 145.65\,\AA\
({\bf right panel}) is present. The 128.73\,\AA\ line is shifted due to
calibration uncertainties. The grey shaded areas indicate the expected spectrum
for high density plasma as found by \cite{dupr93}.}
\end{figure*}

As described in Sect.~\ref{fe21_theo} the Fe\,{\sc xxi} density diagnostics are
essentially based on the appearance of certain lines in high-density plasmas.
Our search for densities exceeding $\log n_e=11$ is therefore based on
detections of these density-sensitive lines. In Table~\ref{fe21} we list the
results for our Fe\,{\sc xxi} line count measurements. Table~\ref{fe21}
reveals that for all stars studied the reference line at 128.73\,\AA\ is the
strongest line, yielding flux ratios below one
for all the stars in our sample. In plasmas with densities exceeding
$n_e = 10^{13.5}$\,cm$^{-3}$, the 121.21\,\AA\ is expected to be the strongest
Fe\,{\sc xxi} line, yet it is detected in none of our sample stars. The ``best''
cases for detection are the LETGS spectra of Algol and Capella (see
Fig.~\ref{fe21_ac} left panel), but the statistical significance of the
``features'' appearing at 121.21\,\AA\ is very low. Even if these features are
taken as real, the measured line fluxes do not imply densities higher than
$10^{12}$\,cm$^{-3}$. We also investigated the Fe\,{\sc xxi} lines at
142.16\,\AA\ and at 145.65\,\AA\ for Capella, which were used by \cite{dupr93}.
They measured flux ratios with EUVE $\lambda 142.16/\lambda 128.73=0.51$ and
$\lambda 145.65/\lambda 128.73=0.24$ implying densities $\log n_e=13.2$ and
12.8, but we are unable to detect any significant flux at these wavelengths
in the LETGS spectrum of Capella. The grey shaded areas in
Fig.~\ref{fe21_ac} show the expected spectrum using the line counts for the
128.73\,\AA\ line and the EUVE flux ratios; it can be seen that such high
densities would have been measurable with the LETGS.

\begin{figure*}[!ht]
 \resizebox{\hsize}{!}{\includegraphics{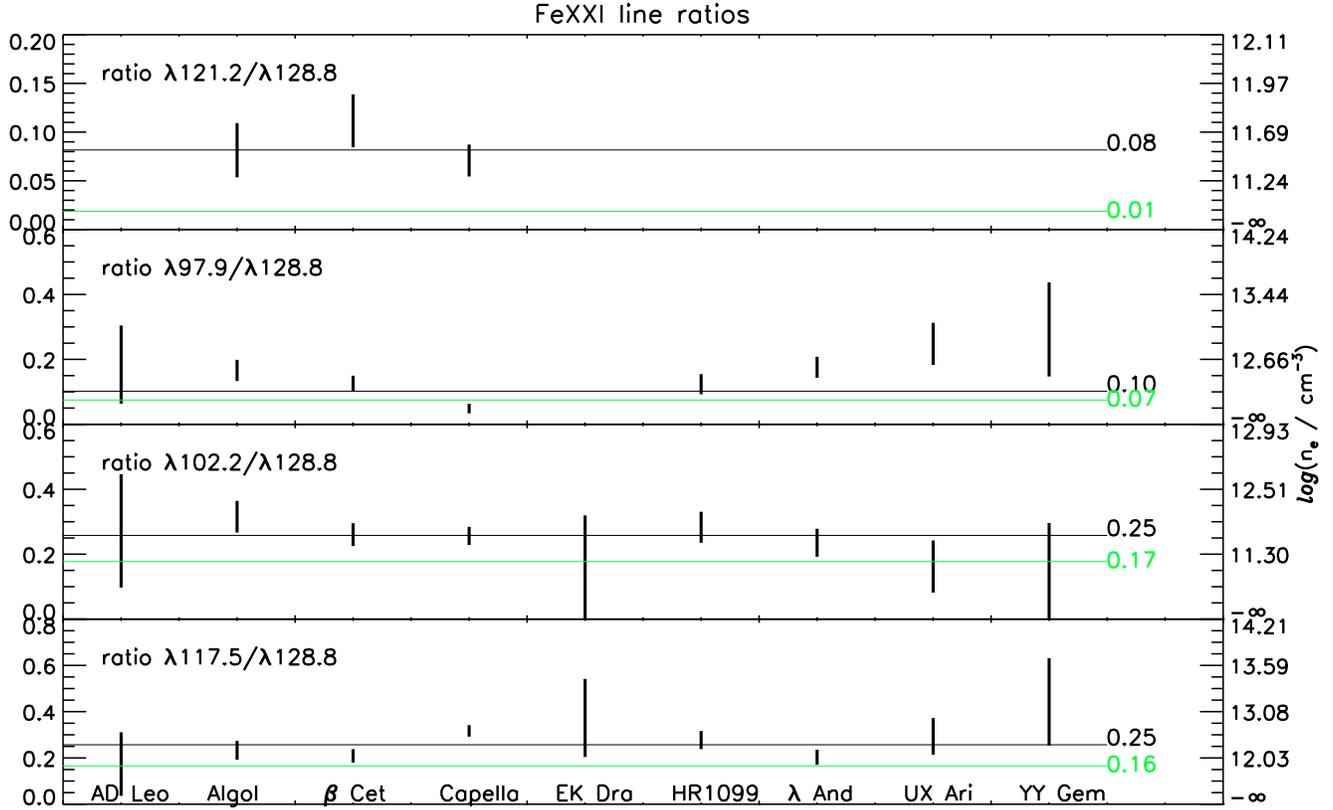}}
\caption[]{\label{fe21rats}Line flux ratios for measured Fe\,{\sc xxi} lines
at 97.9\,\AA, 102.2\,\AA, 117.5\,\AA, 121.21\,\AA, and 128.8\,\AA. The grey
lines represent the theoretical low-density limits. The average ratios,
weighted with the measurement errors, are marked with the black lines.}
\end{figure*}

In Fig.~\ref{fe21rats} we plot the line flux ratios measured for our sample
stars between the detected Fe\,{\sc xxi} lines at 117.5\,\AA, 102.22\,\AA, and
97.87\,\AA, all with respect to the Fe\,{\sc xxi} line at 128.73\,\AA. To
convert line counts into fluxes we used the effective areas and ISM
transmissions listed in Table~\ref{fe21}. Note that we did not consider
individual ISM transmissions
for each star; since this effect is small and differential; the error is
smaller than the statistical error of our measurements. The grey lines in
Fig.~\ref{fe21rats} represent the line flux ratios computed from APEC for
the case of a low density plasma. As can be seen from Fig.~\ref{fe21rats} all
the ratios for 117.5\,\AA/128.73\,\AA\ are above the computed low density
limit, however, all the observed line ratios are consistent with a value of
0.25. Since we consider it unlikely that all observed coronae are above the
low-density limit at the same density, a far more plausible explanation
is that all coronae are in the low-density limit and that a flux ratio of
0.25 is a more appropriate value for the low-density limit than the computed
value of 0.16. Similar conclusions apply to the flux ratios of the
102.2\,\AA/128.73\,\AA\ and 97.87\,\AA/128.73\,\AA\ lines, where the computed
low-density values are 0.17 and 0.07, which has to be compared to the observed
values of 0.25 and 0.10. Again, the most plausible explanation is that all
coronae are in the low density limit, which is consistent with the
non-detection of the Fe\,{\sc xxi} lines at 121.21\,\AA, 142.16\,\AA, and
145.65\,\AA. We also point out that in no case {\it all} Fe\,{\sc xxi} line
ratios yield consistent high densities. Individual deviations from the
low-density limit could be due to unidentified blending or other uncertainties
in the atomic data bases. In particular, inclompleteness of atomic databases
result in a bias towards higher densities. Unknown emission lines could mimic
high densities when unexpectedly showing up where we expect to see
density-sensitive lines.
Our conclusion is that densities above $10^{13}$\,cm$^{-3}$ can definitely be
ruled out, and densities above $5\times 10^{12}$\,cm$^{-3}$ appear highly
improbable.

\subsubsection{Densities from He-like lines}
\label{hedens}

 Our measurements of line fluxes for O\,{\sc vii} and Ne\,{\sc ix} lines are
used to determine f/i ratios which are converted to electron
densities $n_e$ with Eq.~(\ref{fidens}); the derived densities (and 1\,$\sigma$
higher limits) are listed in Tables~\ref{fi}/\ref{fi_rs}. The radiation term $\phi/\phi_c$
describing the contribution to f/i ratios from radiatively induced
f$\rightarrow$i transitions is negligible for Ne\,{\sc ix} and for
O\,{\sc vii} for most of our sources. For O\,{\sc vii} we calculate
$\phi/\phi_c$ values for the stars with the highest effective temperatures
(cf. Table~\ref{sprop}) from IUE measurements at 1630\,\AA. The method is
described in \cite{ness_cap,ness_CS}. For Algol, Capella, and Procyon we
calculate values for $\phi/\phi_c$ of 2.18, 0.003, and 0.01, respectively.
For Algol the source of UV radiation is the companion
B star \citep{ness_alg} and depending on the phase geometry during the
observation $\phi/\phi_c$ can be significantly lower. The density values
listed in Tables~\ref{fi}/\ref{fi_rs} take radiation effects into account.

\begin{figure*}[!ht]
 \resizebox{\hsize}{!}{\includegraphics{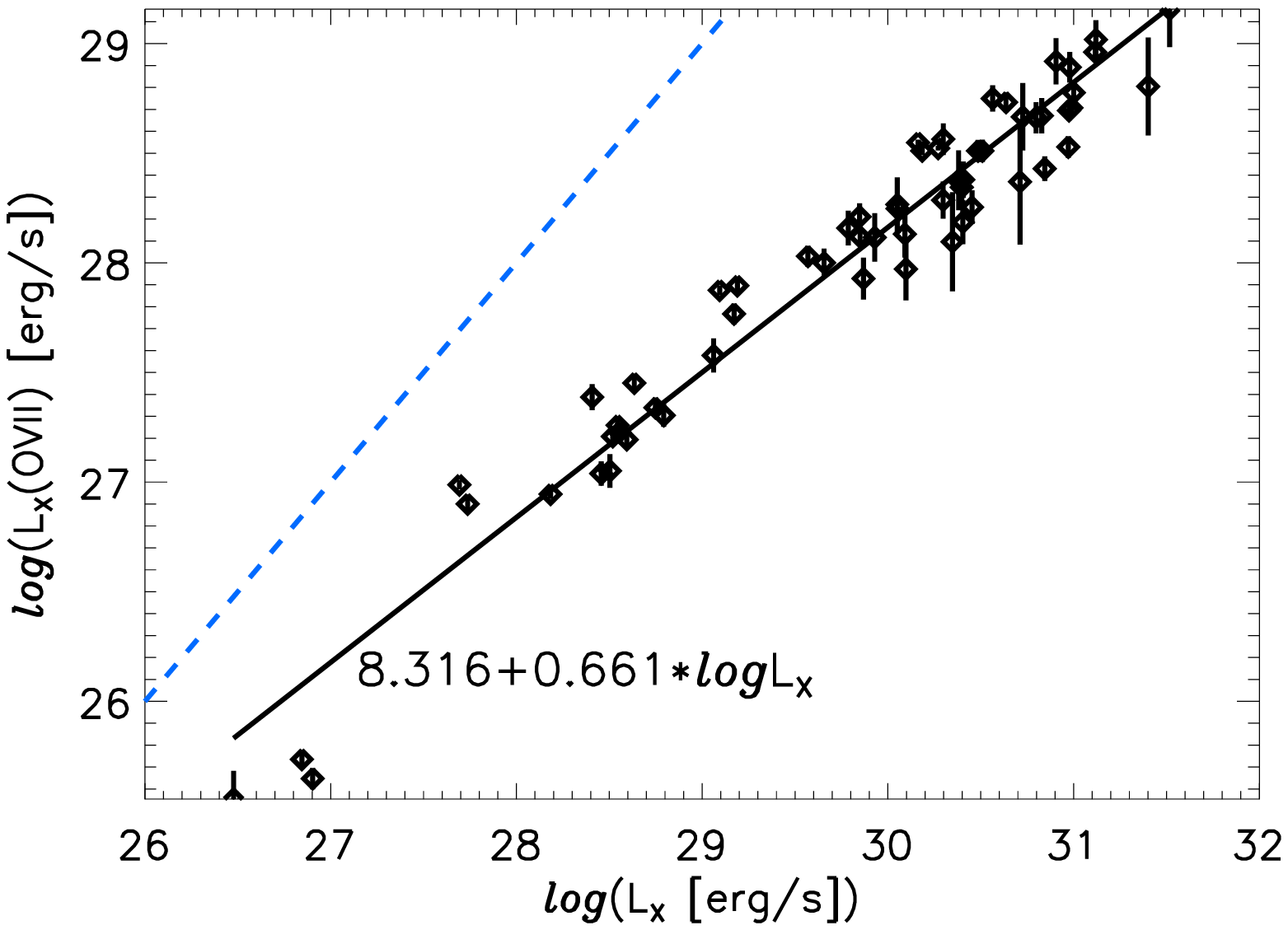}\includegraphics{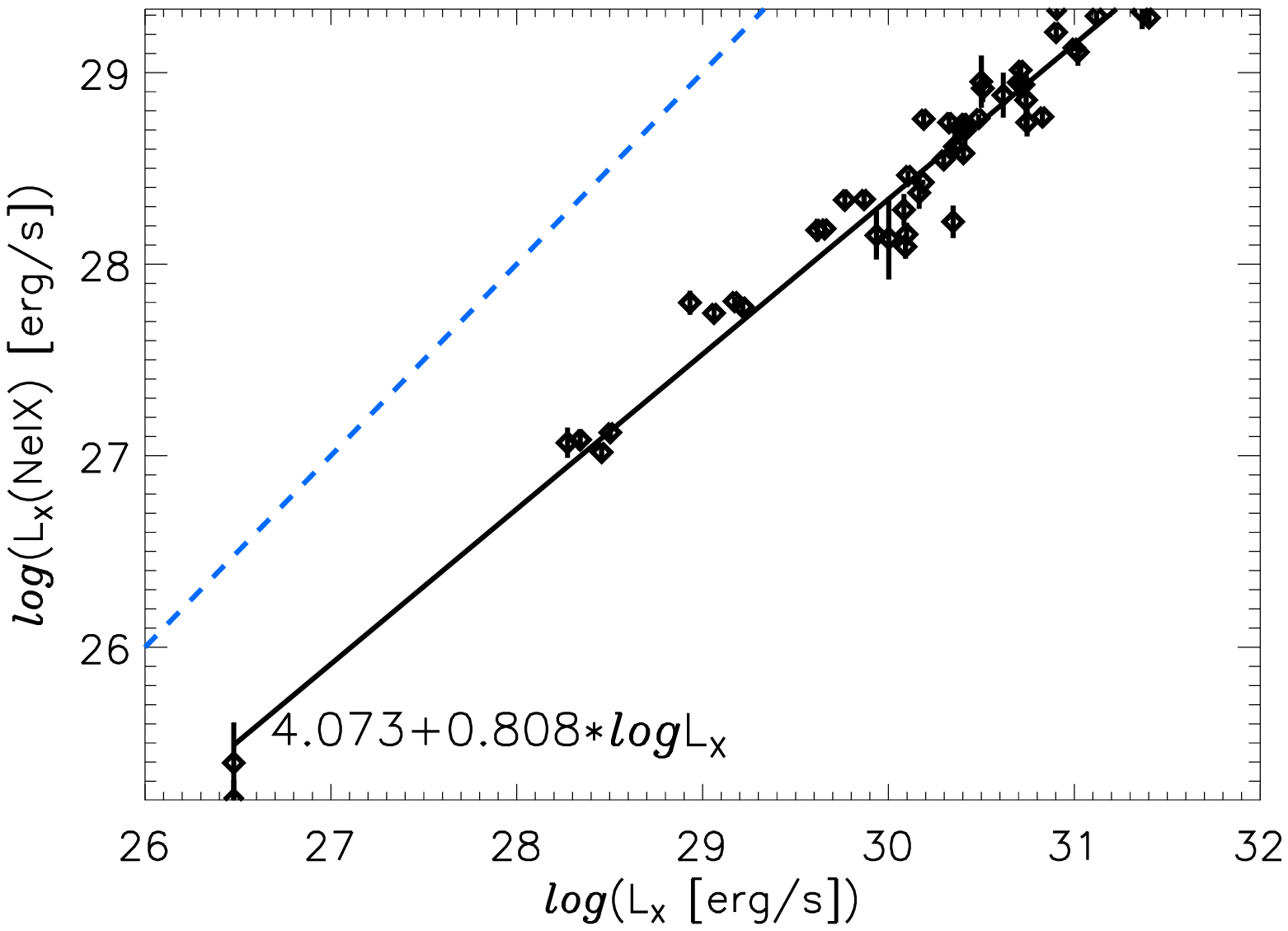}}
\caption[]{\label{lxlo}Comparison of total X-ray luminosity $L_X$ and
O\,{\sc vii} (left) and Ne\,{\sc ix}-specific luminosities. The dashed lines indicate lines of equal luminosity.}
\end{figure*}

\subsection{Blending of Ne\,{\sc ix}}
\label{neblsect}

\begin{figure}[!ht]
 \resizebox{\hsize}{!}{\includegraphics{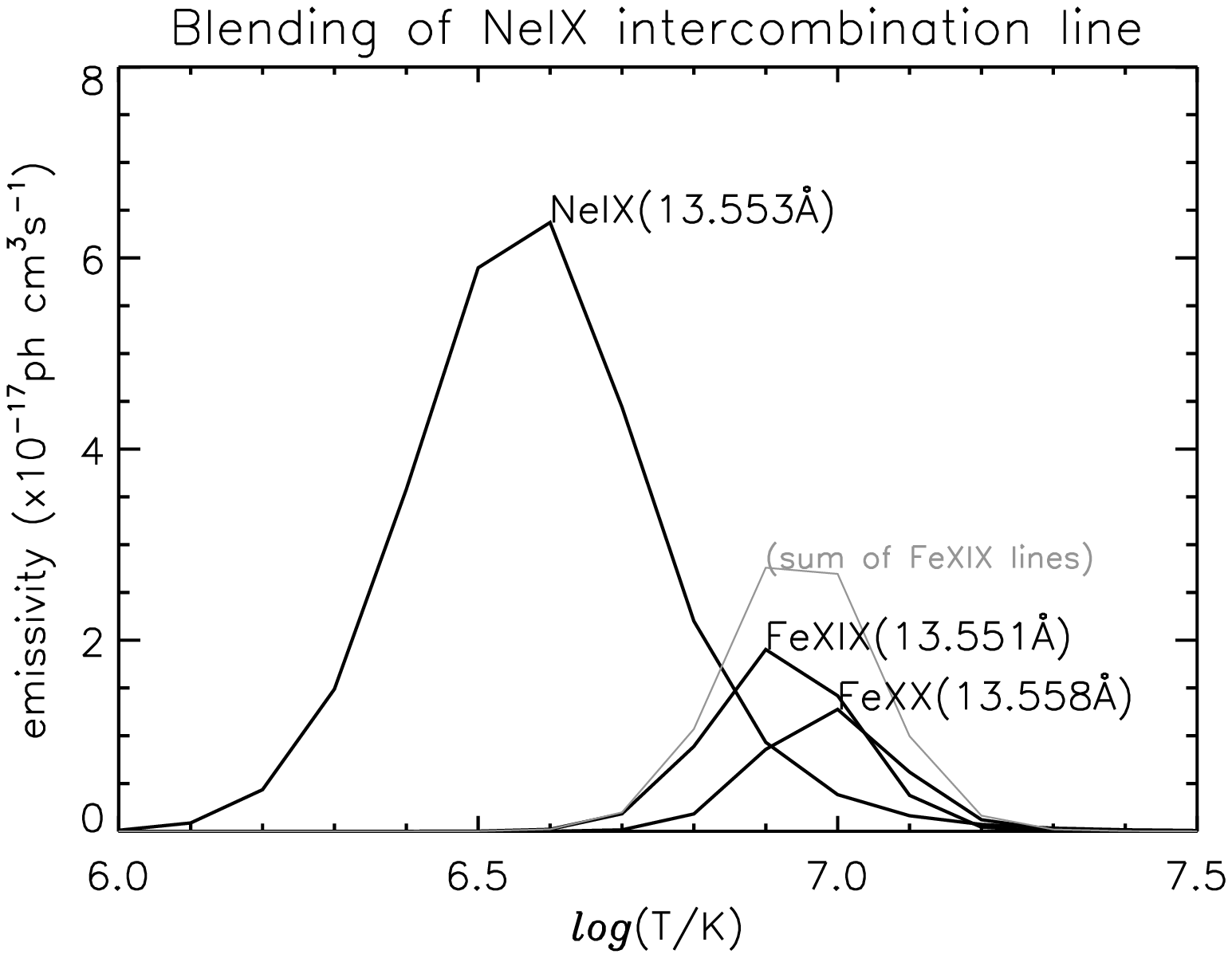}}
\caption[]{\label{neint}Iron lines blending the Ne\,{\sc ix} intercombination
line in high-temperature plasmas predicted by the APEC database assuming
solar elemental abundances.}
\end{figure}

Specific problems in the measurement of the Ne\,{\sc ix} lines arise from
the complicated blending structure, studied by \cite{nebr}
for Capella using the best SNR data available. According to \cite{nebr}
the intercombination line could possibly be blended with an additional line of
Fe\,{\sc xix}. Unfortunately this line is only predicted by the APEC
line database, but it could not even be resolved with the HEG or in any
laboratory measurements. For Capella, \cite{nebr} found
this line to contribute to the measured flux with about a third of the
total flux measured at 13.55\,\AA, thus pushing
the f/i ratio into the low-density limit (for Capella). When inspecting our
Tables~\ref{fi}/\ref{fi_rs} we find systematically higher densities from
Ne\,{\sc ix} line ratios than from O\,{\sc vii} line ratios. Since this is
critical for the assumption of constant pressure in the X-ray
emitting structures, we will test whether systematically higher densities
are still found, if the blending is accounted for.
We extracted the respective line emissivities from the
APEC database and show their temperature dependence in Fig.~\ref{neint}.
According to the APEC, blending can be significant at higher temperatures
and for larger Fe/Ne abundance ratios (which are rather small for most
coronal sources). To assess the amount of expected Fe\,{\sc xix}
contamination we adopt a scaling procedure, measuring the line flux of
supposedly isolated strong Fe\,{\sc xix} lines and scale with the theoretical
ratio of line fluxes.\\
We extracted emissivities for the Fe\,{\sc xix} line at 13.462\,\AA\
and found a value for the ratio of the emissivities and the
blending line at 13.551\,\AA\ of 8. We then measured line
fluxes with the HEG, scaled these with the emissivity ratio 8, and in
Table~\ref{nebl} we list line counts thus predicted for the 13.551\,\AA\ line.
These are contrasted with the original measurements
and for most sources higher f/i ratios, yielding lower densities, are indeed
found. A similar behavior was found when the 13.518\,\AA\ line was used.
In Table~\ref{fi_best} only sources with particularly high
densities are listed, and for Ne\,{\sc ix} we calculate densities from the
corrected f/i ratios. From Table~\ref{fi_best} we conclude that the densities
derived from the Ne\,{\sc ix} triplet are still systematically higher
than those derived from the O\,{\sc vii} triplet even if blending is taken
into account. Hence the O\,{\sc vii}- and Ne\,{\sc ix}-emitting layers cannot
be at the same pressure. We caution, however, that the blending is purely
theoretical and all conclusions rely
on the accuracy of the APEC database. Additional blending is predicted to
occur from an Fe\,{\sc xx} line (see Fig.~\ref{neint}), but we did not make
estimates for the contribution from this line, because no Fe\,{\sc xx} lines
are available for a scaling procedure. For further analysis we use the
non-corrected f/i ratios for Ne\,{\sc ix}.

\begin{table}[!ht]
\begin{flushleft}
\renewcommand{\arraystretch}{1.1}
\caption{\label{nebl}Blending prediction of Ne\,{\sc ix} intercombination line for HEG (first lines) and MEG (second lines).}
\vspace{-.2cm}
{\scriptsize
\begin{tabular}{p{1.1cm}crrrrr}
star & cts$^a$ & i [cts] & f [cts] & new f/i & old f/i\\
         24\,UMa&  1.18&  8.83& 13.64&  1.86&  1.61&\\
\ \ MEG&    4.98&   17.19&   80.49&    6.87&    5.14&\\
         44\,Boo&  4.43& 40.76& 84.48&  2.42&  2.17&\\
\ \ MEG&    6.03&  132.90&  352.56&    2.89&    2.91&\\
         AB\,Dor&  4.14& 33.50& 77.68&  2.76&  2.42&\\
\ \ MEG&   13.09&  137.61&  290.84&    2.43&    2.32&\\
         AD\,Leo&  2.75& 12.77& 26.91&  2.80&  2.20&\\
\ \ MEG&    4.40&   53.71&  171.40&    3.62&    3.50&\\
           Algol&  2.81& 35.73& 53.05&  1.68&  1.55&\\
\ \ MEG&   20.04&  150.82&  236.96&    1.89&    1.72&\\
         AR\,Lac&  2.30& 10.09& 13.90&  1.86&  1.44&\\
\ \ MEG&    4.98&   48.98&  140.80&    3.33&    3.15&\\
         AU\,Mic&  1.02& 24.99& 52.33&  2.27&  2.19&\\
\ \ MEG&    5.81&   78.25&  214.19&    3.08&    3.00&\\
    $\beta$\,Cet&  4.81& 32.41& 64.72&  2.44&  2.09&\\
\ \ MEG&   22.87&  127.79&  211.92&    2.10&    1.82&\\
         Canopus&  0.67&  4.68& 13.03&  3.39&  2.91&\\
\ \ MEG&    3.37&   23.29&   42.20&    2.21&    1.99&\\
         Capella& 35.05&190.47&373.48&  2.50&  2.05&\\
\ \ MEG&  105.42&  709.92& 1373.17&    2.37&    2.12&\\
         ER\,Vul&  4.01& 16.96& 32.71&  2.63&  2.02&\\
\ \ MEG&    9.97&   51.41&  168.72&    4.24&    3.60&\\
         EV\,Lac&  2.73& 24.32& 68.82&  3.32&  2.96&\\
\ \ MEG&    6.75&   85.95&  222.32&    2.92&    2.84&\\
        HR\,1099&  6.13&100.55&228.75&  2.52&  2.38&\\
\ \ MEG&   22.02&  304.42&  834.17&    3.08&    3.01&\\
         II\,Peg&  2.21& 19.99& 46.00&  2.70&  2.41&\\
\ \ MEG&    2.03&  103.95&  268.87&    2.75&    2.84&\\
         IM\,Peg&  1.89& 12.86& 30.64&  2.91&  2.49&\\
\ \ MEG&    7.41&   49.64&  127.99&    3.16&    2.83&\\
  $\lambda$\,And&  2.51& 12.84& 35.64&  3.59&  2.90&\\
\ \ MEG&   11.26&   81.33&  219.34&    3.26&    2.96&\\
      $\mu$\,Vel&  3.13& 24.12& 31.65&  1.57&  1.37&\\
\ \ MEG&    8.75&   50.20&   59.62&    1.50&    1.30&\\
       Prox\,Cen&  0.24&  2.93& 12.81&  4.95&  4.57&\\
\ \ MEG&    0.97&    5.54&   18.02&    4.11&    3.57&\\
 $\sigma^2$\,CrB& 15.30& 88.11&130.60&  1.87&  1.55&\\
\ \ MEG&   28.83&  296.10&  521.80&    2.03&    1.93&\\
     Speedy\,Mic&  0.22&  2.77& 10.59&  4.32&  3.99&\\
\ \ MEG&    0.68&   11.57&   37.35&    3.57&    3.54&\\
         TY\,Pyx&  1.09&  7.09& 16.96&  2.95&  2.50&\\
\ \ MEG&    5.05&   26.51&   72.69&    3.53&    3.01&\\
         UX\,Ari&  1.38&  7.99& 18.57&  2.93&  2.43&\\
\ \ MEG&    9.39&  120.16&  237.38&    2.23&    2.17&\\
       V824\,Ara&  4.82& 29.84& 66.01&  2.75&  2.31&\\
\ \ MEG&   18.54&  157.09&  298.00&    2.24&    2.08&\\
      $\xi$\,UMa&  2.74& 29.59& 81.17&  3.15&  2.87&\\
\ \ MEG&   10.77&  138.57&  354.90&    2.89&    2.81&\\
\hline
\end{tabular}
$^a$prediction of blend in i-line from 13.462\,\AA\ line
}
\renewcommand{\arraystretch}{1}
\end{flushleft}
\end{table}

\subsection{Activity indicators for oxygen and neon}

As an activity indicator specific only for the O\,{\sc vii} and Ne\,{\sc ix} emitting layers
we calculate an ion-specific X-ray
luminosity from the sum of the three He-like line fluxes r+i+f as described
by \cite{ness_opt}. These luminosities are plotted in Fig.~\ref{lxlo} in
comparison with the total X-ray luminosities (also taken from \cite{ness_opt}
and listed in Tables~\ref{lxem}/\ref{lxem_rs}). Clearly, the O\,{\sc vii} and
Ne\,{\sc ix} luminosities strongly correlate with the total X-ray luminosity
and we conclude that these luminosities represent the overall degree of
magnetic activity of the coronae at least as well as the total X-ray luminosities.
This is supported by our finding that the average temperatures derived from
the respective ions correlate with the overall X-ray luminosity. We find
that for the least active stars O\,{\sc vii} emission contributes on
average less than 10\% to the overall luminosity and Ne\,{\sc ix} emission
at most 7\% for stars of intermediate activity. For the active
stars the percentage drops to below 3\% for oxygen and neon.
From the ion-specific luminosities we can also calculate ion-specific emission
measures, but a temperature structure has to be assumed for this procedure. For
simplicity we assume an isothermal plasma at the peak formation temperature for
each ion ($T_{\rm O \mathsc{vii}}=2$\,MK and $T_{\rm Ne \mathsc{ix}}=4$\,MK).
Note that the emission peak around the He-like ions is very narrow.
The ion-specific emission measure is then calculated from
\begin{equation}
\label{lxemeq}
EM_{\rm ion}=\frac{L_{\rm X,ion}}{p_\lambda(T_m)}
\end{equation}
with $p_\lambda(T_m)$ being the emissivity taken from the APEC database at the
peak formation temperature $T_m$. Here we assume solar photospheric
abundances, but for
each element the ion-specific emission measure will linearly increase with
increasing elemental abundance relative to solar. The emission measures thus
obtained are listed in Tables~\ref{lxem}/\ref{lxem_rs}.

\section{Structure of oxygen and neon emitting layers}

\begin{table}[!ht]
\begin{flushleft}
\renewcommand{\arraystretch}{1.1}
\caption{\label{vol}Derived volumes $V_{\rm cor}$ in comparison with available volumes $V_{\rm avail}$.}
\vspace{-.6cm}
{\scriptsize
\begin{tabular}{p{1.cm}p{.5cm}p{.7cm}p{.7cm}p{.7cm}p{.7cm}p{.7cm}p{.7cm}}
&&\multicolumn{3}{c}{O\,{\sc vii}}&\multicolumn{3}{c}{Ne\,{\sc ix}}\\
 star & Instr.& log$V_{\rm cor}^a$ & log$V_{\rm av}^b$ & $f^c$ & log$V_{\rm cor}^a$ & log$V_{\rm av}^b$ &$f^c$\\
&&[cm$^3$]&[cm$^3$]&\%&[cm$^3$]&[cm$^3$]&\%\\
         24\,UMa&   MEG&        29.4&        33.2&        0.01&        30.8&        33.4&        0.22\\
 &   HEG&        --&        --&        --&        28.5&        32.2&        0.02\\
         44\,Boo&   MEG&        30.0&        32.9&        0.11&        30.1&        32.6&        0.38\\
 &   HEG&        --&        --&        --&        29.0&        32.0&        0.11\\
         47\,Cas&  RGS1&        30.5&        33.5&        0.09&        --&        --&        --\\
         AB\,Dor&  RGS1&        30.3&        33.0&        0.16&        --&        --&        --\\
 &   MEG&        30.7&        33.4&        0.18&        29.4&        32.2&        0.18\\
 &   HEG&        --&        --&        --&        29.5&        32.2&        0.23\\
$\alpha$\,Cen\,A&  LETG&        28.7&        31.4&        0.18&        --&        --&        --\\
$\alpha$\,Cen\,B&  RGS1&        31.2&        33.5&        0.51&        --&        --&        --\\
 &  LETG&        29.4&        31.6&        0.62&        --&        --&        --\\
         AD\,Leo&  RGS1&        29.8&        33.5&        0.01&        --&        --&        --\\
 &  LETG&        30.7&        32.6&        1.08&        --&        --&        --\\
 &   MEG&        29.2&        31.9&        0.19&        29.1&        31.5&        0.43\\
 &   HEG&        --&        --&        --&        28.1&        30.8&        0.20\\
           Algol&  RGS1&        30.4&        34.0&        0.02&        --&        --&        --\\
 &  LETG&        31.0&        34.3&        0.05&        --&        --&        --\\
 &   MEG&        30.0&        33.7&        0.02&        29.1&        32.8&        0.02\\
 &   HEG&        --&        --&        --&        28.9&        32.7&        0.01\\
         AT\,Mic&  RGS1&        30.1&        32.5&        0.37&        --&        --&        --\\
         AU\,Mic&  RGS1&        31.8&        34.2&        0.35&        --&        --&        --\\
 &   MEG&        31.0&        34.3&        0.04&        29.1&        31.7&        0.28\\
 &   HEG&        --&        --&        --&        28.7&        31.3&        0.24\\
    $\beta$\,Cet&  RGS1&        31.1&        34.1&        0.10&        --&        --&        --\\
 &  LETG&        30.8&        34.5&        0.01&        --&        --&        --\\
 &   MEG&        30.5&        33.1&        0.22&        29.1&        32.9&        0.01\\
 &   HEG&        --&        --&        --&        29.4&        33.0&        0.02\\
         Canopus&   MEG&        --&        --&        --&        29.7&        33.5&        0.01\\
 &   HEG&        --&        --&        --&        29.0&        33.1&        0.00\\
   $\chi^1$\,Ori&  RGS1&        31.5&        33.3&        1.48&        --&        --&        --\\
         EK\,Dra&  RGS1&        30.4&        33.1&        0.21&        --&        --&        --\\
 &  LETG&        29.7&        32.9&        0.06&        --&        --&        --\\
 $\epsilon$\,Eri&  RGS1&        30.2&        32.1&        1.43&        --&        --&        --\\
 &  LETG&        30.1&        32.3&        0.62&        --&        --&        --\\
         EQ\,Peg&  RGS1&        30.2&        32.3&        0.84&        --&        --&        --\\
         ER\,Vul&   MEG&        29.8&        33.3&        0.02&        30.5&        33.1&        0.28\\
 &   HEG&        --&        --&        --&        29.4&        32.5&        0.07\\
         EV\,Lac&  RGS1&        29.8&        32.1&        0.48&        --&        --&        --\\
 &   MEG&        28.9&        33.0&        0.00&        28.9&        31.3&        0.38\\
 &   HEG&        --&        --&        --&        28.1&        30.8&        0.19\\
      HD\,223460&   MEG&        30.2&        33.4&        0.05&        --&        --&        --\\
   $\kappa$\,Cet&  RGS1&        30.1&        32.6&        0.33&        --&        --&        --\\
      $\mu$\,Vel&   MEG&        30.4&        34.8&        0.00&        28.1&        32.3&        0.00\\
 &   HEG&        --&        --&        --&        28.4&        32.4&        0.01\\
    $\pi^1$\,UMa&  RGS1&        29.9&        32.2&        0.50&        --&        --&        --\\
         Procyon&  LETG&        30.4&        32.3&        1.31&        --&        --&        --\\
       Prox\,Cen&   MEG&        27.6&        32.9&        0.00&        25.5&        28.7&        0.06\\
 &   HEG&        --&        --&        --&        25.4&        28.7&        0.05\\
     Speedy\,Mic&   MEG&        29.0&        29.9&        11.5&        29.2&        32.4&        0.05\\
 &   HEG&        --&        --&        --&        27.8&        31.6&        0.01\\
       V471\,Tau&  LETG&        29.8&        32.9&        0.08&        --&        --&        --\\
         VW\,Cep&  LETG&        29.7&        32.8&        0.08&        --&        --&        --\\
      $\xi$\,UMa&   MEG&        30.1&        34.1&        0.01&        29.6&        32.1&        0.30\\
 &   HEG&        --&        --&        --&        28.8&        31.7&        0.10\\
         YY\,Gem&  LETG&        30.5&        33.0&        0.31&        --&        --&        --\\
         YZ\,CMi&  RGS1&        29.5&        31.9&        0.36&        --&        --&        --\\
\hline
\end{tabular}
\\
$^a$emitting coronal volumes from Eq.~\ref{vcor}\ \ \ \
$^b$Available volumes from Eq.~\ref{vavail}\\
$^c$Filling factor $f=V_{\rm cor}/V_{\rm avail}$\\
}
\renewcommand{\arraystretch}{1}
\end{flushleft}
\end{table}

\begin{table}[!ht]
\begin{flushleft}
\renewcommand{\arraystretch}{1.1}
\caption{\label{vol_rs}Same as Table~\ref{vol} for RS\,CVn systems in our sample.}
\vspace{-.6cm}
{\scriptsize
\begin{tabular}{p{1.cm}p{.5cm}p{.7cm}p{.7cm}p{.7cm}p{.7cm}p{.7cm}p{.7cm}}
&&\multicolumn{3}{c}{O\,{\sc vii}}&\multicolumn{3}{c}{Ne\,{\sc ix}}\\
 star & Instr.& log$V_{\rm cor}^a$ & log$V_{\rm av}^b$ & $f^c$ & log$V_{\rm cor}^a$ & log$V_{\rm av}^b$ &$f^c$\\
&&[cm$^3$]&[cm$^3$]&\%&[cm$^3$]&[cm$^3$]&\%\\
         AR\,Lac&  RGS1&        31.3&        33.0&        1.78&        --&        --&        --\\
 &   MEG&        --&        --&        --&        30.3&        33.2&        0.13\\
 &   HEG&        --&        --&        --&        29.1&        32.5&        0.04\\
         Capella&  LETG&        32.8&        35.0&        0.64&        --&        --&        --\\
 &   MEG&        32.3&        34.9&        0.26&        29.3&        32.9&        0.02\\
 &   HEG&        --&        --&        --&        29.2&        32.8&        0.02\\
        HR\,1099&  RGS1&        31.3&        34.1&        0.19&        --&        --&        --\\
 &  LETG&        31.2&        34.2&        0.08&        --&        --&        --\\
 &   MEG&        32.0&        32.7&        18.8&        31.4&        33.8&        0.40\\
 &   HEG&        --&        --&        --&        30.4&        33.2&        0.17\\
         II\,Peg&   MEG&        30.6&        34.3&        0.02&        31.1&        33.6&        0.32\\
 &   HEG&        --&        --&        --&        30.3&        33.2&        0.10\\
         IM\,Peg&   MEG&        29.0&        33.5&        0.00&        30.2&        33.8&        0.02\\
 &   HEG&        --&        --&        --&        29.6&        33.4&        0.01\\
  $\lambda$\,And&  RGS1&        30.7&        33.9&        0.06&        --&        --&        --\\
 &  LETG&        31.2&        34.5&        0.04&        --&        --&        --\\
 &   MEG&        30.4&        34.4&        0.01&        29.9&        33.0&        0.08\\
 &   HEG&        --&        --&        --&        28.9&        32.5&        0.02\\
 $\sigma^2$\,CrB&  RGS1&        31.7&        34.1&        0.37&        --&        --&        --\\
 &   MEG&        30.8&        33.4&        0.23&        29.5&        32.5&        0.08\\
 &   HEG&        --&        --&        --&        29.0&        32.1&        0.08\\
         TY\,Pyx&   MEG&        30.3&        33.5&        0.07&        29.8&        32.9&        0.08\\
 &   HEG&        --&        --&        --&        28.7&        32.5&        0.01\\
         UX\,Ari&  RGS1&        31.1&        34.1&        0.08&        --&        --&        --\\
 &  LETG&        32.2&        34.8&        0.25&        --&        --&        --\\
 &   MEG&        30.7&        33.4&        0.17&        30.2&        33.1&        0.13\\
 &   HEG&        --&        --&        --&        29.0&        32.5&        0.03\\
       V824\,Ara&   MEG&        30.2&        33.7&        0.03&        29.5&        32.6&        0.08\\
 &   HEG&        --&        --&        --&        29.9&        32.7&        0.15\\
         VY\,Ari&  RGS1&        31.8&        34.3&        0.34&        --&        --&        --\\
\hline
\end{tabular}
\\
$^a$emitting coronal volumes from Eq.~\ref{vcor}\ \ \ \
$^b$Available volumes from Eq.~\ref{vavail}\\
$^c$Filling factor $f=V_{\rm cor}/V_{\rm avail}$\\
}
\renewcommand{\arraystretch}{1}
\end{flushleft}
\end{table}

\begin{figure}[!ht]
 \resizebox{\hsize}{!}{\includegraphics{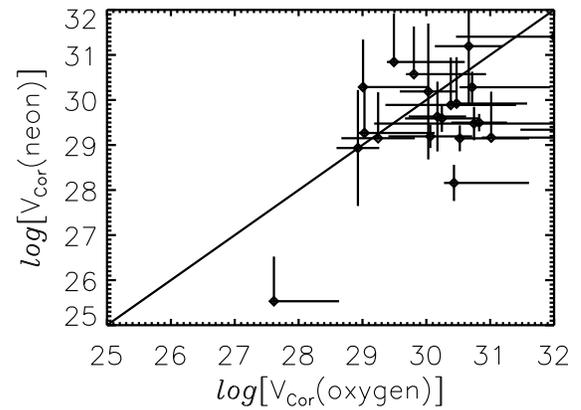}}
\caption[]{\label{vcmp}Comparison of coronal volumes obtained from O\,{\sc vii}
and Ne\,{\sc ix} (only MEG measurements). The solid line marks the line of
equal volumes.}
\end{figure}

Fig.~\ref{ox} (for O\,{\sc vii}) and Fig.~\ref{ne} (for Ne\,{\sc ix}) show the
central results of our measurements in graphical form.
In the upper panels of these figures we show the measured f/i
ratios and the densities derived with these measurements with Eq.~(\ref{fidens})
vs. the ion-specific luminosities ($L_{\rm X, O\,\mathsc{vii}}$ and $L_{\rm X,
Ne\,\mathsc{ix}}$). The low-density limit $R_0$ is marked with a vertical
dotted line in the upper left panels, only measurements with f/i-ratios $<R_0$
yield actual density measurements. Therefore all f/i measurements resulting in
low-density limits have been marked by light colors. In the bottom left panels
we plot the densities versus the emission measure obtained from the ion-specific
luminosities calculated with Eq.~(\ref{lxemeq}). In the plot we include lines of
equal emitting (coronal) volumes $V_{\rm cor}$ derived from
\begin{equation}
\label{vcor}
 EM_{\rm ion} = 0.85\,n_e^2V_{\rm cor}\,.
\end{equation}
The coronal volumes derived from the O\,{\sc vii} densities and the Ne\,{\sc ix}
densities are consistent with each other. This is demonstrated in
Fig.~\ref{vcmp}, where the bulk of measured volumes does not depart
significantly from the line of equal volumes. 

\begin{figure}[!ht]
 \resizebox{\hsize}{!}{\includegraphics{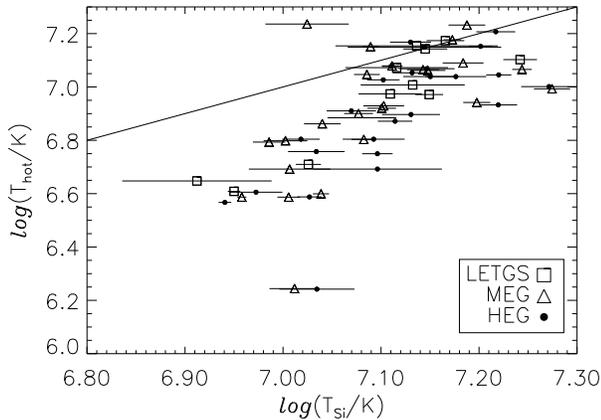}}
\caption[]{\label{tsi}Comparison of temperatures derived from
Eq.~(\ref{thot_scal}) and from Si H-like and He-like line ratios representing
the hot component for those stars where HETGS and LETGS spectra are available.}
\end{figure}

\begin{figure*}[!ht]
 \resizebox{\hsize}{!}{\includegraphics{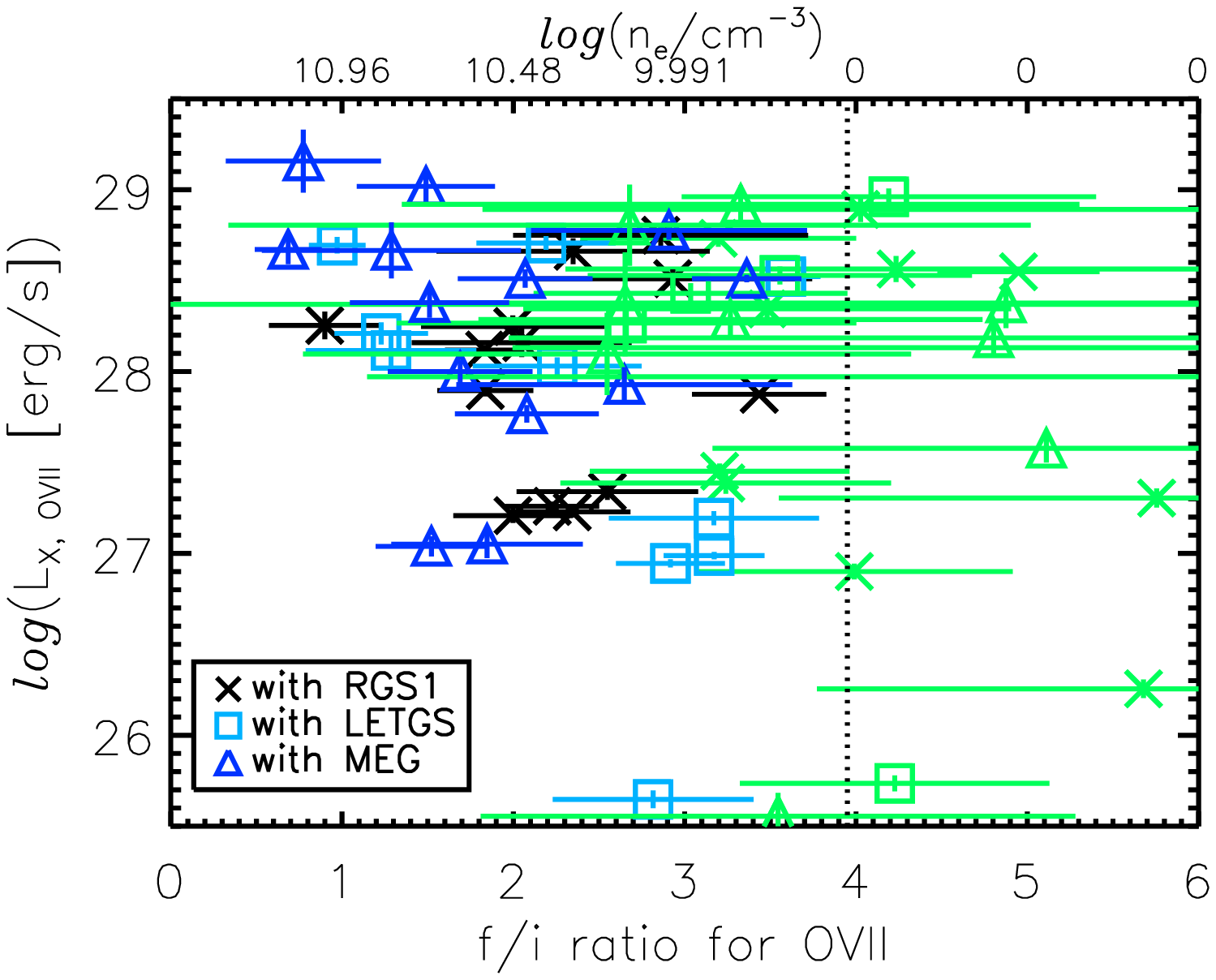}\includegraphics{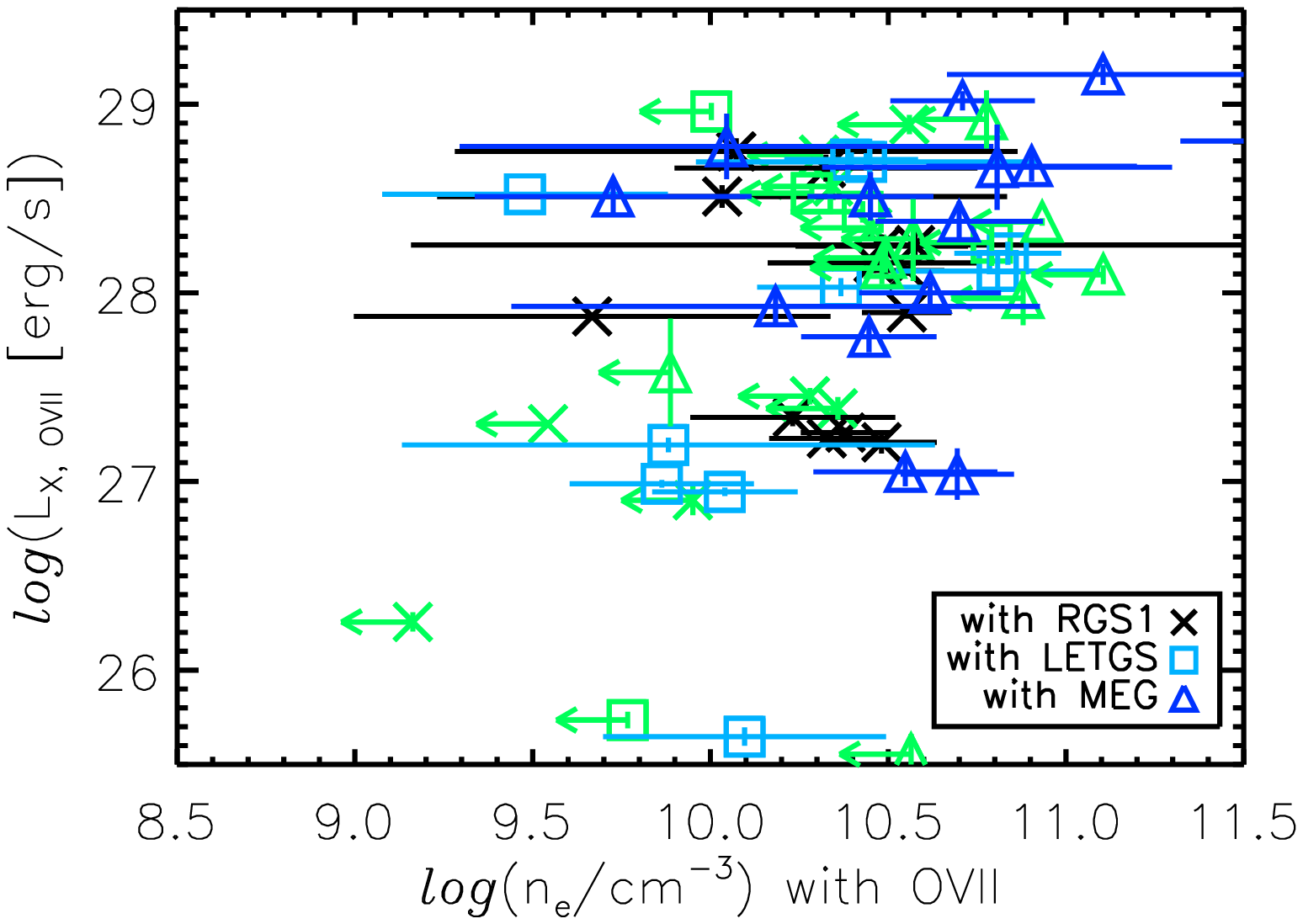}}
 \resizebox{\hsize}{!}{\includegraphics{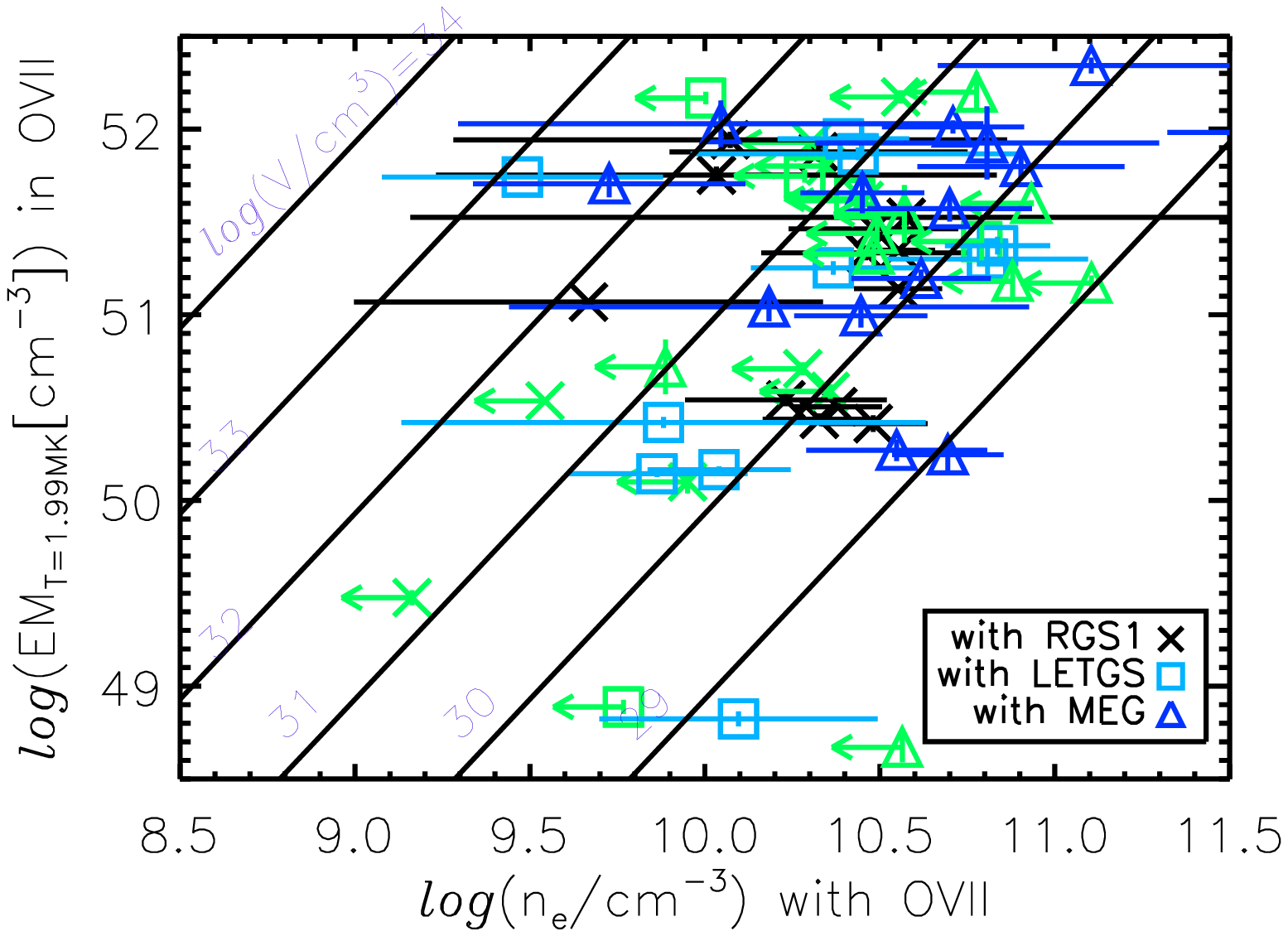}\includegraphics{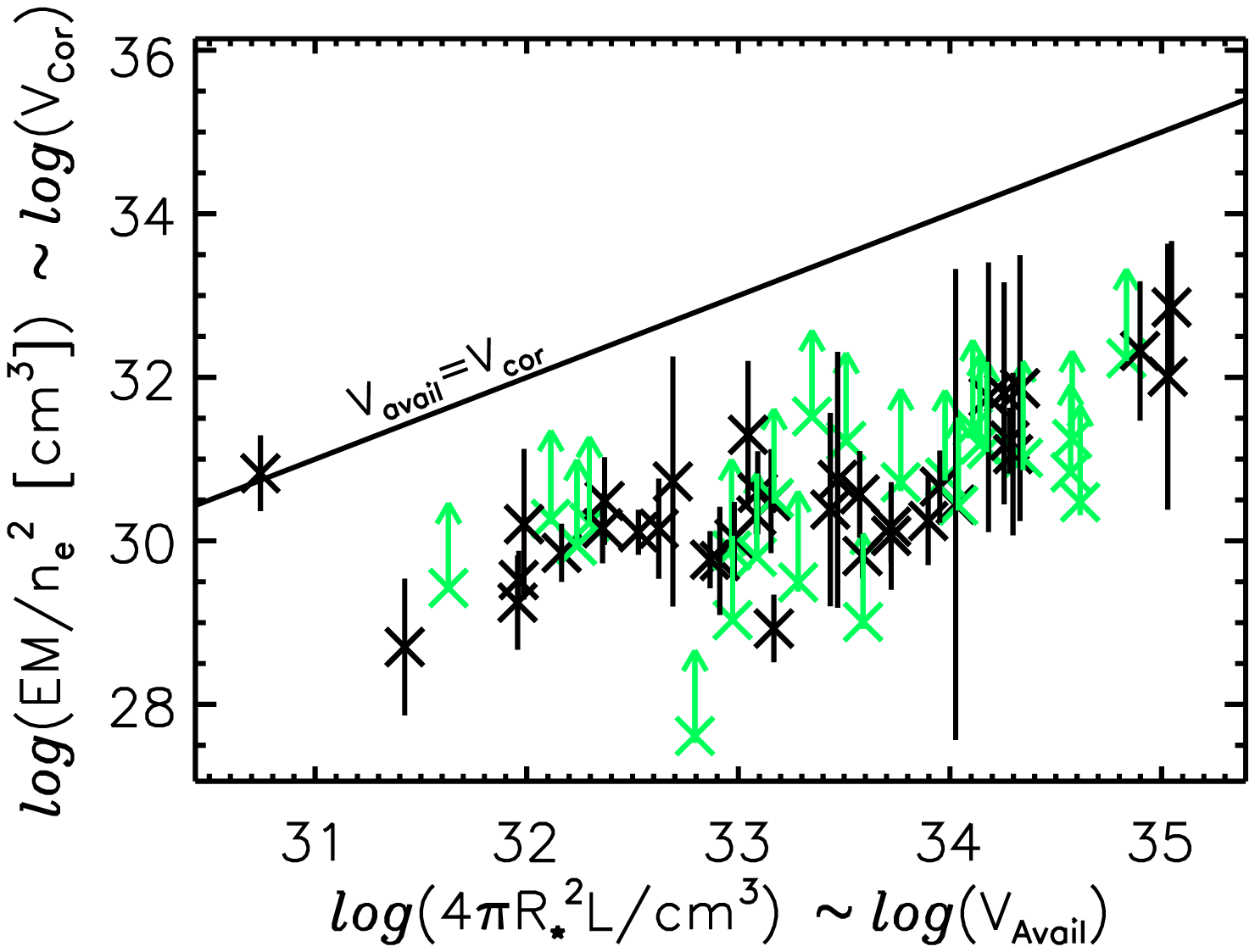}}
\caption[]{\label{ox}Measured f/i ratios and derived densities versus the
O\,{\sc vii} specific X-ray luminosities (upper two panels). The low-density
limit is marked with a vertical dotted line and measurements with only upper
limits are marked by light symbols. {\bf Lower left panel}: densities versus
O\,{\sc vii} specific emission measure at T=2\,MK. {\bf Lower right}: emitting
volume versus available volume (Eq.~\ref{vavail} with $T_{\rm hot}$ from
Eq.~\ref{thot_scal}; numbers are listed in
Tables~\ref{fi}/\ref{fi_rs}, \ref{lxem}/\ref{lxem_rs}, and \ref{vol}/\ref{vol_rs}).}
\end{figure*}

\begin{figure*}[!ht]
 \resizebox{\hsize}{!}{\includegraphics{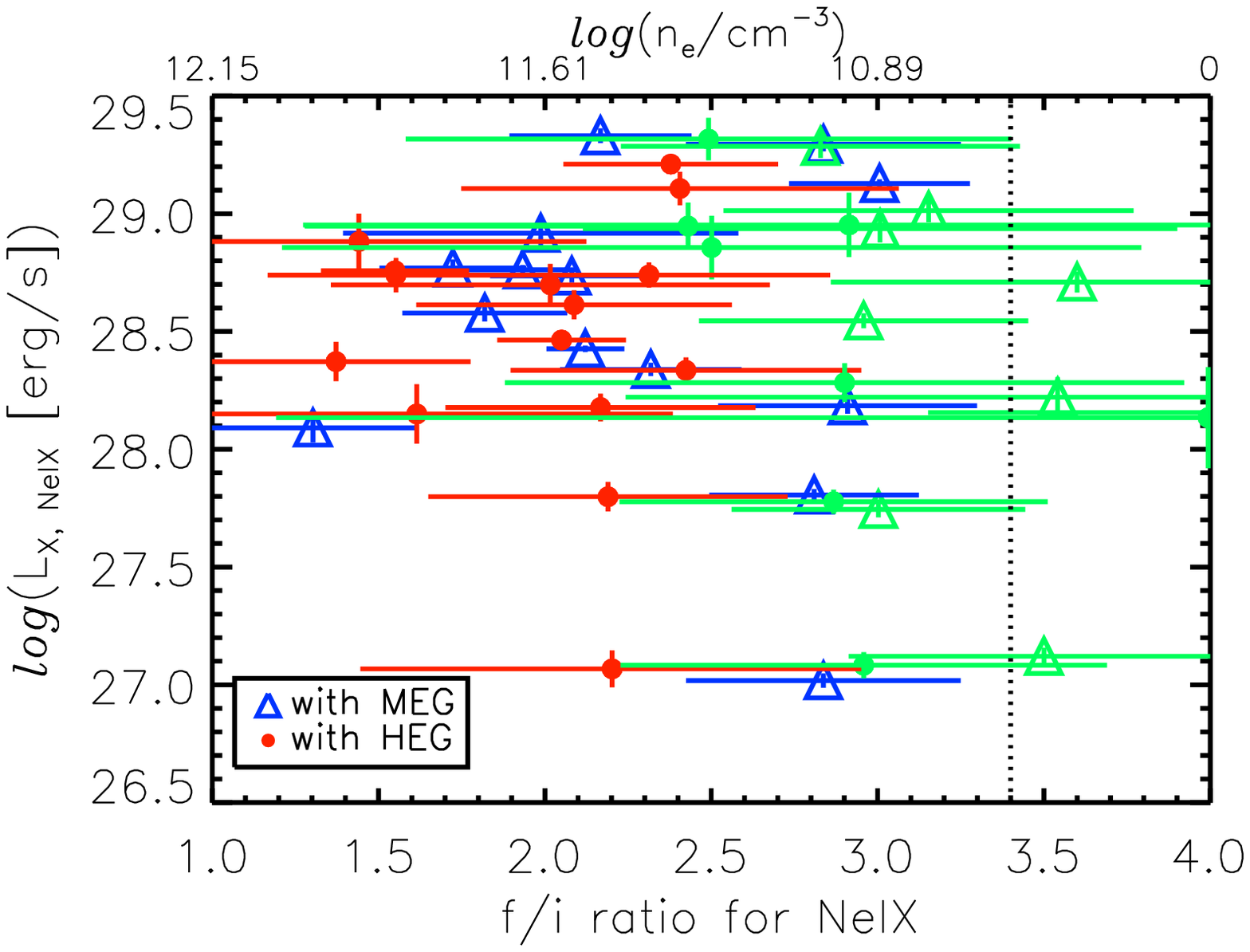}\includegraphics{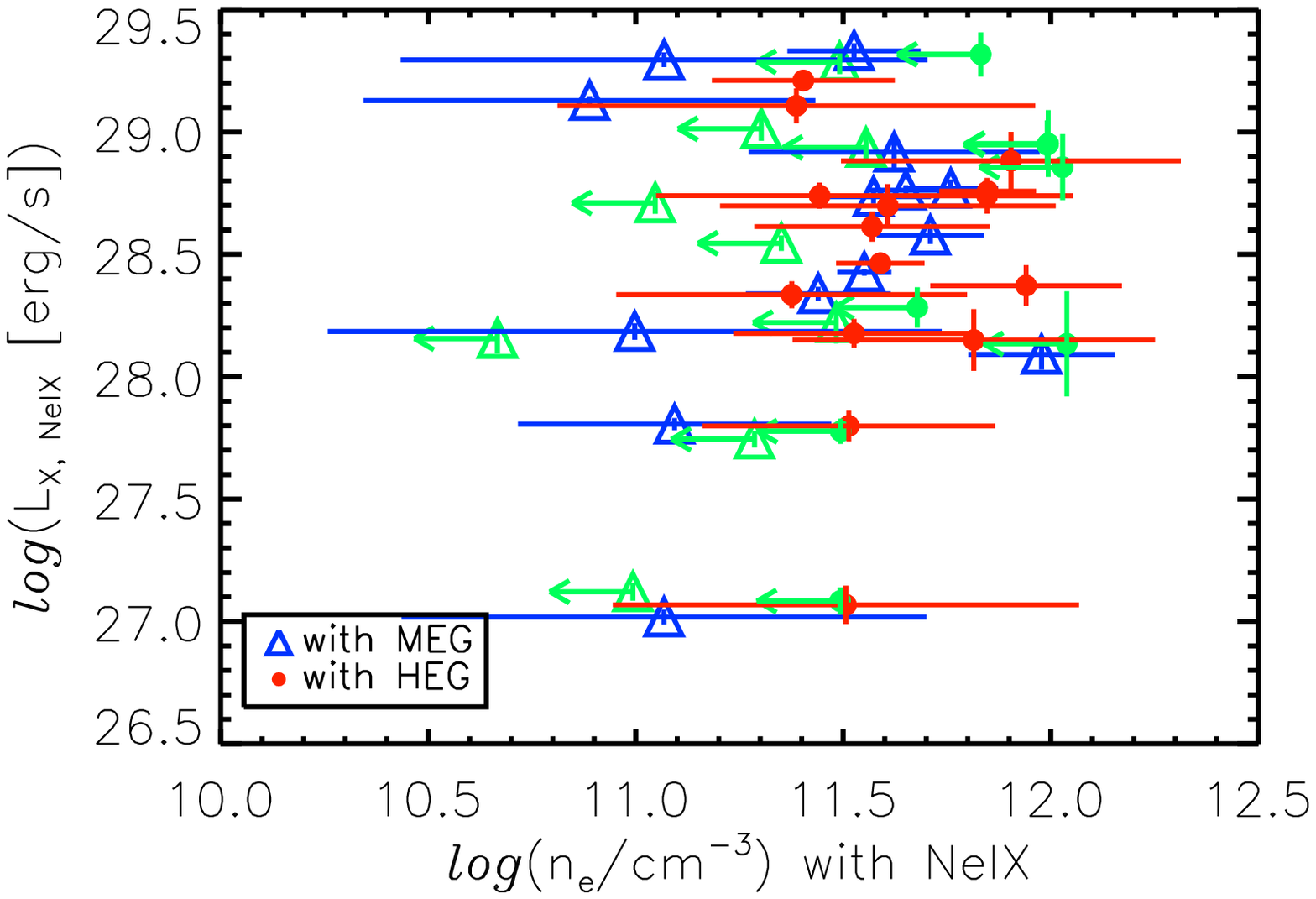}}
 \resizebox{\hsize}{!}{\includegraphics{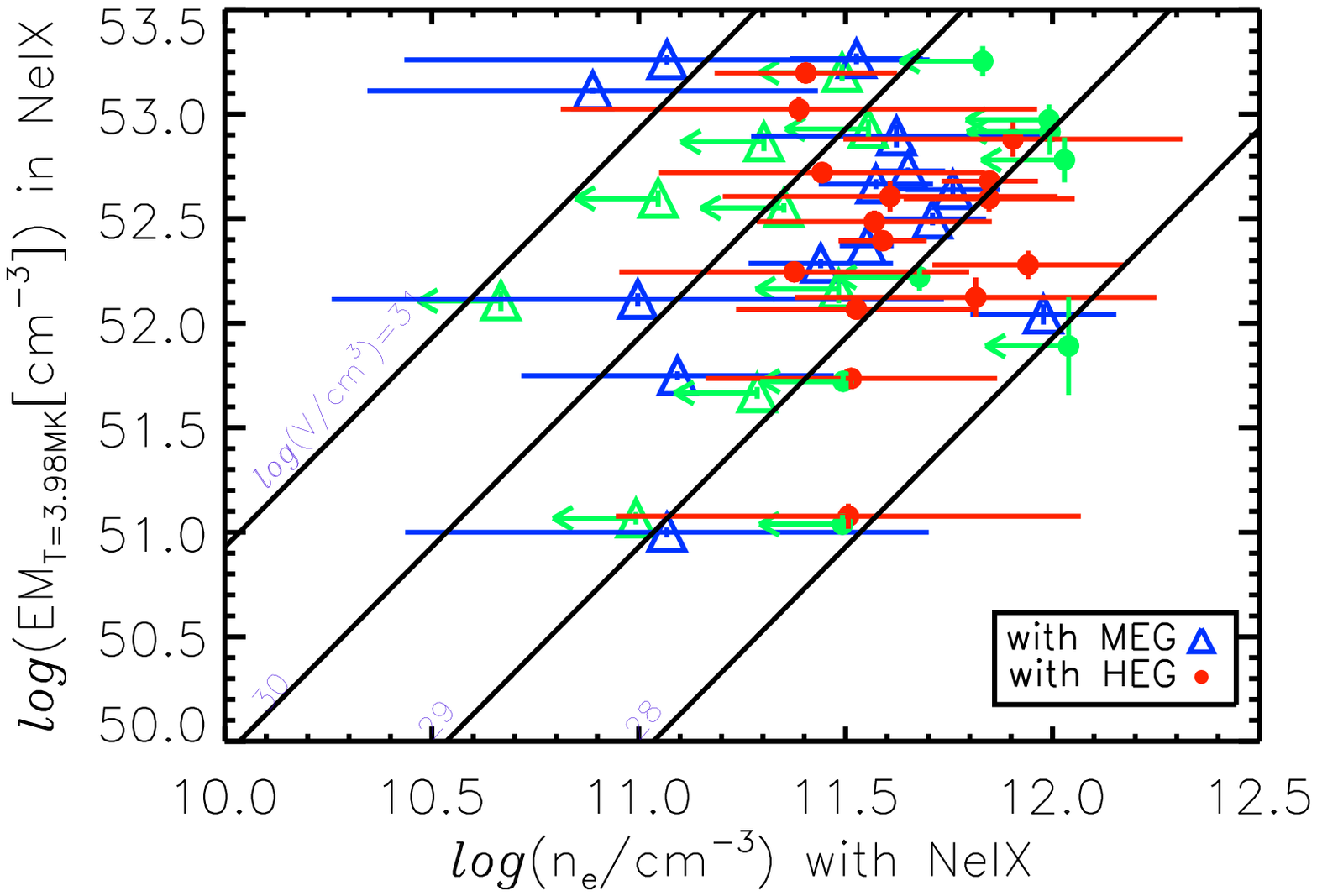}\includegraphics{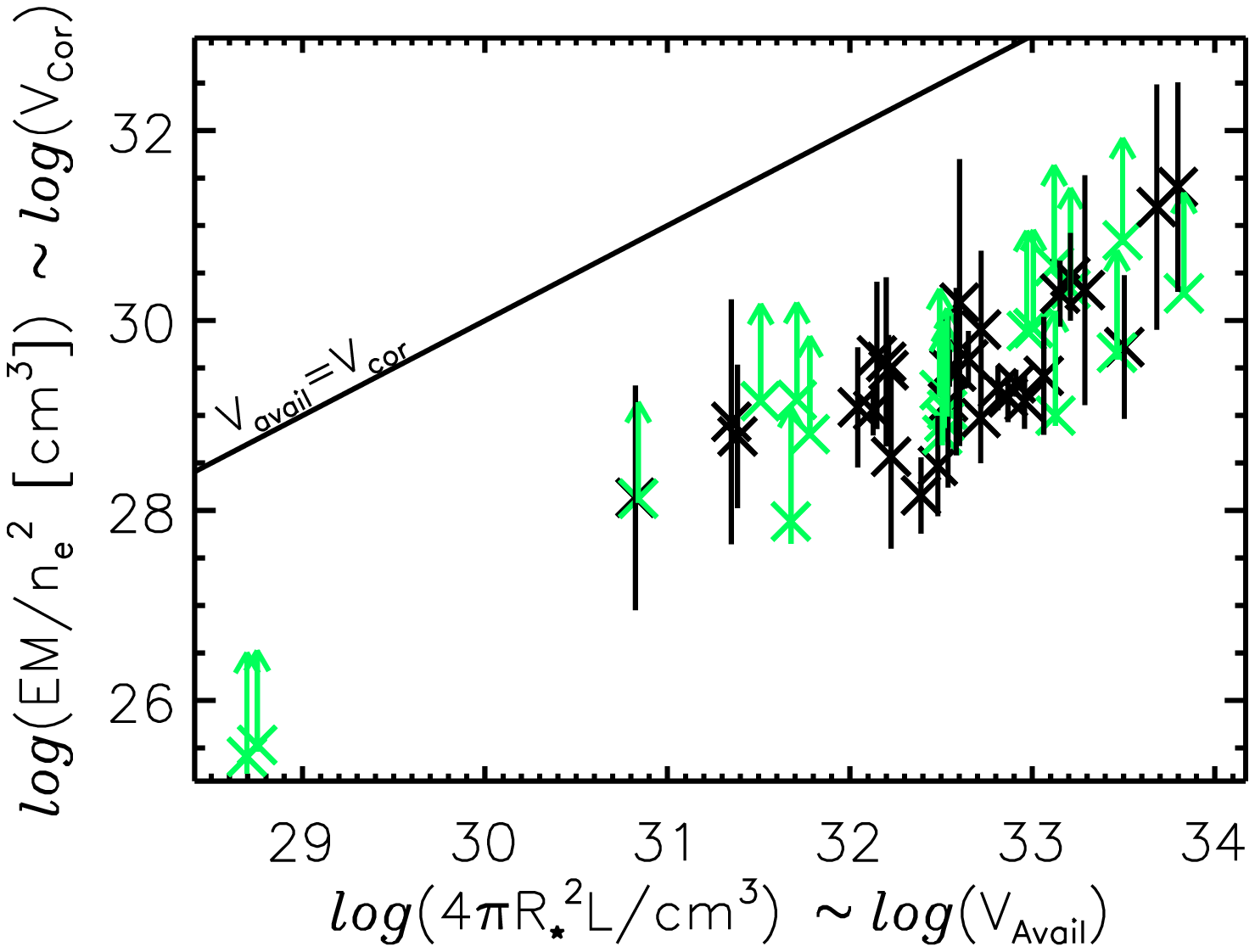}}
\caption[]{\label{ne}Same as Fig.~\ref{ox} for Ne\,{\sc ix}.}
\end{figure*}

For each ion we estimate available volumes $V_{\rm avail}$, which can
potentially be filled with coronal plasma. We use stellar radii (cf.
Table~\ref{sprop}) and a scale height $H$: $V_{\rm avail}=4\pi R_\star^2H$
(terms of $R_\star H^2$ and $H^3$ are neglected) and compare $V_{\rm avail}$
with the ion specific emitting volumes.
For an estimate of the height we assume the plasma to be confined in a
uniform distribution of loop structures obeying the loop scaling law by
\citet[][ RTV]{rtv}:
\begin{equation}
\label{rtv_eq}
n_{e, {\rm hot}} L=1.3\times 10^6 T_{\rm hot}^{\,2}
\end{equation}
with the electron density $n_{e, {\rm hot}}$ and the plasma temperature $T_{\rm
hot}$ at the loop top.
While emitting volumes can be derived only from the densities
(Eq.~\ref{vcor}), we
additionally need the plasma temperature to estimate loop lengths $L$.
As a first approach we use the temperatures derived from the H-like and He-like
lines of the respective ions from which the densities are derived. This implies
that the ion-specific temperature estimate represents the overall corona, which
is certainly not true for the more active stars. Nevertheless, the loop lengths
derived from the oxygen and neon temperatures can be regarded as lower limits.
Our second approach is based on finding a loop-top temperature $T_{\rm hot}$
that scales with the X-ray luminosity.
We regard the loop-top temperature as equivalent to
the hotter component of a two-temperature distribution as found by
\cite{guedel97}. The temperature $T_{\rm hot}$ we therefore derive from the
X-ray luminosity using Eq.~(\ref{thot}). Since \cite{guedel97} had only solar
like stars in their sample we scale the relation to the different stellar
radii, thus using surface fluxes instead of luminosities:
\begin{equation}
\label{thot_scal}
T_{\rm hot}^4=\frac{L_X}{55}\left(\frac{R_\star}{R_\odot}\right)^{-2}\,.
\end{equation}
For binary stars we use the stellar radius of the component assumed to be more
active. We tested the consistency of these temperatures with line-based hot
temperatures determined from the H-like and He-like Si lines measured with the
HETGS and the LETGS and estimated plasma temperatures with the APEC line
database. From Fig.~\ref{tsi} it can be seen that in particular for the most
active of our sample stars these temperatures agree quite well, while for
somewhat less active stars some emission measure is still present at high
temperature and the bulk of the emission measure is at a lower temperature.
The density at the loop-top (i.e., of the hot component) $n_{e, {\rm hot}}$ we
derive from the measured densities $n_e$(O\,{\sc vii}) and $n_e$(Ne\,{\sc ix})
assuming pressure equilibrium, i.e.,
$n_e T_{\rm ion}=n_{e, {\rm hot}}T_{\rm hot}$.

\begin{figure*}[!ht]
 \resizebox{\hsize}{!}{\includegraphics{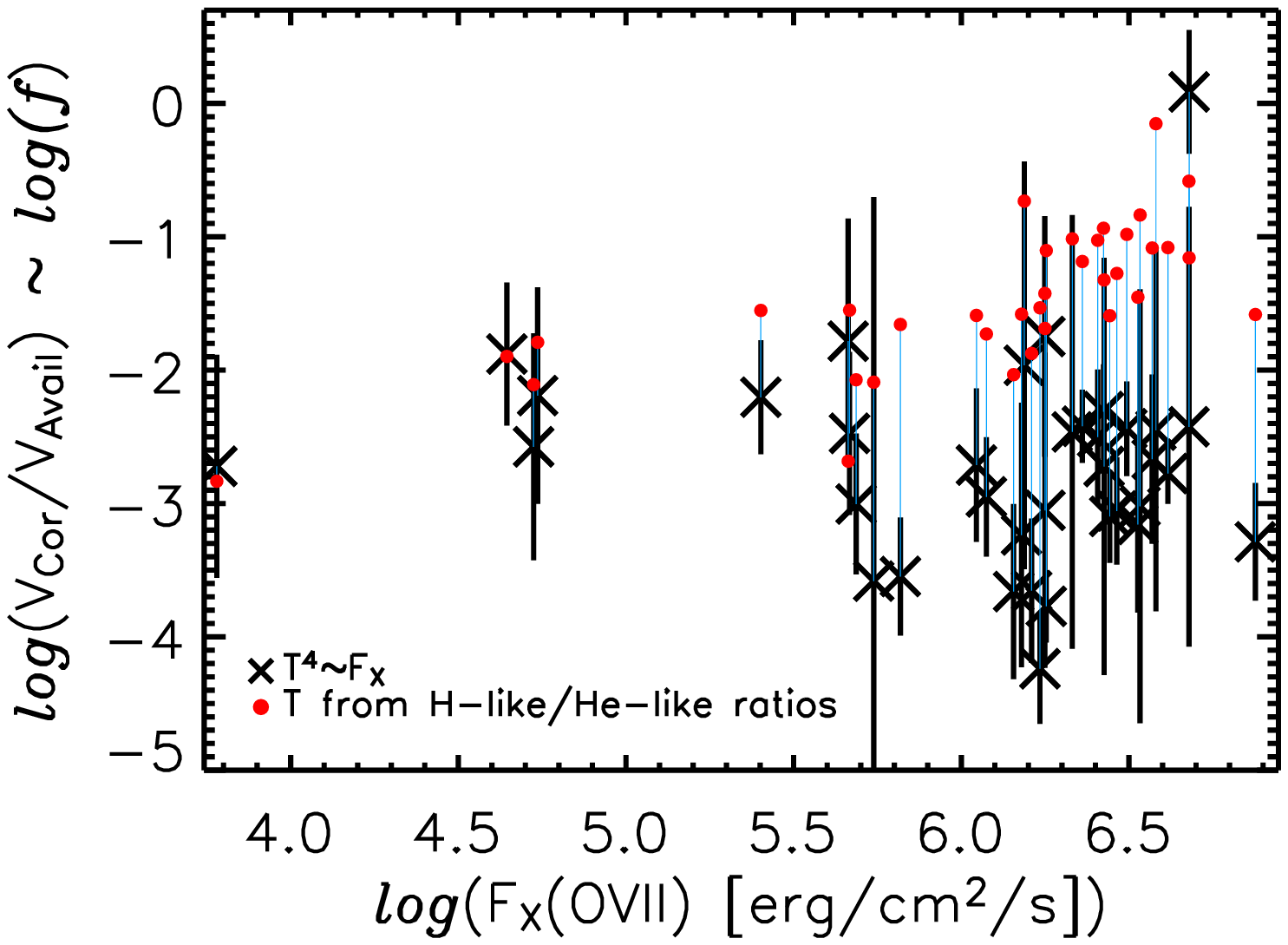}\includegraphics{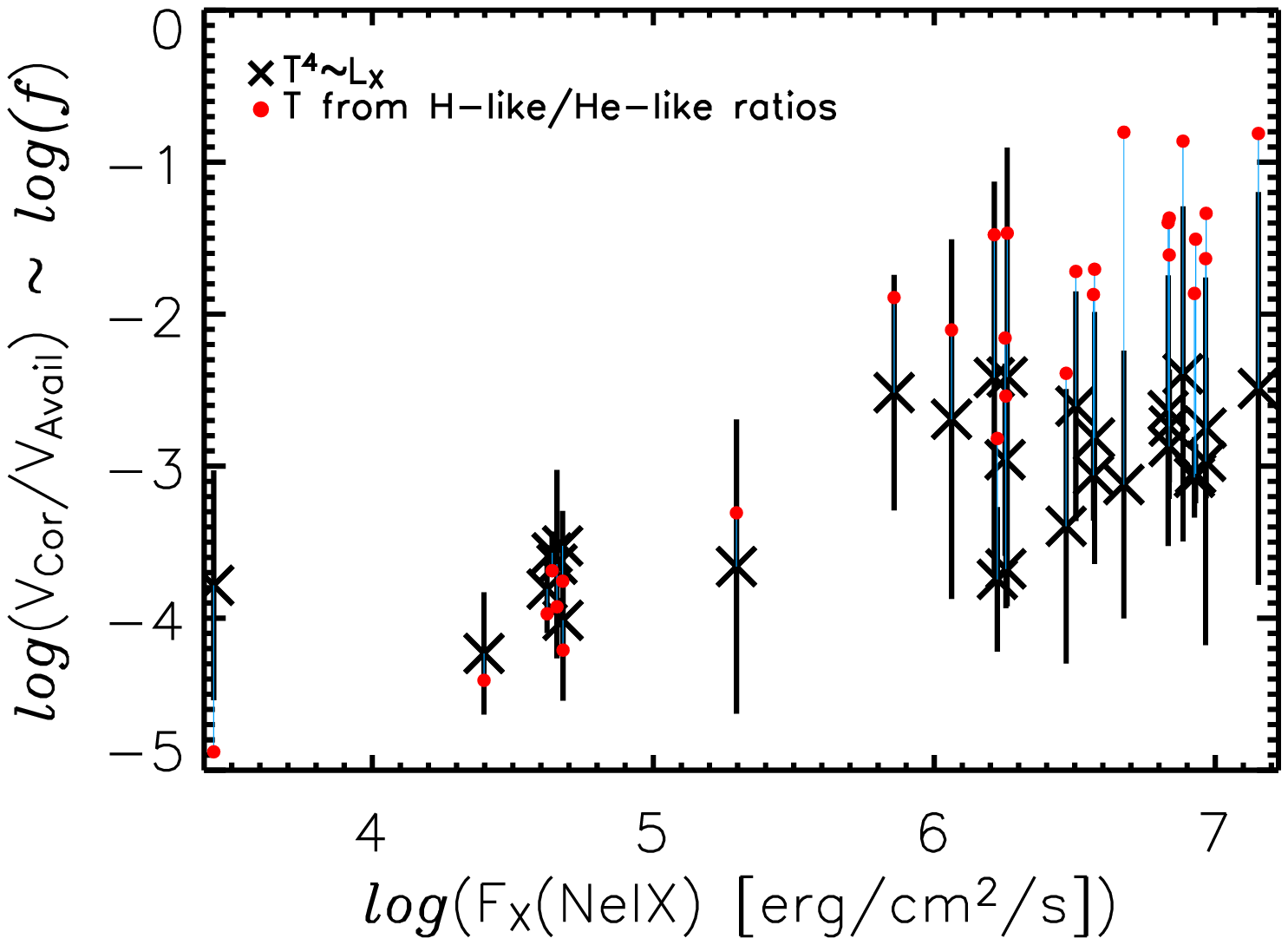}}
\caption[]{\label{fill}Filling factors obtained from ratios of coronal volumes
(derived from O\,{\sc vii} and Ne\,{\sc ix} densities) and assumed available
volumes. The latter depend on assumed loop-top temperatures, which we derive
from the H-like and He-like line ratios (marked with red bullets) and from $L_X$
(Eq.~\ref{thot_scal}). As activity indicator we use here the surface fluxes
$F_X$ from $L_X({\rm ion})/(4\pi R_\star^2)$.}
\end{figure*}

From these considerations we derive the available volume
\begin{equation}
\label{vavail}
V_{\rm avail}\stackrel{(a)}{=}4\pi R_\star^2 L\stackrel{(b)}{=}\frac{4\pi\,1.3\times 10^6 R_\star^{1/2}R_\odot^{3/2}}{n_e T_{\rm ion}}\left(\frac{L_X}{55}\right)^{3/4}
\end{equation}
in cgs units. The values we derived for $V_{\rm cor}$ and $V_{\rm avail}$ are
listed in Tables~\ref{vol}/\ref{vol_rs} and we plot these volumes in the bottom
right panels of Figs.~\ref{ox} and \ref{ne}. In addition we plot a line of equal
volumes and it can be seen that the derived coronal volumes are significantly
lower than the available volumes. One has to keep in mind that here only the
O\,{\sc vii} and Ne\,{\sc ix} emitting regions contribute to the coronal
volume while the available volume represents the maximal extent of the corona
derived from its degree of activity.
The ratio of these volumes is defined as the filling factor
\begin{equation}
f=\frac{V_{\rm cor}}{V_{\rm avail}},
\end{equation}
and we plot the derived filling factors versus activity in Fig.~\ref{fill}.
The values obtained for the filling factors are given in
Tables~\ref{vol}/\ref{vol_rs}. Upper estimates of filling factors are
calculated by using the temperatures derived from the ratios of H-like and
He-like lines in Eq.~(\ref{rtv_eq}) as $T_{\rm hot}$. Since these temperatures
are lower than the temperatures derived from Eq.~(\ref{thot_scal}), the estimated
loop lengths and thus the available volumes as calculated with Eq.~(\ref{vavail})
(a) will be smaller. Therefore an upper limit is found for the filling factors,
and those are marked with red bullets in Fig.~\ref{fill}, while all other
quantities are based on the temperatures from Eq.~(\ref{thot_scal}).

\section{Discussion and Conclusions}
\label{discussion}

For inference of sizes in stellar coronae the scaling laws derived
for the solar corona can be used to link physical properties with spatial
sizes under the assumption of loop-like geometries as identical building blocks.
The physical parameters required for the application of the scaling laws are
the plasma temperature, emission measure, and density. While good estimates of
plasma temperatures and emission measures are available from
low-resolution spectroscopy, estimates of plasma density do require
high-resolution spectroscopy. The He-like triplets provide a direct way to
measure densities by using
collisionally induced reduction of the so-called f/i line flux ratio with
increasing plasma density. The density-sensitive regimes of the different
He-like ions increase with atomic number and the best measurable density range
is provided by the O\,{\sc vii} triplet and the Ne\,{\sc ix} triplet. The
N\,{\sc vi} and C\,{\sc v} triplets are also good density tracers; however,
the lines of both are quite faint.
C\,{\sc v} can only be measured with the LETGS and can be blended with
Ne\,{\sc ix}/Fe\,{\sc xix} third order lines for the more active stars
\citep{ness_cap}. The ions of
C\,{\sc v} and N\,{\sc vi} are produced at rather low temperatures and
therefore probe only the cooler coronal plasma. The O\,{\sc vii} triplet is very
prominent and can be measured by all grating instruments with high precision,
but again it represents only the low-temperature regions of a multi-temperature
plasma. The Ne\,{\sc ix} triplet is sensitive to somewhat
higher densities and is produced at higher temperatures than O\,{\sc vii}.
However, the measurement of Ne\,{\sc ix} is difficult because of Fe\,{\sc xix}
lines blending with the Ne\,{\sc ix} lines,
particularly the important intercombination line \citep{nebr}. The most
reliable results on the Ne\,{\sc ix} lines can therefore be derived
from the HETGS data for the Ne\,{\sc ix} f/i ratios.\\

The results of our density measurements can be summarized as follows: First,
for O\,{\sc vii} the measured f/i-ratios are in the range $\approx 1$ to the
low density limit of 3.95. The coronal source with the hitherto lowest measured
O\,{\sc vii} f/i-ratio is Algol, which yields f/i$\sim 1$ for RGS, LETGS, and
MEG data, i.e., in three independent measurements. While Algol is indeed very
active, it is in our opinion very likely that these low f/i-values are affected
by the radiation field of Algol's primary. This is supported by the lack of
such low f/i-ratios for Ne\,{\sc ix} \citep[for higher-Z He-like ions the
effects from UV radiation fields become lower; c.f. Fig.~8 in][]{ness_10}.
Second, for Ne\,{\sc ix} the measured f/i-ratios are in the range $\approx 1$
to the low density limit of 3.4. A handful of stars like 44 Boo, Algol,
AR Lac, and EV Lac have the lowest values, but discrepancies appear between
measurements with different instruments and even between simultaneous
measurements (in MEG and HEG). Third, in no case do we have significant density
measurements from Fe\,{\sc xxi}. Although all measured line ratios should yield
consistent densities, we found that none of our spectra returned consistently
high densities. We further found no detections for lines that typically appear
in high density plasmas. From the upper limits
of our Fe\,{\sc xxi} density estimates the typically encountered densities
in coronal plasmas are definitely not higher than
$5\times 10^{12}$\,cm$^{-3}$. Therefore the Si\,{\sc xiii} and the Mg\,{\sc xi}
triplets will yield only low-density limits in coronal plasmas.
Recent systematic studies of Si and Mg He-like f/i ratios by \cite{testa04}
indeed revealed only low-density limits for Si\,{\sc xiii} for all stars in
their sample,
but some density measurements for Mg\,{\sc xi} are also reported. Deviations
from the Si f/i low-density limit (systematically higher f/i values as expected)
were argued to imply too low a theoretical low-density f/i value.
\cite{testa04} point out that densities in stellar coronae do not exceed
$\log n_e=13$ as reported by, e.g., \cite{jsf03AB} for AB\,Dor. The
deviations found for Mg f/i ratios were found to be related to the ratio of
X-ray luminosity and bolometric luminosity, but no discernible trend with the
X-ray surface flux was found. A particular difficulty was that the Mg\,{\sc xi}
lines are blended with lines of the Ne Ly series ($n>5$), increasing the formal
measurement errors. \cite{testa04} successfully disentangled the lines, but
admit that residual Ne blending might still be present.

It is instructive to inspect the ``low'' f/i-ratios for O\,{\sc vii} and
Ne\,{\sc ix} where measurements with good SNR and high resolution (i.e., MEG)
are available. In Table~\ref{fi_best} we list only those measurements
where the oxygen f/i-ratios are below 2.2 (within the errors) and the neon
f/i-ratios are below 2.0. The peculiar role of Algol becomes apparent; its low
O\,{\sc vii} f/i-ratio stands out, while the Ne\,{\sc ix} f/i-ratio does not.
The densities derived from O\,{\sc vii} and Ne\,{\sc ix} usually differ, the
Ne\,{\sc ix} densities being higher than the corresponding O\,{\sc vii}
densities. However, using the MEG values the densities for 44 Boo,
EV Lac, and II Peg are consistent, while they are inconsistent
for the RS CVn stars Capella and $\sigma$ CrB. The case of Capella appears
especially striking: While three different measurements (with RGS, LETGS, MEG)
yield consistent ``high'' values of the O\,{\sc vii} f/i-ratio, both MEG and
HEG yield consistent ``low'' values for the Ne\,{\sc ix} f/i-ratio. The
Ne\,{\sc ix} measurement for Capella has been discussed in great detail by
\cite{nebr} and it was found that the intercombination line could be blended
with an additional Fe\,{\sc xix} line that cannot be resolved with MEG and HEG.
In the case of Capella it turned out that accounting for the predicted amount
of blending leads to the low-density limit, thus to densities consistent with the
densities obtained from O\,{\sc vii}. We note that the densities (for Capella)
derived from Ne\,{\sc ix} are fully consistent with the upper limits derived
from Fe\,{\sc xxi} (even without accounting for theoretically
predicted blending); at any rate, the case of Capella (and possibly that of
$\sigma$ CrB) appears somewhat peculiar.

\begin{table}[!ht]
\begin{flushleft}
\renewcommand{\arraystretch}{1.1}
\caption{\label{fi_best}Coronae with lowest measured f/i ratios (oxygen: f/i $<2.2$, neon: f/i $<2.0$). (H=HEG, M=MEG); for neon the predicted blending is taken into account (Sect.~\ref{neblsect})}
\vspace{-.4cm}
{\scriptsize
\begin{tabular}{p{.8cm}p{.1cm}p{1.3cm}p{1.3cm}ccc}
&&\multicolumn{2}{c}{O\,{\sc vii}}&\multicolumn{2}{c}{Ne\,{\sc ix}}\\
 star & Instr.& f/i & log($n_e$) & f/i& log($n_e$)\\
&&&[cm$^{-3}$]&&[cm$^{-3}$]\\
         44\,Boo&     M&\mbox{$   1.69\,\pm\,0.42$}&\mbox{  $10.6\,\pm\,0.19$}&$   2.89\,\pm\,0.39$&  $11.0\,\pm\,0.50$\\
 &     H&   --&                 --&   2.42\,$\pm$\,   0.47&  $11.4\,\pm\,0.32$\\
           Algol&     M&\mbox{$   0.69\,\pm\,0.19$}&\mbox{  $10.9\,\pm\,0.29$}&$   1.89\,\pm\,0.22$&  $11.7\,\pm\,0.11$\\
 &     H&   --&                 --&   1.68\,$\pm$\,   0.39&  $11.8\,\pm\,0.19$\\
         EV\,Lac&     M&\mbox{$   1.52\,\pm\,0.33$}&\mbox{  $10.7\,\pm\,0.15$}&$   2.92\,\pm\,0.41$&  $11.0\,\pm\,0.60$\\
 &     H&   --&                 --&   3.32\,$\pm$\,   0.73&          $<\,11.5$\\
         II\,Peg&     M&\mbox{$   1.49\,\pm\,0.40$}&\mbox{  $10.7\,\pm\,0.20$}&$   2.75\,\pm\,0.41$&  $11.2\,\pm\,0.40$\\
 &     H&   --&                 --&   2.70\,$\pm$\,   0.66&  $11.2\,\pm\,0.83$\\
      $\mu$\,Vel&     M&\mbox{$   7.24\,\pm\,5.24$}&          $<\,11.5$&$   1.50\,\pm\,0.31$&  $11.8\,\pm\,0.16$\\
 &     H&   --&                 --&   1.57\,$\pm$\,   0.40&  $11.8\,\pm\,0.21$\\
 $\sigma^2$\,CrB&     M&\mbox{$   2.07\,\pm\,0.39$}&\mbox{  $10.4\,\pm\,0.17$}&$   2.03\,\pm\,0.17$&  $11.6\,\pm\,0.08$\\
 &     H&   --&                 --&   1.87\,$\pm$\,   0.22&  $11.7\,\pm\,0.11$\\
         TY\,Pyx&     M&\mbox{$   1.29\,\pm\,0.76$}&\mbox{  $10.8\,\pm\,0.49$}&$   3.53\,\pm\,0.89$&          $<\,12.2$\\
 &     H&   --&                 --&   2.95\,$\pm$\,   1.29&          $<\,12.0$\\
       V824\,Ara&     M&\mbox{$   1.51\,\pm\,0.46$}&\mbox{  $10.7\,\pm\,0.23$}&$   2.24\,\pm\,0.25$&  $11.5\,\pm\,0.13$\\
 &     H&   --&                 --&   2.75\,$\pm$\,   0.54&  $11.2\,\pm\,0.63$\\
\hline
\end{tabular}
}
\renewcommand{\arraystretch}{1}
\end{flushleft}
\end{table}

With the measured densities of the cool and hot
plasma component we computed the emitting volumes of these plasma components
(cf. Tables~\ref{vol} and \ref{vol_rs}). Surprisingly, these volumes
are rather small, for the ``best'' data sets with the smallest
errors one finds volumes between 10$^{29} - 10^{30.5}$\,cm$^3$
for O\,{\sc vii} and somewhat smaller values for Ne\,{\sc ix}.
For example, for EV Lac we find V$_{cor} = 10^{28.9}$\,cm$^3$ for oxygen
and neon; assuming a filling factor of unity, one would obtain a coronal
scale height of 100\,km, which appears pathologically small.
We therefore conclude that the filling factor is far away from unity,
and that conclusion is substantiated by more sophisticated calculations
of the coronal volume. We first calculated the volume of a maximum corona
consisting of hydrostatic loops with isothermal temperatures derived from
the Ly$_\alpha$/He-like line ratios for the respective ions and
the measured density, and second, computed maximal coronal
volumes from the relationship between the temperature of the hotter
plasma and the total stellar X-ray luminosity reported by \cite{guedel97} and,
again, find small overall filling factors.

So far, the discussion assumed isolated temperature components. Under
reasonable physical conditions one expects the pressure to stay
approximately constant along any magnetic field line since the high observed
temperatures imply large pressure scale heights. Under this assumption
the high-density Ne\,{\sc ix} emission regions cannot be magnetically connected
with the high-density O\,{\sc vii} emission regions because the Ne densities
would have to be below the O densities, the opposite of what is observed.
On the other hand, assuming isobaric loops, any loop emitting
in Ne\,{\sc ix} will also emit in the O\,{\sc vii} lines; also, the density of
these O\,{\sc vii} layers will be even larger than those of the Ne\,{\sc ix}
emission layers.
Therefore the observed O\,{\sc vii} emission would have to be composed of at
least two components, a high-temperature, high-density component, which
contributes rather little to the observed flux in the forbidden
O\,{\sc vii} line, and a lower-temperature, lower-density component, which
contributes to the bulk of the forbidden O\,{\sc vii} line.
A detailed modeling in terms of physically consistent loops is beyond the
scope of this paper; here, we just consider
the following numerical example of EV Lac. The observed ratios
of the fluxes of the He-r line of Ne\,{\sc ix} and the Ly$_{\alpha}$ lines
imply temperatures of about log$T \approx 6.7$, the observed
Ne f/i-ratio of 2.92 implies densities of about $10^{11.1}$\,cm$^{-3}$.
The O\,{\sc vii} emission would then be located at a
density of $10^{11.6}$\,cm$^{-3}$, which in turn would lead to an
O\,{\sc vii} f/i-ratio of 0.29 for this material. Decomposing the observed
O\,{\sc vii} f and i emission into a ``hot'' component (with f/i$ = 0.29$) and
a ``cool'' component (with an assumed low-density limit of f/i$ = 3.95$),
so that the overall f/i-ratio is equal to the
observed f/i-ratio of 1.52, leads to the following numbers:
f$_c = 46.6$, i$_c = 11.8$, f$_h = 8.3$, i$_h = 28.5$, i.e., a situation
where the f-line is dominated by the low-pressure component, but the
i-line by the high-pressure component. The filling factor of the
low-pressure component is undetermined and could be large, while the
filling factor of the high-pressure component, which contributes
most of the flux, would definitely be quite small and dominant
by virtue of its high density.

What, then, do we learn about coronal structure? First, an inactive
corona is generally dominated by cool plasma (1--4\,MK), and this plasma
never covers a large fraction of the surface as is the case for the
non-flaring Sun. For active stars, cool plasma might actually
cover a larger fraction of the stellar surface, but more strikingly,
a hot component appears. This hot component has been known since the
earliest stellar X-ray observations, but its nature has been debated.
Its characteristic temperature seems to be correlated with the activity
level, and in extremely active stars it reaches temperatures that on the
Sun are known exclusively during flares. In our investigation, we cannot
confine the extent of the hot plasma because we do not find conclusive
indications for a definitive density of this plasma component. We rather
argue that we measure upper limits. Yet we have added two further pieces
of information: First, the cooler plasma component cannot cover a large
fraction of the stellar surface. And second, there must be spatially
separate plasma components at different temperatures (e.g., those
detected by the O\,{\sc vii} and the Ne\,{\sc ix} density analysis). We
tentatively argue that the hotter plasma loops fill the space between
cooler loops until much of the corona is dominated by the hot plasma.
Why, then, do hot loops become progressively more important as the stellar
activity increases? As the magnetic activity level and consequently 
the surface magnetic filling factor increases, the coronal magnetic fields
become denser, leading to increased interactions between neighboring field
lines, which leads to increased heating \citep{guedel97}. The increased
heating rate inevitably drives chromospheric material into the loops until
an equilibrium is attained. The X-ray luminosity of the hot plasma thus
rapidly increases as we move to more active stars. In their most extreme
form, such interactions lead to increased levels of flaring, again resulting
in increased amounts of hot, luminous plasma. In this picture, the cooler
loops are post-flare loops that are still over-dense while returning to 
their equilibrium state. 

Whatever the cause for the increased heating, a relatively cool component,
corresponding to typical active regions as seen on the Sun, appears to be
present in all stars at a similar level of surface coverage. Our survey has
shown that this component reveals densities that may exceed
$10^{10}$\,cm$^{-3}$ but a clear systematic trend with the overall activity
level does not seem to be present.

\begin{acknowledgements}
This work is based on observations obtained with Chandra and XMM-Newton, an
ESA science mission with instruments and contributions directly funded by ESA
Member States and the USA (NASA).\\
This research has made use of the SIMBAD database, operated at CDS, Strasbourg, France\\
J.-U.N. acknowledges financial support from Deutsches Zentrum f\"ur Luft- und
Raumfahrt e.V. (DLR) under 50OR98010. AT and MG acknowledge support from the Swiss National Science Foundation (grant no. 2000-066875).

\end{acknowledgements}

\bibliographystyle{aa}
\bibliography{jhmm,astron,jn}

\end{document}